\documentclass[10pt]{article} 
\usepackage{amssymb,euscript,latexsym,amsmath} 
\usepackage{graphicx} 
\usepackage{epstopdf}
\usepackage{bm} 
\usepackage{enumerate} 
\usepackage{color} 
\usepackage{setspace} 
\usepackage{subfigure} 
\usepackage{hyperref}
\usepackage{array}
\usepackage{cancel}

\begin{document}
\title{Stability of Underwater Periodic Locomotion}
\author{Fangxu Jing and Eva Kanso}
\maketitle

\begin{abstract}
Most aquatic vertebrates swim by lateral flapping of their bodies and caudal fins. While much effort has been devoted to understanding the flapping kinematics and its influence on the swimming efficiency, little is known about the stability  (or lack of) of periodic swimming. 
It is believed that 
stability limits maneuverability and body designs/flapping motions that are adapted for stable swimming are not suitable for high maneuverability and vice versa.
In this paper, we consider a simplified model of a planar elliptic body undergoing prescribed periodic heaving and pitching in potential flow. We show that periodic locomotion can be achieved due to the resulting hydrodynamic forces, and its value depends on several parameters including the aspect ratio of the body, the amplitudes and phases of the prescribed flapping. We obtain closed-form solutions for the locomotion and efficiency for small flapping amplitudes, and numerical results for finite flapping amplitudes. We then study the stability of the (finite amplitude flapping) periodic locomotion using Floquet theory. We find that stability depends nonlinearly on all parameters. Interesting trends of switching between stable and unstable motions emerge and evolve as we continuously vary the parameter values. This suggests that, for live organisms that control their flapping motion, maneuverability and stability need not be thought of as disjoint properties, rather the organism may manipulate its motion in favor of one or the other depending on the task at hand. 
\end{abstract}

\section{Introduction} 
\label{sec:intro}

A large proportion of fish species are characterized by elongated bodies that swim forward by flapping sideways. These sideways oscillations produce periodic propulsive forces that cause the fish to swim along time-periodic trajectories,~\cite{Lighthill1970}. The kinematics of the flapping motion and the resulting swimming performance, as well as their relationship to the swimmer's morphology, have been the subject of numerous studies, see, for example,~\cite{Wu2011, Eloy2013}. However, little attention has been given to the stability of underwater locomotion. The importance of motion stability and its mutual influence on body morphology and behavior is noted in the work of Weihs, see~\cite{Weihs2002} and references therein. Weihs uses clever arguments and simplifying approximations founded on a deep understanding of the equations governing underwater locomotion to obtain ``educated estimates" of the stability of swimming fish without ever solving the complicated set of equations. 

The swimming motion is said to be unstable if a perturbation in the conditions surrounding the swimmer's body result in forces and moments that tend to increase the perturbation, and it is stable if these emerging forces tend to reduce such perturbations or keep them bounded so that the fish returns to or stays near its original periodic swimming. 

Stability may be achieved actively or passively. Active stabilization requires neurological control that activate musculo-skeletal components to compensate for external perturbations acting against stability. On the other hand, passive stability of the locomotion gaits requires no additional energy input by the fish. In this sense, one can argue that stability reduces the energetic cost of locomotion. 
Therefore, from an evolutionary perspective, it seems reasonable to conjecture that stability would have a positive selection value in behaviors such as migration over prolonged distances and time. However, stability limits maneuverability and body designs/flapping motions that are adapted for stable swimming are not suitable for high maneuverability and vice versa,~\cite{Weihs2002, Weihs1993}. 

In this work, we study stability of periodic swimming using a simple model consisting of a planar elliptic body undergoing prescribed flapping motion in unbounded potential flow. By flapping motion, we mean periodic heaving and pitching of the body as shown in Figure~\ref{fig:model}. We formulate the equations of motion governing the resulting locomotion and examine its efficiency. We then investigate the stability of this motion using Floquet theory (see~\cite{JoSm2007}). We find that stability depends in a non-trivial way on the body geometry (aspect ratio of the ellipse) as well as on the amplitudes and phases of the flapping motion. Most remarkable is the ability of the system to transition from stability to instability and back to stability as we vary some of these parameters. 

This model is reminiscent of the  three-link swimmer used by Kanso {\em et al.} to examine periodic locomotion in potential flow, see~\cite{KaMaRoMe2005}. The three-link swimmer undergoes periodic shape deformations that result in coupled heaving, pitching and locomotion. Here, we ignore body deformations for the sake of simplicity and prescribe the heaving and pitching motion directly. Note that the three-link swimmer was also used by Jing \& Kanso to study the effect of body elasticity on the stability of the coast motion of fish (motion at constant speed). They found that elasticity of the body may lead to passive stabilization of the (otherwise unstable) coast motion, see~\cite{Jing2011,JiKa2012}. The present model consisting of a single elliptic body is mostly similar to the system studied by Spagnolie {\em et al.} (2010) both experimentally and numerically, see~\cite{SpMoShZh2010}. In the latter, an elliptic body undergoes passive pitching (via a torsional spring) subject to prescribed periodic heaving in viscous fluid, whereas in our model both the pitching and heaving motions are prescribed and the fluid medium is inviscid. Despite these differences, the two models exhibit qualitatively similar behavior as discussed in Section~\ref{sec:discussion_and_conclusion}.

The paper is organized as follows. In Section~\ref{sec:problem_setup}, we formulate the equations of motion
governing the locomotion of a periodically flapping body in unbounded potential flow. 
We analyze the body's locomotion and efficiency when subject to small amplitude flapping motion in Section~\ref{sec:small_amplitude_actuations}, and consider the more general case of finite amplitude flapping in Section~\ref{sec:locomotion}.  In Section~\ref{sec:stability_of_periodic_solution}, we assess the stability of the periodic  locomotion using Floquet theory. The main findings and their relevance to biolocomotion are discussed in Section~\ref{sec:discussion_and_conclusion}.

\section{Problem Formulation}
\label{sec:problem_setup}

\begin{figure}[!t]
	\centering
		\includegraphics[width=0.4\textwidth]{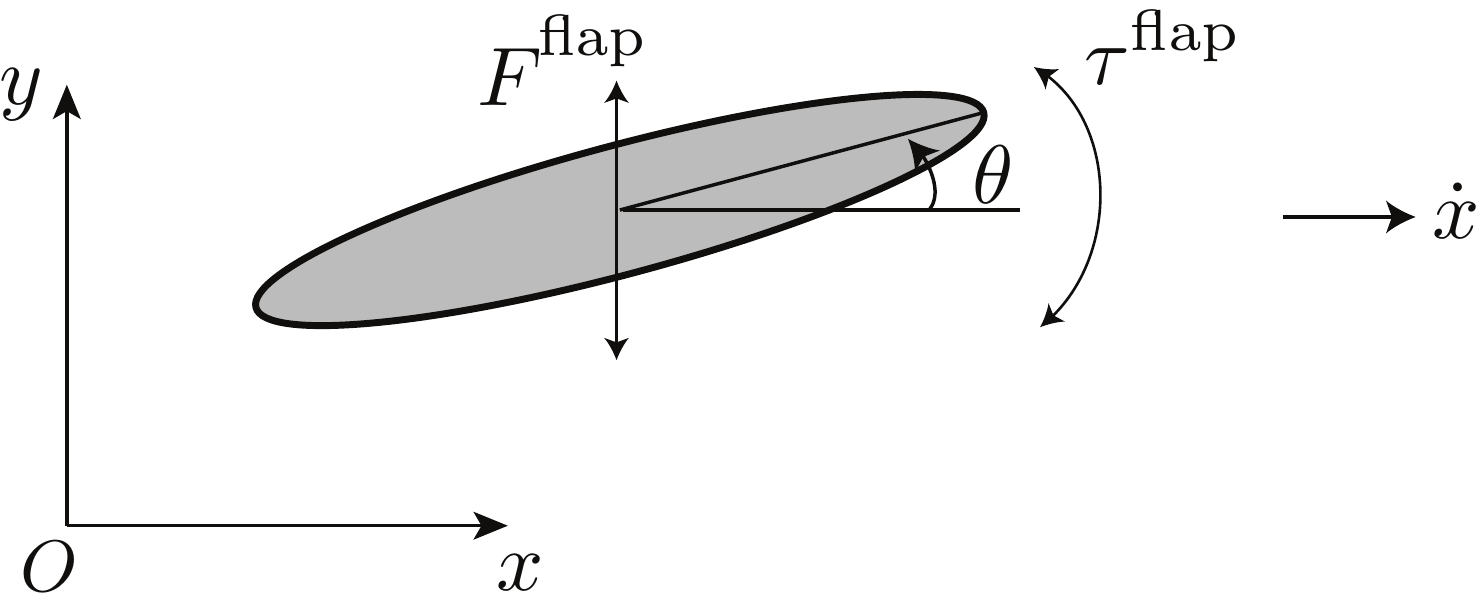}
	\caption{\footnotesize Model of flapping fish: An ellipse with semi-axes $a$ and $b$ is submerged in unbounded potential fluid. Motion is observed in inertial frame with position of mass center given by $(x, y)$ and orientation by $\theta$. Fish flaps in $y$- and $\theta$-directions, and propels in $x$-direction.}
	\label{fig:model}
\end{figure}

\par\noindent
Consider a planar elliptic body with semi-major axis $a$ and semi-minor axis $b$, submerged in unbounded  potential flow that is at rest at infinity. The elliptic body is neutrally buoyant, that is to say, the body and fluid densities are equal to $\rho$. Its mass is given by $m_b = \rho \pi a b$, and its moment of inertia about the center of mass $C$ is $J_b = m_b(a^2 + b^2)/4$. Let $(x , y)$ denote the position of the mass center with respect to a fixed inertial frame and let $\theta$ denote the orientation angle of the ellipse measured from the positive $x$-direction to the ellipse's major axis, see Figure~\ref{fig:model}. The linear and angular velocities are given by $(\dot{x} , \dot{y})$ and $\dot{\theta}$, respectively, where the dot $\dot{()}$ correspond to derivative with respect to time $t$.

In order to emulate the flapping motion of a swimming body, we assume that $y$ and $\theta$ vary periodically in time due to some periodic flapping force $F^{\rm flap}$ and flapping moment $\tau^{\rm flap}$ generated by the swimming body. Note that in the case of a body swimming by deforming itself, $y$ and $\theta$ are a result of the body deformation. Here, we do not account for the body deformation but rather prescribe $y(t)$ and $\theta(t)$ as periodic functions of time. Namely, we set
\begin{equation}
	y(t) = A_y \sin (\omega t + \phi_y),\qquad 
	\theta(t) = A_{\theta} \sin(\omega t + \phi_{\theta}). \label{eq:prescribed}
\end{equation}
and solve for the resulting locomotion in the $x$-direction. 

The equations governing the motion of the flapping body are basically Kirchhoff's equations expressed in inertial frame and subject to forcing $F^{\rm flap}$ and $\tau^{\rm flap}$ in the $y$- and $\theta$-directions, that is,
\begin{equation}	
		m_b\, \ddot{x} = F_x, \label{eq:periodiceomX}
\end{equation}
and
\begin{equation}	
		m_b\, \ddot{y}   =  F_y + F^{\rm flap},  \qquad
	 	J_b\, \ddot{\theta}   = \tau + \tau^{\rm flap}, \label{eq:periodiceom}
\end{equation}
where $F_x$, $F_y$ and $\tau$ are the hydrodynamic forces and moment acting on the body. For motions in potential flow,
 $F_x$, $F_y$ and $\tau$ can be obtained using a classic procedure,
 \begin{equation}\label{eq:hydroforcestorque}
	\begin{split}
		F_x & = \frac{1}{2}\left[ - (m_1+m_2) + (m_2-m_1) \cos2\theta\right]\ddot{x} + \frac{1}{2}(m_2-m_1) \ddot{y} \sin2\theta - (m_2-m_1)  (\dot{x} \sin 2 \theta - \dot{y}\cos 2\theta)\dot{\theta},\\[1ex]
		F_y & =   \frac{1}{2}\left[ -(m_1+m_2) - (m_2-m_1)  \cos2\theta\right]\ddot{y}  + \frac{1}{2}(m_2-m_1)  \ddot{x} \sin2\theta + (m_2-m_1)  (\dot{x} \cos 2 \theta + \dot{y} \sin 2\theta)\dot{\theta},\\[1ex]
		\tau & = -J \ddot{\theta}  + \frac{1}{2}(m_2- m_1) \left(\dot{x}^2\sin 2\theta - \dot{y}^2 \sin 2\theta - 2 \dot{x}\dot{y} \cos 2\theta \right).
	\end{split}
\end{equation}
Here  $m_1 = \rho \pi b^2$, $m_2=\rho \pi a^2$ are, respectively, the added mass of the elliptic body along its major and minor directions and $J = \rho \pi (a^2-b^2)^2/8$ is the added moment of inertia, see, e.g.,~\cite{Newman1977}. Substituting~\eqref{eq:hydroforcestorque} into~\eqref{eq:periodiceomX} and~\eqref{eq:periodiceom}, one  can use~\eqref{eq:periodiceomX} to solve for $x(t)$, and~\eqref{eq:periodiceom} to compute the forcing $F^{\rm flap}$ and $\tau^{\rm flap}$ needed in order for the body to achieve the prescribed flapping motion in~\eqref{eq:prescribed}.

The total angular momentum $h$ of the body-fluid system is given by $h =  (J_b + J)\dot{\theta}$ whereas the total linear momentum can be written as
\begin{equation}
	\begin{split}
		 \begin{pmatrix}
			p_x\\
			p_y
		\end{pmatrix} & = m_b \begin{pmatrix}
			\dot{x}\\
			\dot{y}
		\end{pmatrix} + \begin{pmatrix}
			\cos\theta & -\sin\theta\\
			\sin\theta & \cos\theta
		\end{pmatrix}
		\begin{pmatrix}
			m_1 & 0\\
			0 & m_2
		\end{pmatrix}
		\begin{pmatrix}
				\cos\theta & \sin\theta\\
				-\sin\theta & \cos\theta
			\end{pmatrix}
			\begin{pmatrix}
					\dot{x}\\
					\dot{y}
				\end{pmatrix}.
	\end{split}\label{eq:momenta}
\end{equation}
The momentum  $p_x$  is conserved since there is no external forcing applied in the $x$ direction. Therefore, one has $p_x(t)
=p_x(t = 0)$, which yields
\begin{equation}
	\dot{x}(t) = \frac{m_1 - m_2}{2\left(m_b + m_1 \cos^2\theta + m_2 \sin^2\theta\right)} \left[(\dot{y} \sin2\theta )_{t = 0} - (\dot{y} \sin2\theta)\right]. \label{eq:xdot}
\end{equation}
That is to say, equation~\eqref{eq:periodiceomX} admits an integral of motion whose value is given by the above equation.
The distance traveled by the body's center of mass in one period of flapping, $T = 2\pi/\omega$, is given by
\begin{equation}
	d = \left|\int_0^T \dot{x}\, \text{d}t\right| = |x(T) - x(0)|. \label{eq:d}
\end{equation}
The total kinetic energy $E$ of the body-fluid system is given by
\begin{equation}
	E = \frac{1}{2} (\dot{x}\, p_x +  \dot{y}\, p_y) + \frac{1}{2}(J_b + J)\dot{\theta}^2.\label{eq:energy}
\end{equation}
By the work-energy theorem, the time derivative of the kinetic energy is  equal to the total power input by the flapping force $F^{\rm flap}$ and moment $\tau^{\rm flap}$.
Thus, the work done by flapping is equivalent to the kinetic energy $E$. To this end, the average work done in one period is given by
\begin{equation}
	\bar{E} = \frac{1}{T}\int_0^T E(t) \text{d}t. \label{eq:aveE}
\end{equation}
We define the {\em cost of locomotion} $e$ as the average work divided by the average distance over one period, namely
\begin{equation}
	e = \dfrac{\bar{E}}{d}. \label{eq:e}
\end{equation}
Hence, smaller $e$ means less energy expenditure for a fixed distance traveled. It is convenient to denote the {\em efficiency} of the system $\eta$ as the inverse of the cost of locomotion, that is $\eta \equiv 1/e = d/\bar{E}$.

Before we proceed to examining the locomotion and efficiency of such swimmer, we non-dimensionalize the system by scaling time with $T$, length with $\sqrt{ab}$ and mass with $m_b = \rho \pi ab$. The variables  are subsequently written in dimensionless form.
The important parameters for this system are: aspect ratio $\gamma \equiv a/b$, as well as the flapping amplitudes $A_y$, $A_\theta$, and phases $\phi_y$ and $\phi_\theta$.

\section{Small Flapping Amplitudes}
\label{sec:small_amplitude_actuations}

\begin{figure}[!htb]
	\centering
		\includegraphics[width=0.55\textwidth]{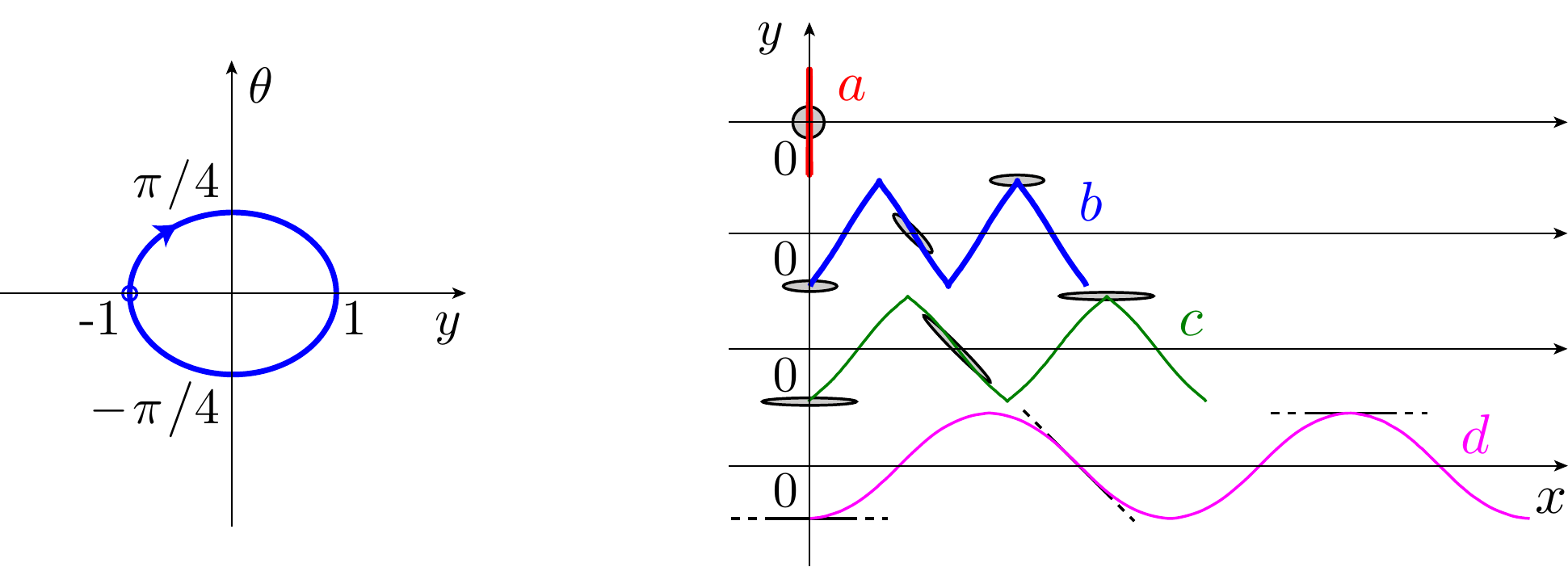}
	\caption{\footnotesize (color online) Left: Prescribed flapping in $(y, \theta)$ plane. Initial points are marked by $\circ$. Right: Trajectories of mass center in $(x,y)$ plane with snapshots of body in motion overlaid. Simulations are for $A_y = 1, A_\theta = \frac{\pi}{4}, \phi_y = -\frac{\pi}{2}, \phi_\theta = 0$ and various aspect ratios: (a) $\gamma = 1.01$, (b) $\gamma = 4$, (c) $\gamma = 8$, (d) $\gamma = 1000$.}
	\label{fig:periodicmotionGamma}
\end{figure}
\begin{figure}[!tb] 
	\centering
		\includegraphics[width=0.55\textwidth]{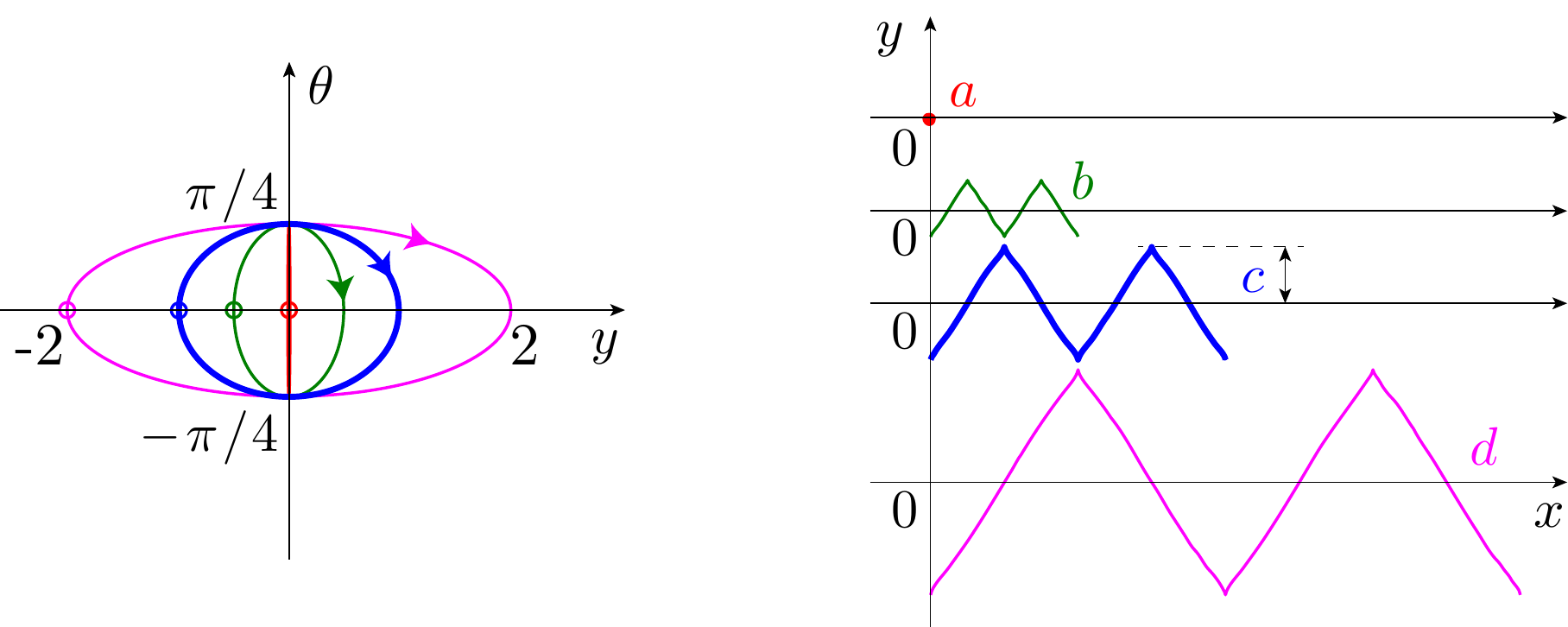}
	\caption{\footnotesize (color online) Left: Prescribed flapping in $(y, \theta)$ plane. Initial points are marked by $\circ$. Right: Trajectories of mass center in $(x,y)$ plane. Simulations are for $A_\theta = \frac{\pi}{4}, \gamma = 4, \phi_y = -\frac{\pi}{2}, \phi_\theta = 0$ and various heaving amplitudes: (a) $A_y = 0.01$, (b) $A_y = 0.5$, (c) $A_y = 1$, (d) $A_y = 2$.}
	\label{fig:periodicmotionAy}
\end{figure}
\begin{figure}[!tb] 
	\centering
		\includegraphics[width=0.45\textwidth]{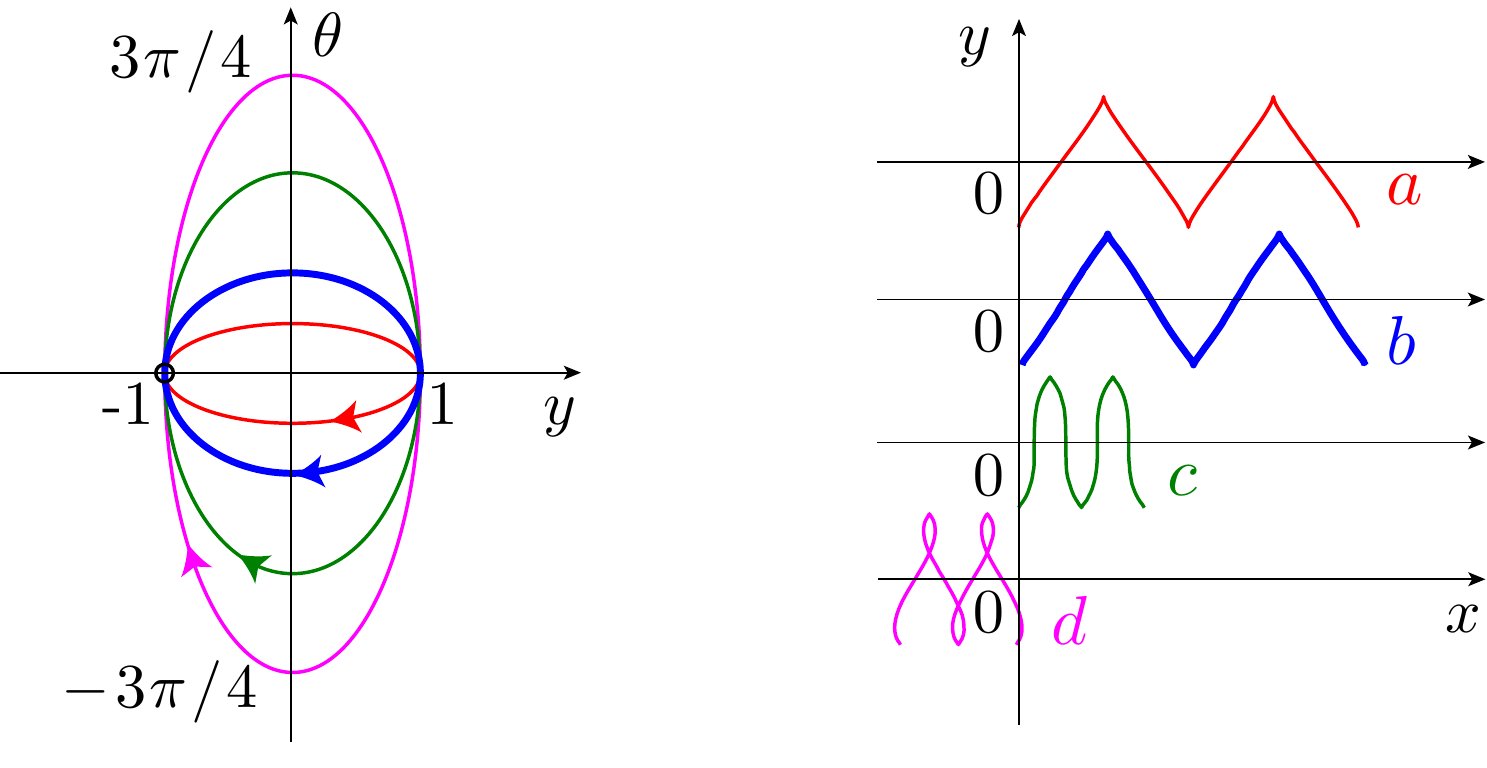}
	\caption{\footnotesize (color online) Left: Prescribed flapping in $(y, \theta)$ plane. Initial points are marked by $\circ$. Right: Trajectories of mass center in $(x,y)$ plane. Simulations are for $A_y = 1, \gamma = 4, \phi_y = -\frac{\pi}{2}, \phi_\theta = 0$ and various pitching amplitudes: (a) $A_\theta = \pi/8$, (b) $A_\theta = \pi/4$, (c) $A_\theta = \pi/2$, (d) $A_\theta = 3\pi/4$.}
	\label{fig:periodicmotionAtheta}
\end{figure}
\begin{figure}[!tb] 
	\centering
		\includegraphics[width=0.75\textwidth]{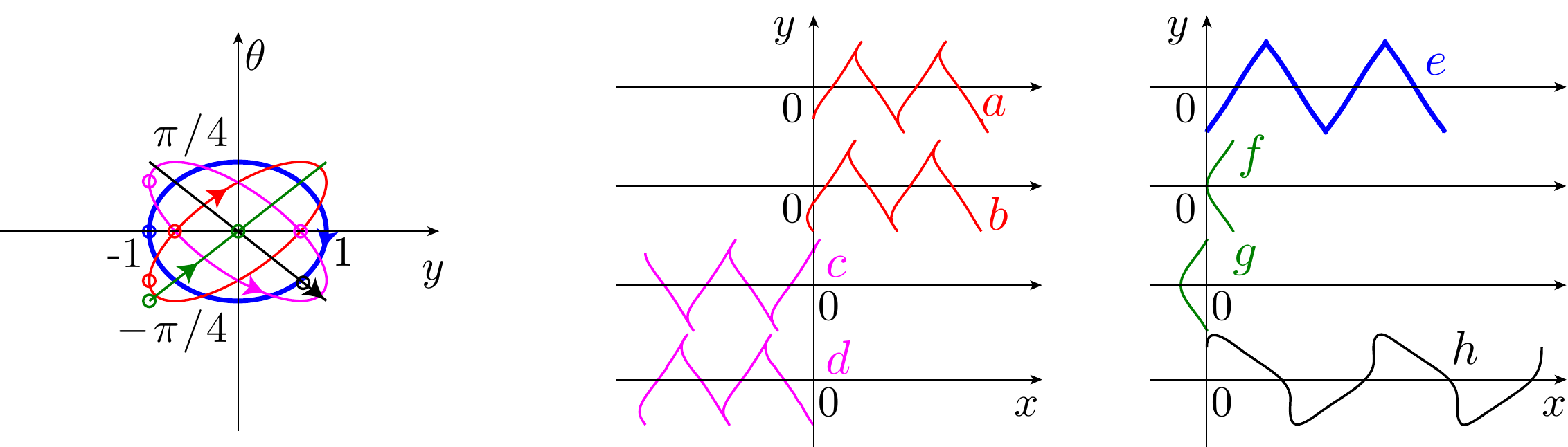}
	\caption{\footnotesize (color online) Left: Prescribed flapping in $(y, \theta)$ plane. Initial points are marked by $\circ$. Middle and right: Trajectories of mass center in $(x,y)$ plane. Simulations are for $A_y = 1, A_\theta = \frac{\pi}{4}, \gamma = 4$ and various combinations of phases: (a) $(\phi_y , \phi_\theta) = (-\frac{\pi}{4} , 0)$, (b) $(\phi_y , \phi_\theta) = (-\frac{\pi}{2} , -\frac{\pi}{4})$, (c) $(\phi_y , \phi_\theta) = (\frac{\pi}{4} , 0)$, (d) $(\phi_y , \phi_\theta) = (-\frac{\pi}{2} , -\frac{3\pi}{4})$, (e) $(\phi_y , \phi_\theta) = (-\frac{\pi}{2} , 0)$, (f) $(\phi_y , \phi_\theta) = (0 , 0)$, (g) $(\phi_y , \phi_\theta) = (-\frac{\pi}{2} , -\frac{\pi}{2})$, (h) $(\phi_y , \phi_\theta) = (\frac{\pi}{4} , -\frac{3\pi}{4})$.}
	\label{fig:periodicmotionPhi}
\end{figure}

Consider the case with small flapping amplitudes $A_y$ and $A_\theta$. Let
$A_y \equiv \epsilon_y \ll 1$ and $A_\theta \equiv \epsilon_\theta \ll 1$ where both $\epsilon_y$ and $\epsilon_\theta$ are of the same order of magnitude.
One gets $y, \dot{y}, \ddot{y} \sim O(\epsilon_y)$ and $\theta, \dot{\theta}, \ddot{\theta} \sim O(\epsilon_\theta)$,
but $\omega, \phi_y$ and $\phi_\theta$ are not necessarily small. Use the approximation $\cos\theta \approx 1$ and $\sin\theta \approx \theta$ and substitute into~\eqref{eq:xdot} to obtain
\begin{equation}
	\dot{x} \approx \frac{1}{2} \left(\gamma - 1\right) \omega \epsilon_y \epsilon_\theta \left[\sin(2\omega t + \phi_y + \phi_\theta) - \sin(\phi_y + \phi_\theta)\right]. \label{eq:xlin}
\end{equation}
Clearly, the velocity in the $x$ direction depends on the aspect ratio $\gamma$ and $\phi_y + \phi_\theta$. This suggests that as long as $\phi_y + \phi_\theta = 2n\pi + \text{constant}$, $\dot{x}$ is the same function of time. 
Its magnitude  is of order $\sim O(\epsilon_y \epsilon_\theta)$. In other words, for small amplitude flapping, the motion in $x$ direction is small compared to the flapping motion in $y$ and $\theta$. 
Approximate expressions of $F^{\rm flap}$ and $\tau^{\rm flap}$ are obtained by substituting~\eqref{eq:prescribed} and~\eqref{eq:xlin} into~\eqref{eq:periodiceom},
\begin{equation}
	F^{\rm flap} \approx (m_b + m_{2}) \ddot{y}, \qquad \tau^{\rm flap} \approx (J_b + J) \ddot{\theta}. \label{eq:forcelin}
\end{equation}
For small amplitude flapping, we can express the cost of locomotion in closed form
\begin{equation}
	e \approx 
	\dfrac{\pi \left[\gamma(\gamma + 1)\epsilon_y^2 + 2(\gamma^2 + 1)\epsilon_\theta^2\right]}{2\gamma(\gamma - 1) \epsilon_y\epsilon_\theta |\sin(\phi_y + \phi_\theta)|}. \label{eq:elin}
\end{equation}
Hence, to minimize $e$ (or, equivalently, to maximize efficiency $\eta$), one needs
\begin{equation}
\gamma \rightarrow \infty, \qquad \text{and} \qquad \phi_y + \phi_\theta = \left(n + \frac{1}{2}\right)\pi, \quad n = 0, \ \pm 1, \ \pm 2, ...
\end{equation}
The closed form expressions do not hold for large amplitudes $A_y$ and $A_\theta$ where the efficiency needs to be analyzed numerically, as done in the next section.

\section{Locomotion and Efficiency} 
\label{sec:locomotion}

We examine the swimming trajectories and their dependence on the following parameters: aspect ratio $\gamma$, amplitudes $A_y$ and $A_\theta$, and phases $\phi_y$ and $\phi_\theta$. The swimming motion is given by~\eqref{eq:prescribed} and~\eqref{eq:xdot}, where the latter is integrated numerically to get $x(t)$. 

Consider the case where  $A_y = 1$, $A_\theta = \frac{\pi}{4}$, $\phi_y = -\frac{\pi}{2}$, $\phi_\theta = 0$ and consider various aspect ratios
$\gamma=1.01, 4, 8,1000$, as shown in Figure~\ref{fig:periodicmotionGamma}.  Note that as we vary the aspect ratio, the total area of the elliptic body remains constant (this is guaranteed by the way we non-dimensionlize length using $\sqrt{ab}$). As expected, the net locomotion is almost zero when the elliptic body is close to a circular shape ($\gamma = 1.01$) and it reaches a maximum as the elliptic body approaches a flat plate ($\gamma = 1000$).

In Figure~\ref{fig:periodicmotionAy}, $\gamma$ is set to 4 and $A_y$ is varied. One can see that the net locomotion $d$ depends linearly on $A_y$, which is also evident from~\eqref{eq:xdot}. In Figure~\ref{fig:periodicmotionAtheta}, different cases of $A_\theta$ are shown. The net locomotion depends nonlinearly on $A_\theta$. Interestingly, the trajectories that correspond to $A_\theta = \pi/4$ and $A_\theta = \pi/8$ are almost identical, whereas for $A_\theta = 3\pi/4$ the locomotion is in the negative $x$ direction. 

Motions for various phases $\phi_y$ and $\phi_\theta$ are shown in Figures~\ref{fig:periodicmotionPhi}. Notice the shape and orientation of the closed path in the $(y, \theta)$ parameter space depend on the difference in phase $\phi_y - \phi_\theta$. This can be readily verified by 
eliminating $t$ from~\eqref{eq:prescribed} and expressing the closed path in the $(y, \theta)$ plane as
\begin{equation}
	\left(\frac{y}{A_y}\right)^2 - 2  \frac{y}{A_y} \frac{\theta}{A_\theta}\cos(\phi_y-\phi_\theta) + \left(\frac{\theta}{A_\theta}\right)^2  = \sin^2(\phi_y - \phi_\theta).\label{eq:closedpath}
\end{equation}
As $\phi_y - \phi_\theta$ varies, the closed path in the $(y,\theta)$ plane is elliptic, except for $\phi_y - \phi_\theta = n\pi$ ($n=0,\pm 1, \pm 2, \ldots$) in which case it is a segment of the straight line given by $\theta = (-1)^n(A_\theta /A_y)y$.  From~\eqref{eq:xdot}, one has that $x(y,\theta)$ possesses the following symmetries
\begin{equation}\label{eq:symmetryx}
	\begin{split}
		x(-y,-\theta) = x(y,\theta), \quad x(-y, \theta) = - x(y, \theta),  \quad x(y, \theta) = -x(y,\theta).
	\end{split}
\end{equation}
whereas the flapping motion in~\eqref{eq:prescribed} has the following symmetries
\begin{equation}\label{eq:symmetryflap1}
	\begin{split}
		y (t; \phi_y + \pi)  =  -y(t; \phi_y),  \qquad 
		y(-t; \phi_y ) = -y(t; -\phi_y),  \qquad  \\
		\theta( t; \phi_\theta + \pi) = -\theta (t; \phi_\theta), \qquad 
				\theta( -t; \phi_\theta) = -\theta (t; -\phi_\theta).\qquad 
		\end{split}
\end{equation}
Based on these symmetries, one can immediately conclude that, when all other parameters are held fixed,  motions that correspond to $(\phi_y , \phi_\theta)$ and $(-\phi_y , -\phi_\theta)$ are mirror images of each other: their distances and energies are the same, 
as seen from~\eqref{eq:symmetryx}. For $(\phi_y, \phi_\theta)$ with $\phi_y - \phi_\theta = 2n\pi + \text{constant}$, one gets the same path in the $(y, \theta)$ space. When tracing the same path in the $(y, \theta)$-plane (but starting at different initial points), the resulting trajectories in the $(x,y)$ plane are similar (with different initial positions). 
Note that, in general, flapping motions that trace a straight line in the $(y , \theta)$-plane do not correspond to zero net locomotion in the $(x,y)$-plane, except for $\phi_y = \phi_\theta = 0$ and $\phi_y = \phi_\theta = -\pi/2$. This is evident  from the example of $\phi_y = \pi/4, \phi_\theta = -3\pi/4$ in Figure~\ref{fig:periodicmotionPhi}$(h)$. The locomotion here is not a result of a geometric phase but a dynamic phase, see~\cite{KaMaRoMe2005}.

\begin{figure}[!tb]
	\centering
	\begin{subfigure}[$e$ vs. $\gamma$]{
		\includegraphics[width=0.28\textwidth]{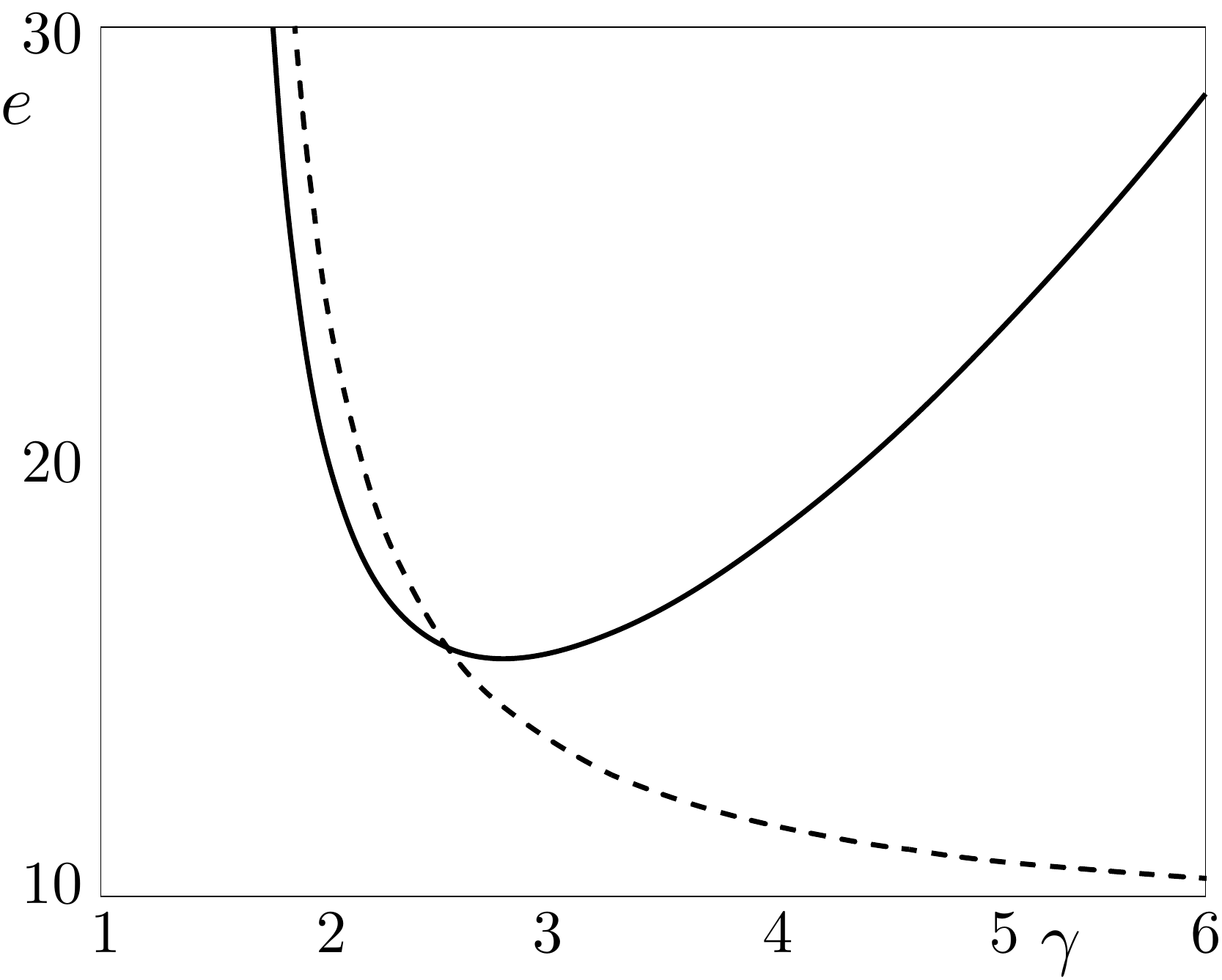}}		
	\end{subfigure}
	\begin{subfigure}[$e$ vs.  $A_y$]{
		\includegraphics[width=0.28\textwidth]{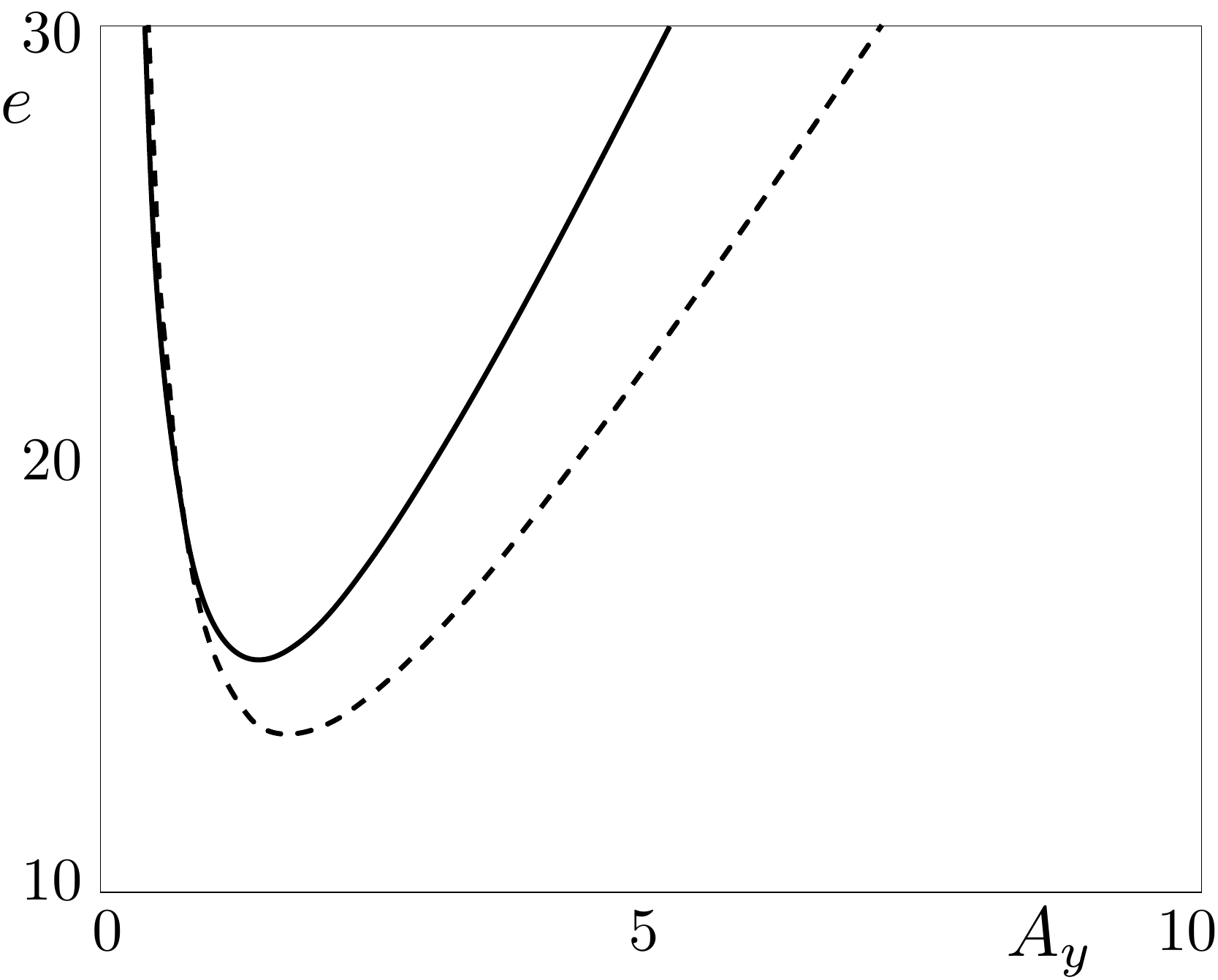}}		
	\end{subfigure}
	\begin{subfigure}[$e$ vs.  $A_\theta$]{
		\includegraphics[width=0.28\textwidth]{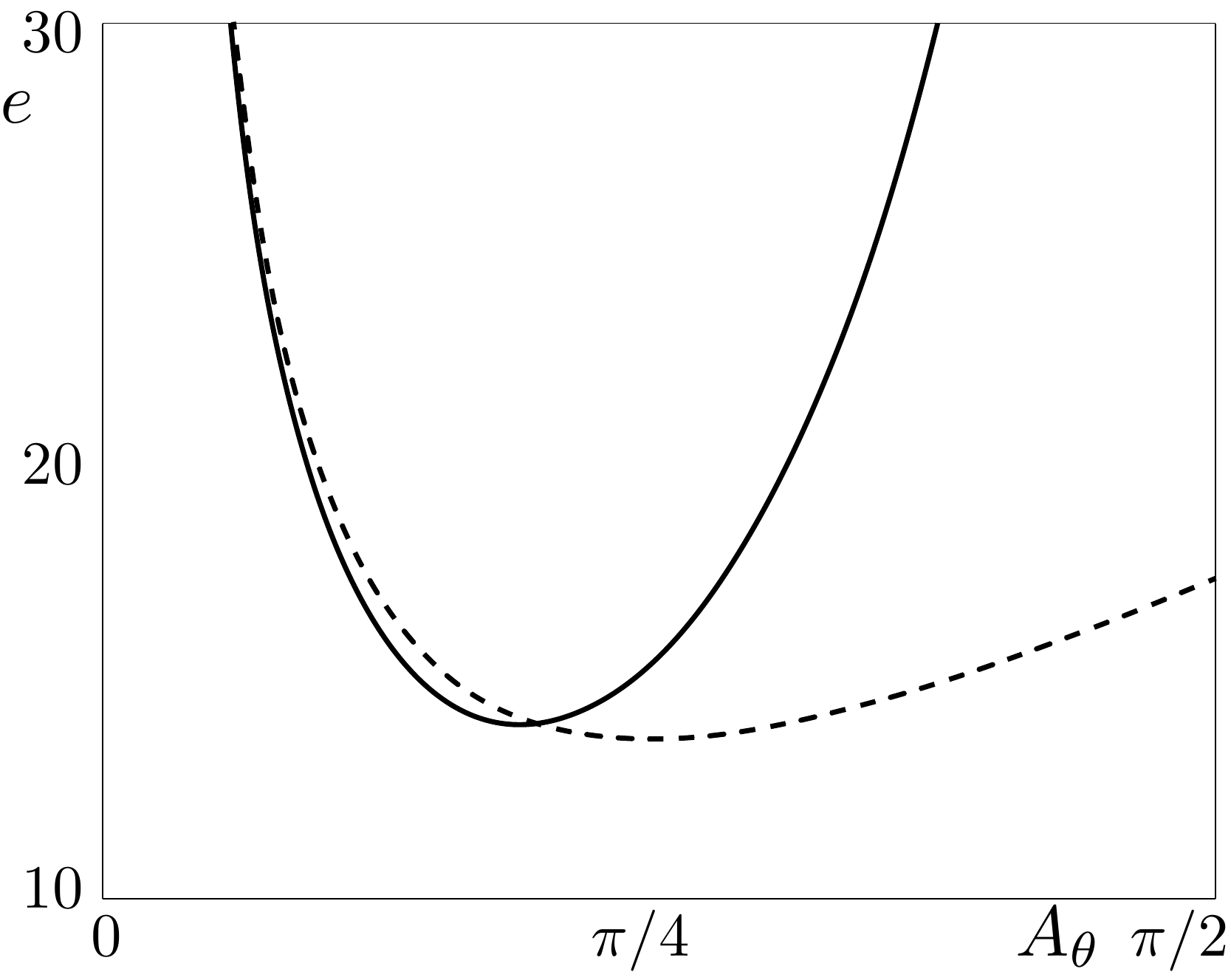}}		
	\end{subfigure}
	\caption{\footnotesize Cost of locomotion $e$ as a function of: (a) aspect ratio $\gamma$, (b) heaving amplitude $A_y$, (c) pitching amplitude $A_\theta$. The base parameter values are set to  $\phi_y = -\pi/2, \phi_\theta = 0$, $\gamma = 4$, $A_\theta = \pi/4$, $A_y = 1$. Solid lines are nonlinear numerical solutions, while dashed lines are based on small amplitude approximation given in~\eqref{eq:elin}.}
	\label{fig:elinnonlin}
\end{figure}

We now compute the average work $\bar{E}$ and cost of locomotion $e=\bar{E}/d$. Ideally, one would like to find optimal parameter values that minimize $e$ (maximize efficiency $\eta$) and/or maximize $d$ (see, for example~\cite{Eloy2013, KeKo2006}). Instead of minimizing $e$ over the five dimensional parameter space, we study the dependence of $e$ on the system's parameters by varying one parameter at a time. 
In Figure~\ref{fig:elinnonlin},  we set $\phi_y= -\frac{\pi}{2}, \phi_\theta=0$ and vary $\gamma, A_y$ and $A_\theta$, respectively. Solid lines correspond to the numerical nonlinear solutions and dashed lines are obtained by substituting the parameters into~\eqref{eq:elin}. Figure~\ref{fig:elinnonlin}$(a)$ shows that, for $A_y = 1, A_\theta = \pi/4$, there exist a optimal value of $\gamma \approx 2.9$, whereas the small amplitude approximation in~\eqref{eq:elin} predicts that $e$ is a decreasing function of $\gamma$. Figure~\ref{fig:elinnonlin}$(b)$ shows an optimal value of $A_y \approx 1.3$ and that the small amplitude results qualitatively follows  the nonlinear behavior of $e$. This is because the work $E$ depends quadratically on $A_y$, the displacement $d$ depends linearly on $A_y$ and that the small amplitude approximation in~\eqref{eq:elin} preserves the form of dependence on this parameter. However, Figure~\ref{fig:elinnonlin}$(c)$ shows that when varying $A_\theta$, the small amplitude results provide good approximation of the nonlinear efficiency only up to $A_\theta \approx \frac{3\pi}{16}$. 

\begin{figure}[!tb]
	\centering
	\begin{subfigure}[$\gamma = 1.01$]{
		\includegraphics[width=0.16\textwidth]{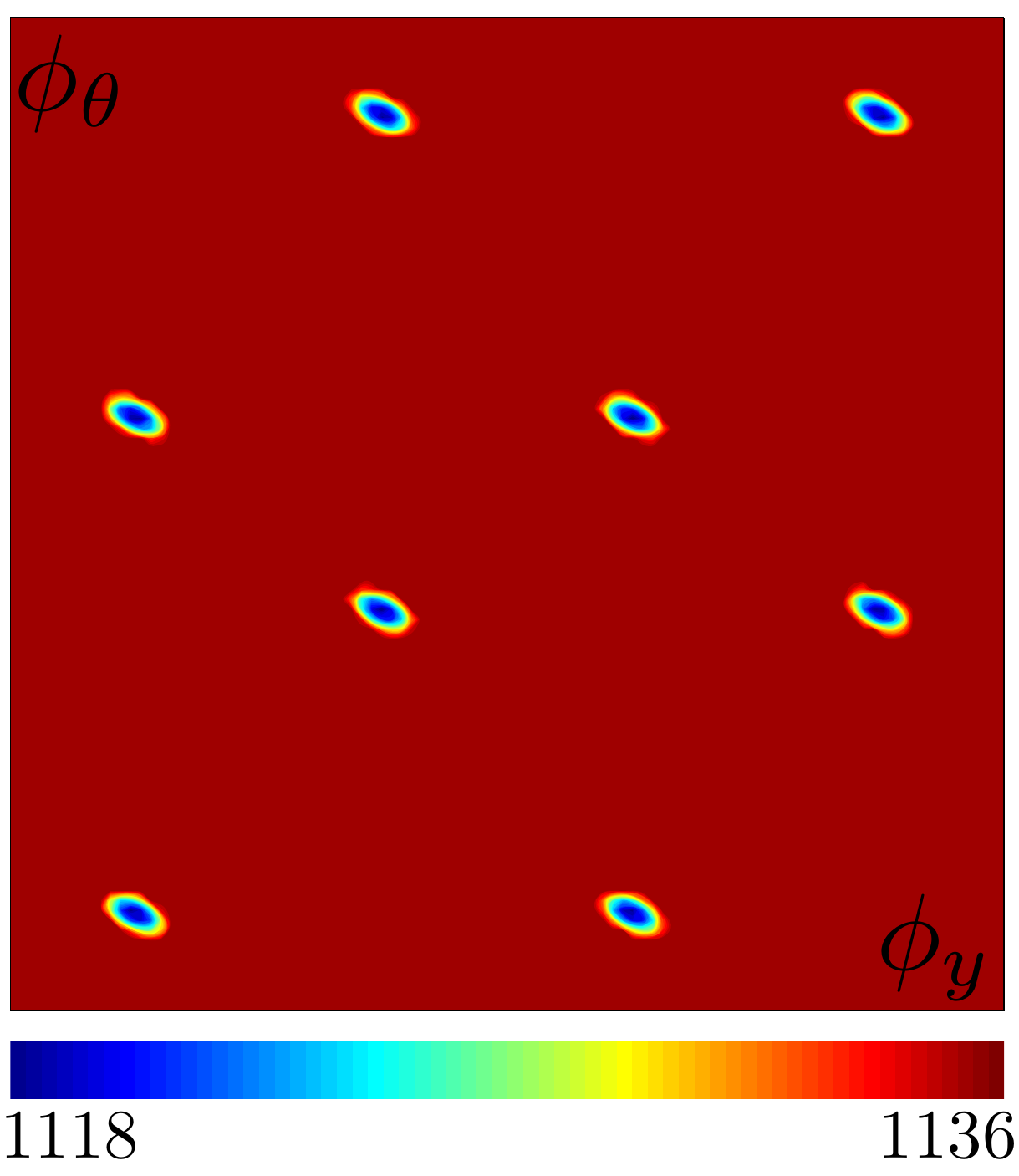}}		
	\end{subfigure}
	\begin{subfigure}[$\gamma = 2$]{
		\includegraphics[width=0.16\textwidth]{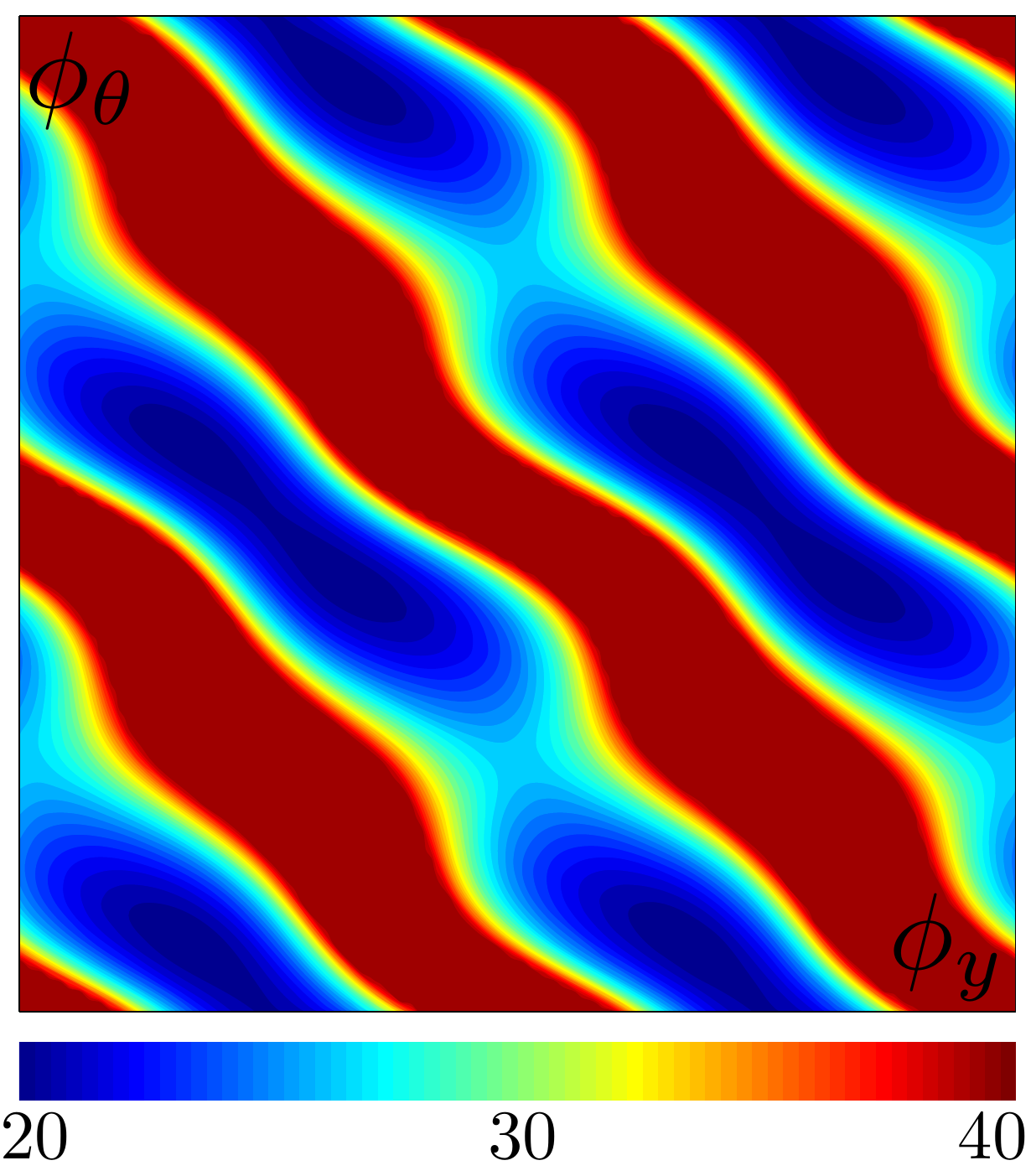}}		
	\end{subfigure}
	\begin{subfigure}[$\gamma = 4$]{
		\includegraphics[width=0.16\textwidth]{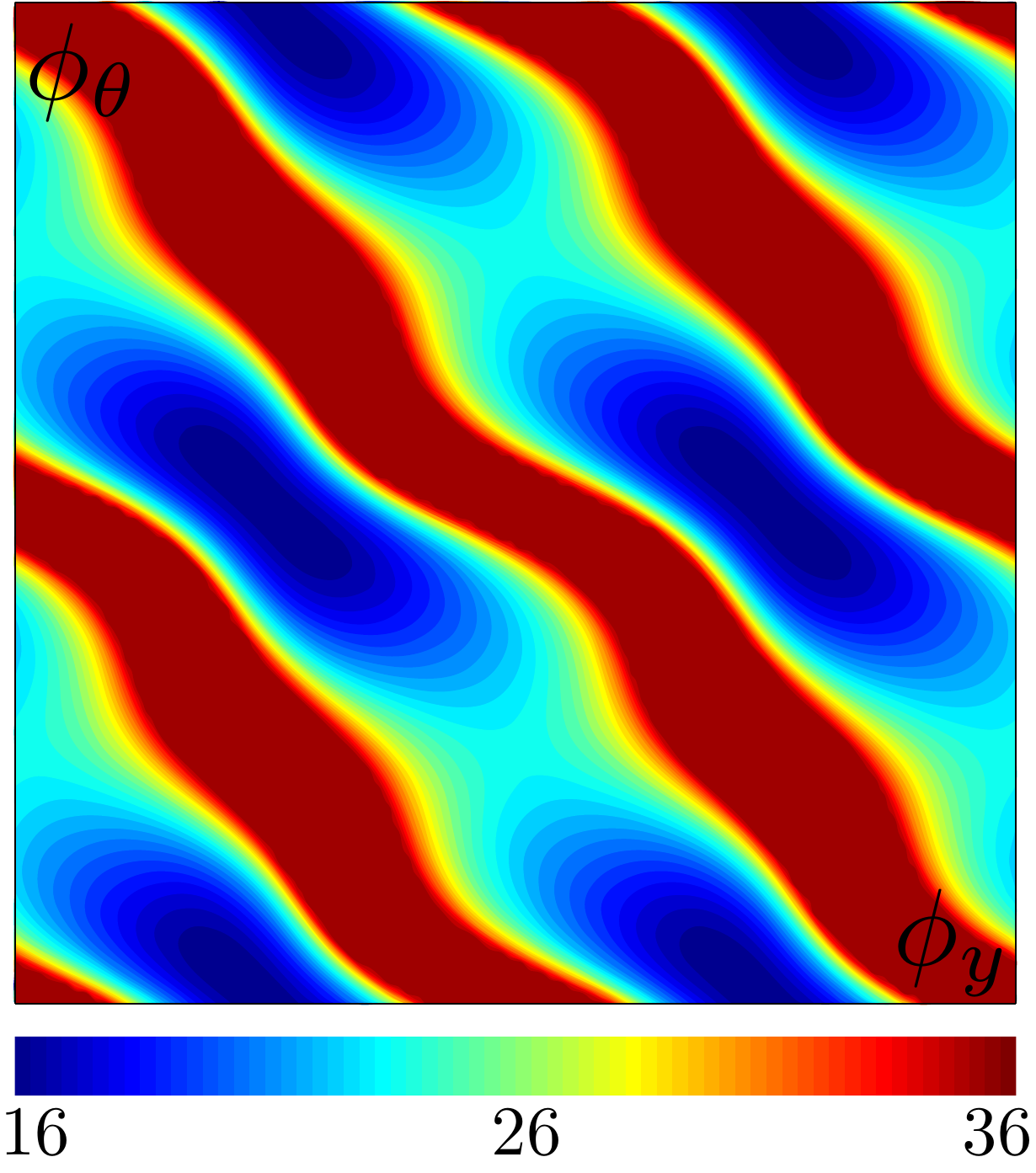}}		
	\end{subfigure}
	\begin{subfigure}[$\gamma = 8$]{
		\includegraphics[width=0.16\textwidth]{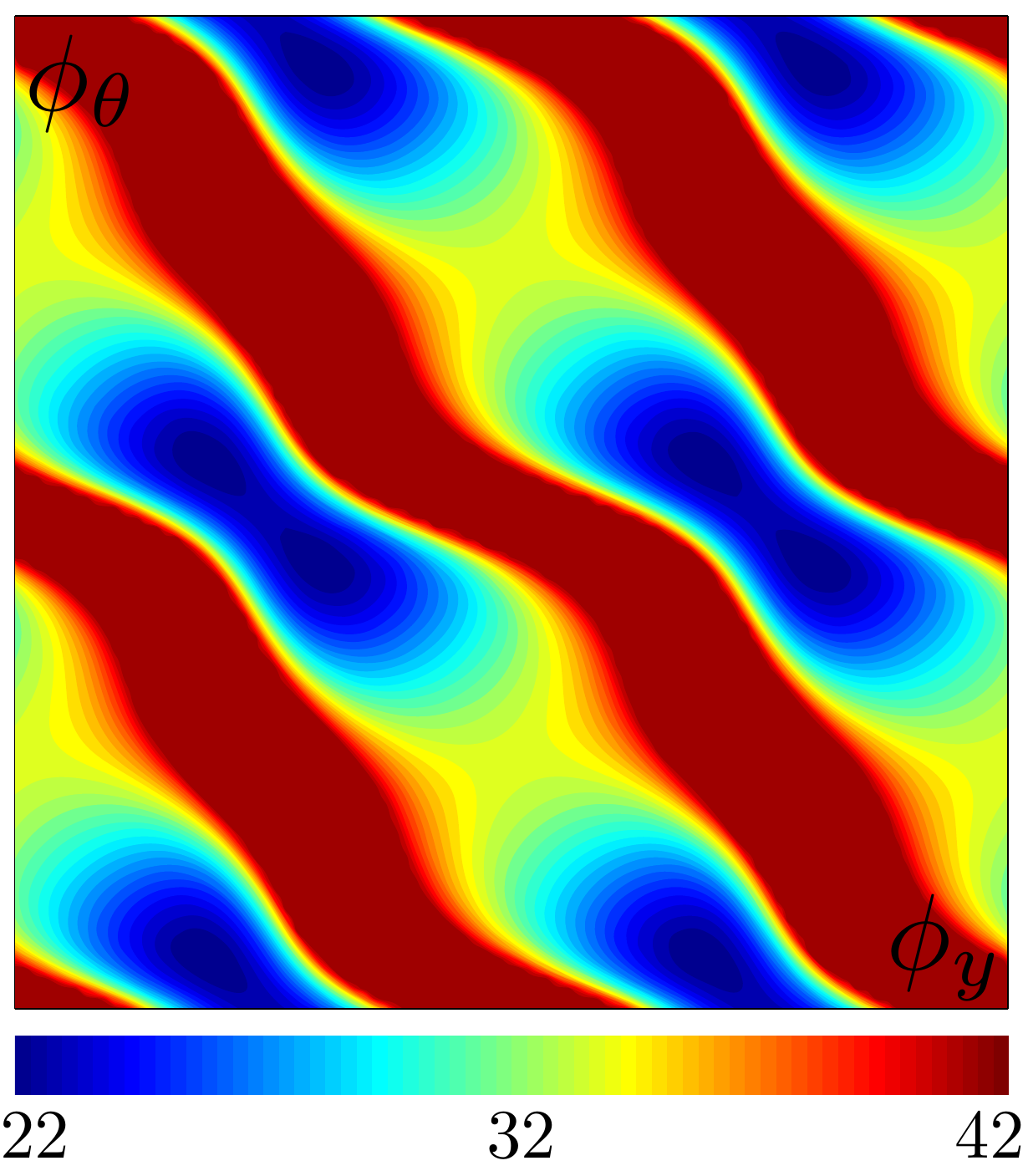}}		
	\end{subfigure}
	\begin{subfigure}[$\gamma = 16$]{
		\includegraphics[width=0.16\textwidth]{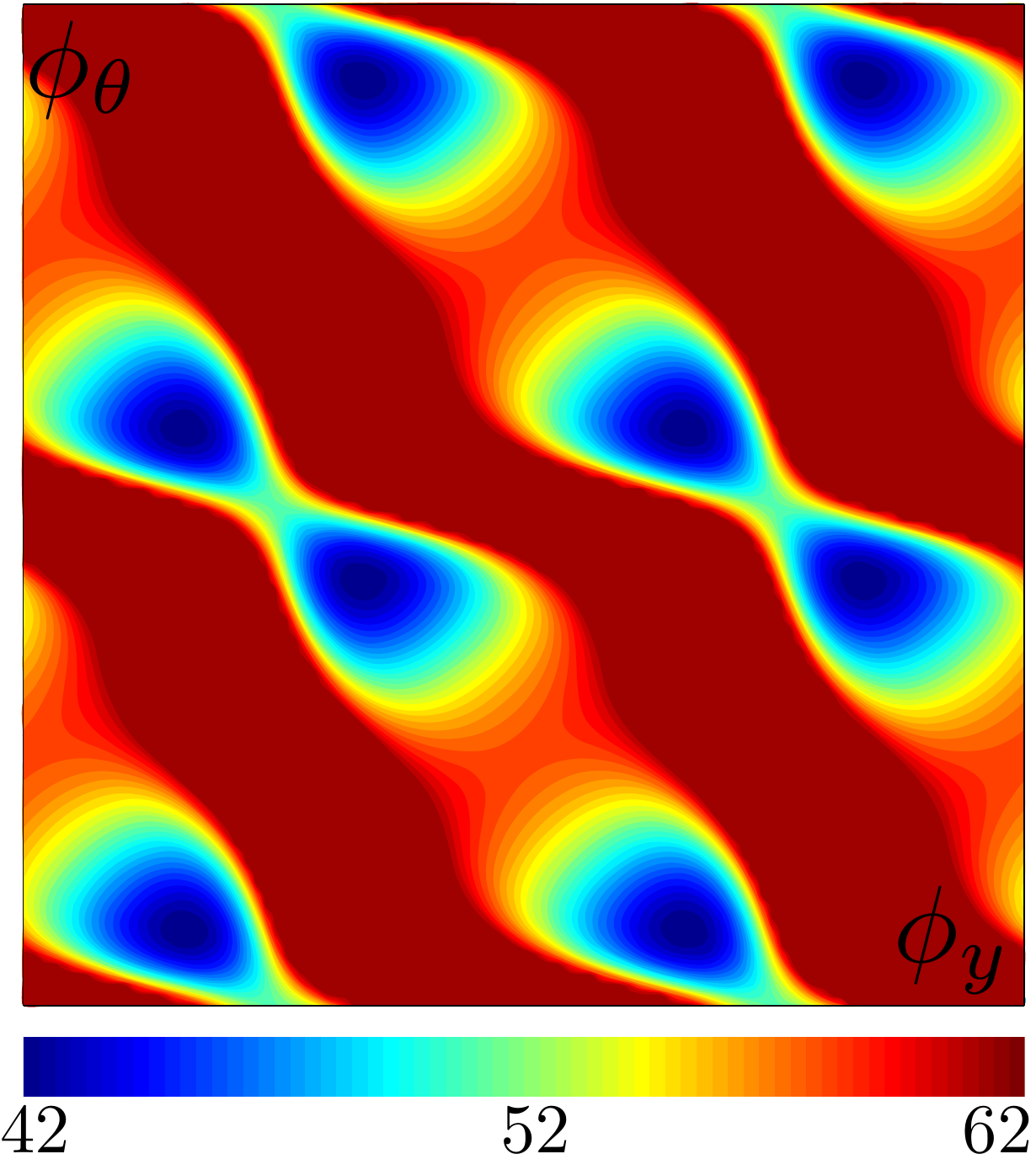}}		
	\end{subfigure}
	\caption{\footnotesize Contour plots of cost of locomotion $e$ for the cases $A_y = 1, A_\theta = \pi/4$ and various aspect ratio $\gamma$. Each plot is evaluated on a $201 \times 201$ mesh in $[-\pi \,,\, \pi] \times [-\pi \,,\, \pi]$ in $(\phi_y , \phi_\theta)$ plane. Lower value of $e$ corresponds to higher efficiency.}
	\label{fig:effvarygamma}
\end{figure}

\begin{figure}[!tb]
	\centering
	\begin{subfigure}[$A_y = 0.01$]{
		\includegraphics[width=0.16\textwidth]{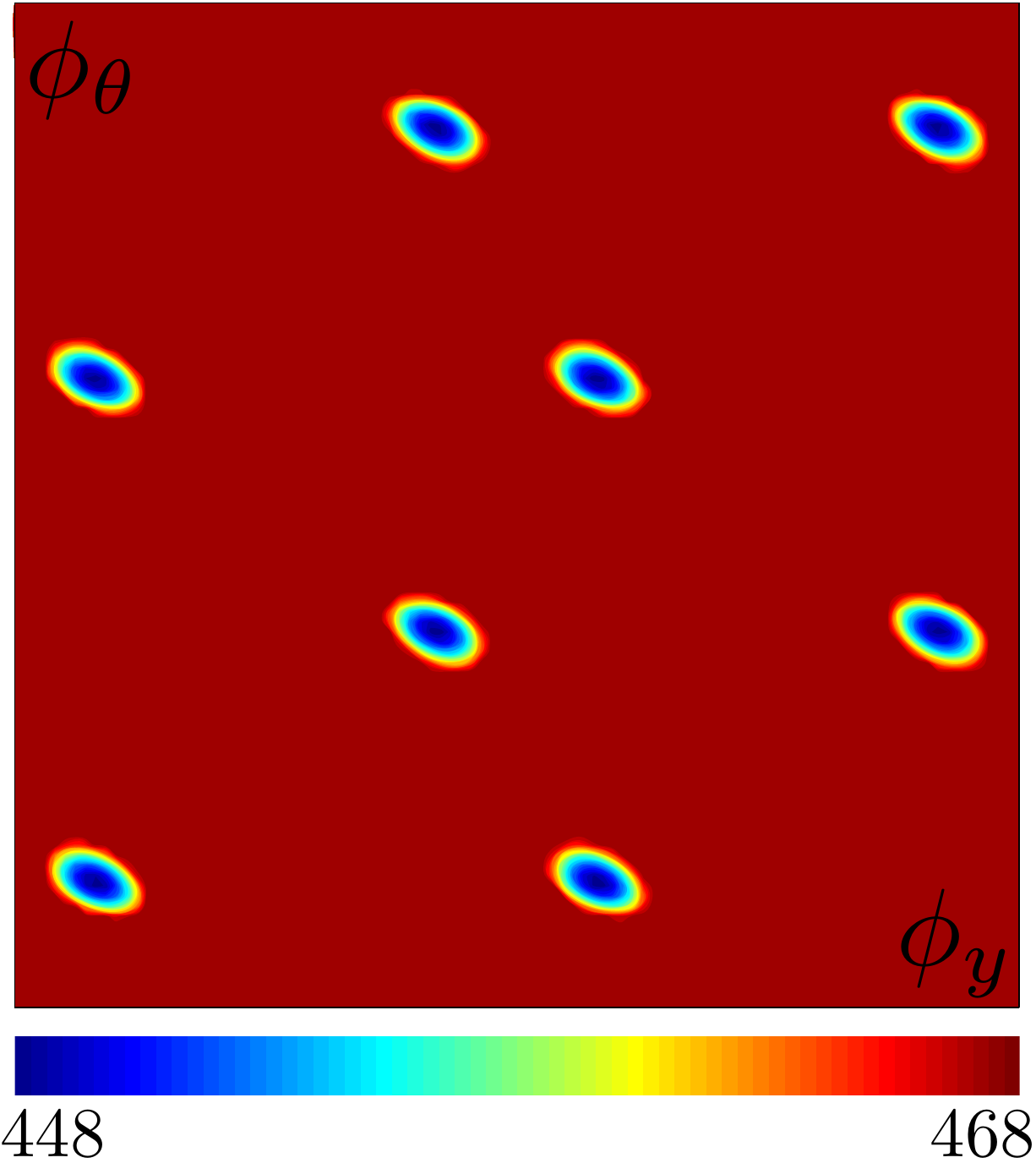}}		
	\end{subfigure}
	\begin{subfigure}[$A_y = 0.5$]{
		\includegraphics[width=0.16\textwidth]{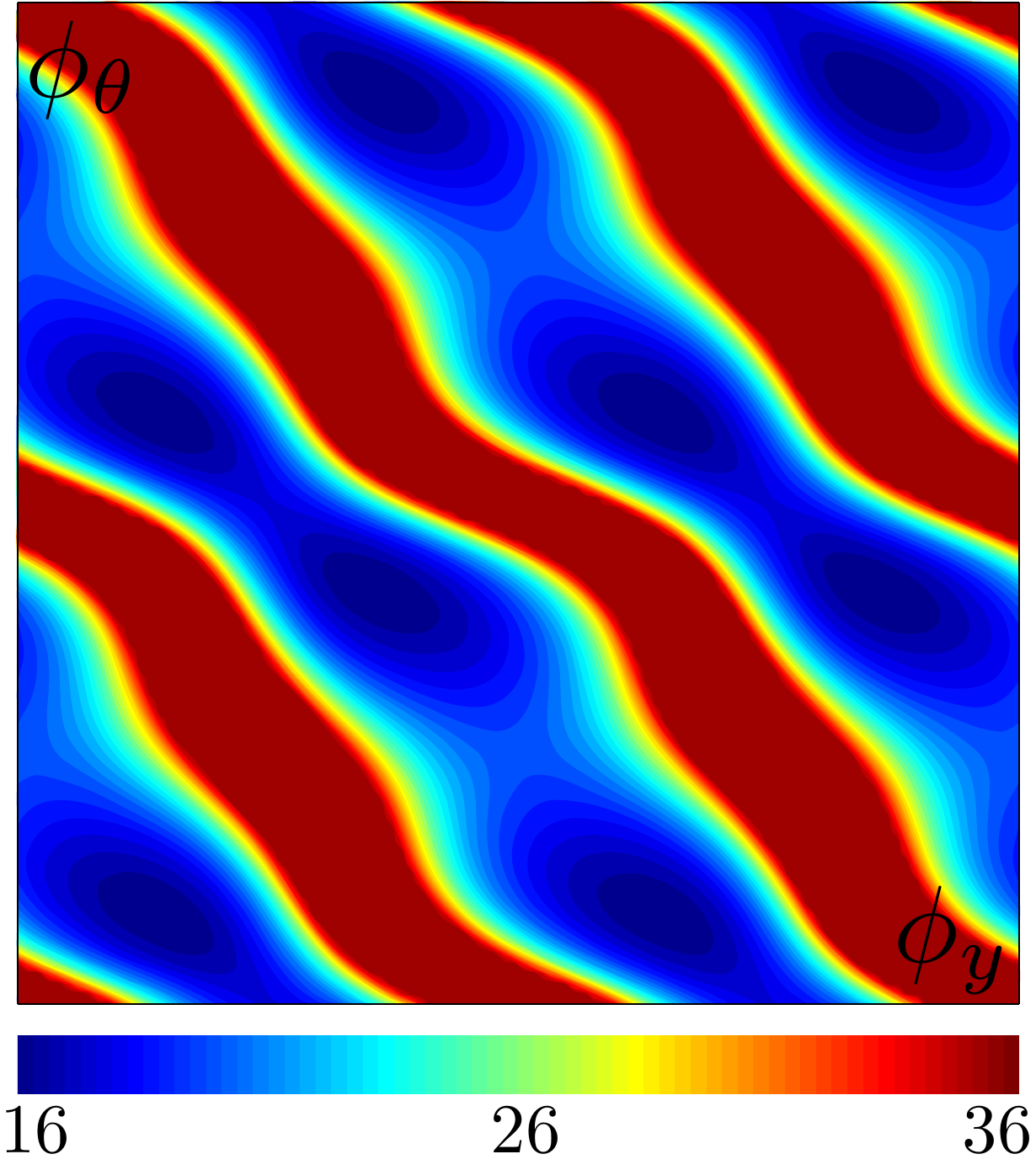}}		
	\end{subfigure}
	\begin{subfigure}[$A_y = 1$]{
		\includegraphics[width=0.16\textwidth]{eff_4_1_pi4.pdf}}		
	\end{subfigure}
	\begin{subfigure}[$A_y = 2$]{
		\includegraphics[width=0.16\textwidth]{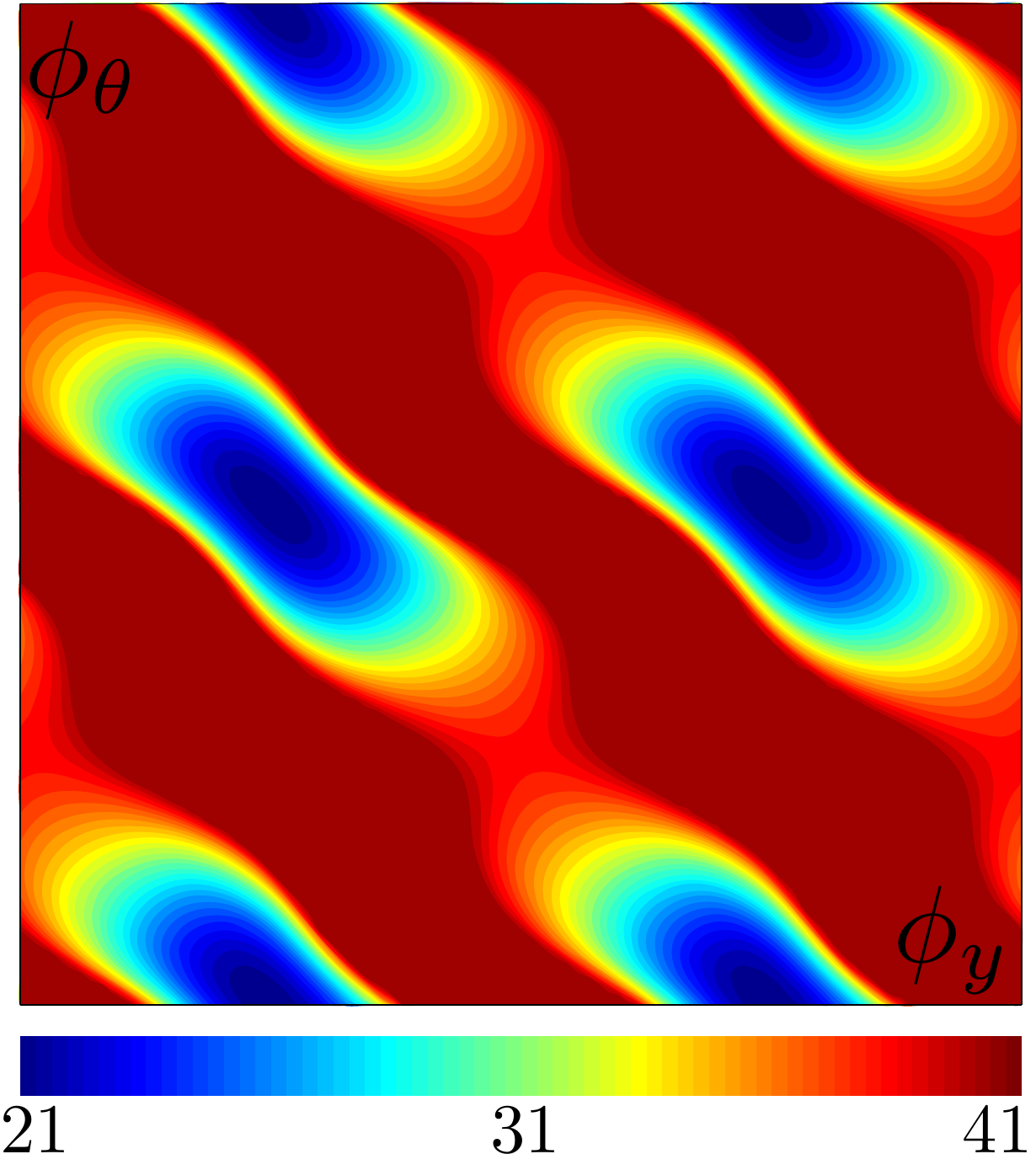}}		
	\end{subfigure}
	\begin{subfigure}[$A_y = 6$]{
		\includegraphics[width=0.16\textwidth]{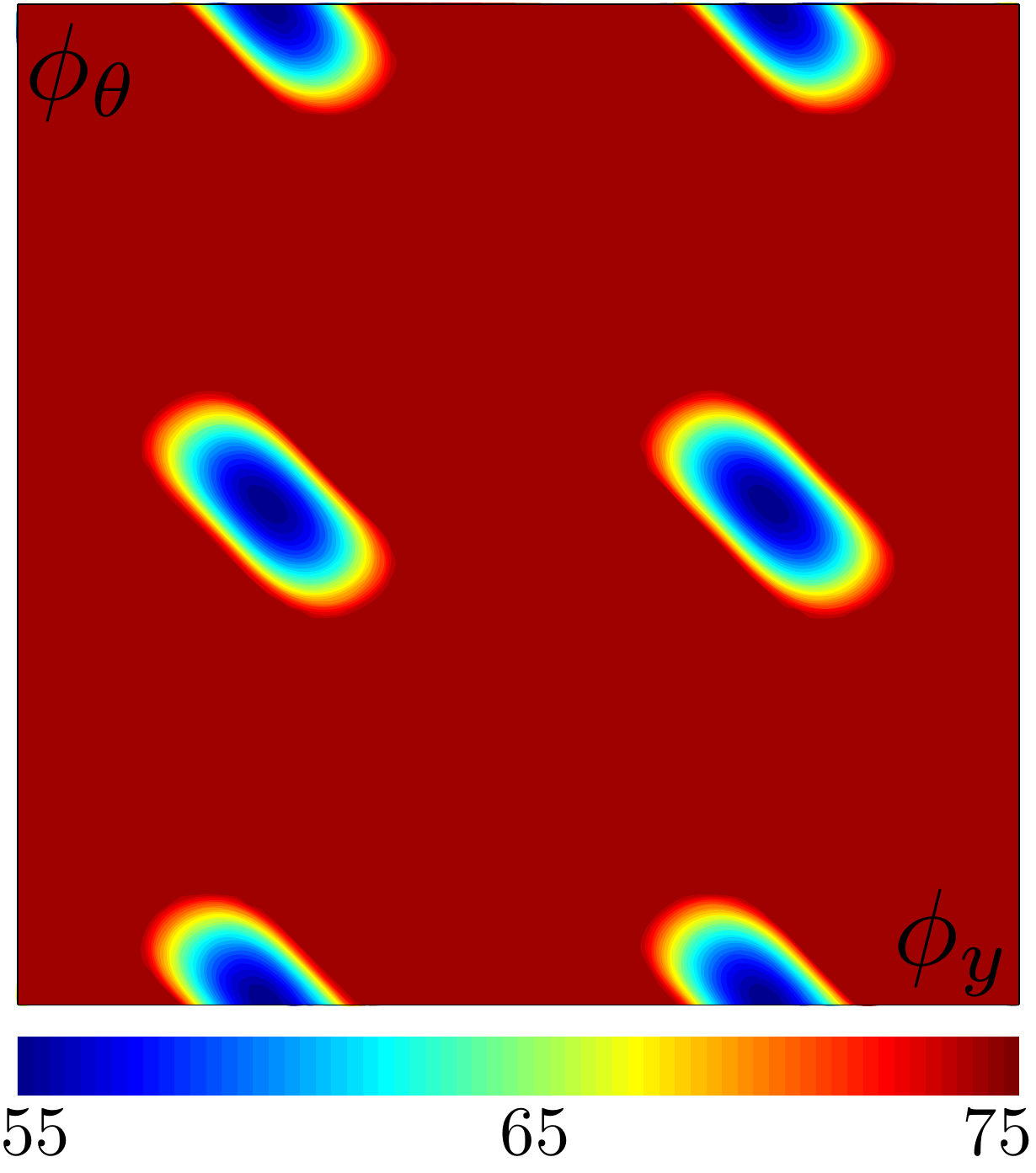}}		
	\end{subfigure}
	\caption{\footnotesize Contour plots of cost of locomotion $e$ for the cases $A_\theta = \pi/4$, $\gamma = 4$ and various heaving amplitude $A_y$. Each plot is evaluated on a $201 \times 201$ mesh in $[-\pi \,,\, \pi] \times [-\pi \,,\, \pi]$ in $(\phi_y , \phi_\theta)$ plane. Lower value of $e$ corresponds to higher efficiency.}
	\label{fig:effvaryAy}
\end{figure}

\begin{figure}[!tb]
	\centering
	\begin{subfigure}[$A_\theta = 0.01$]{
		\includegraphics[width=0.16\textwidth]{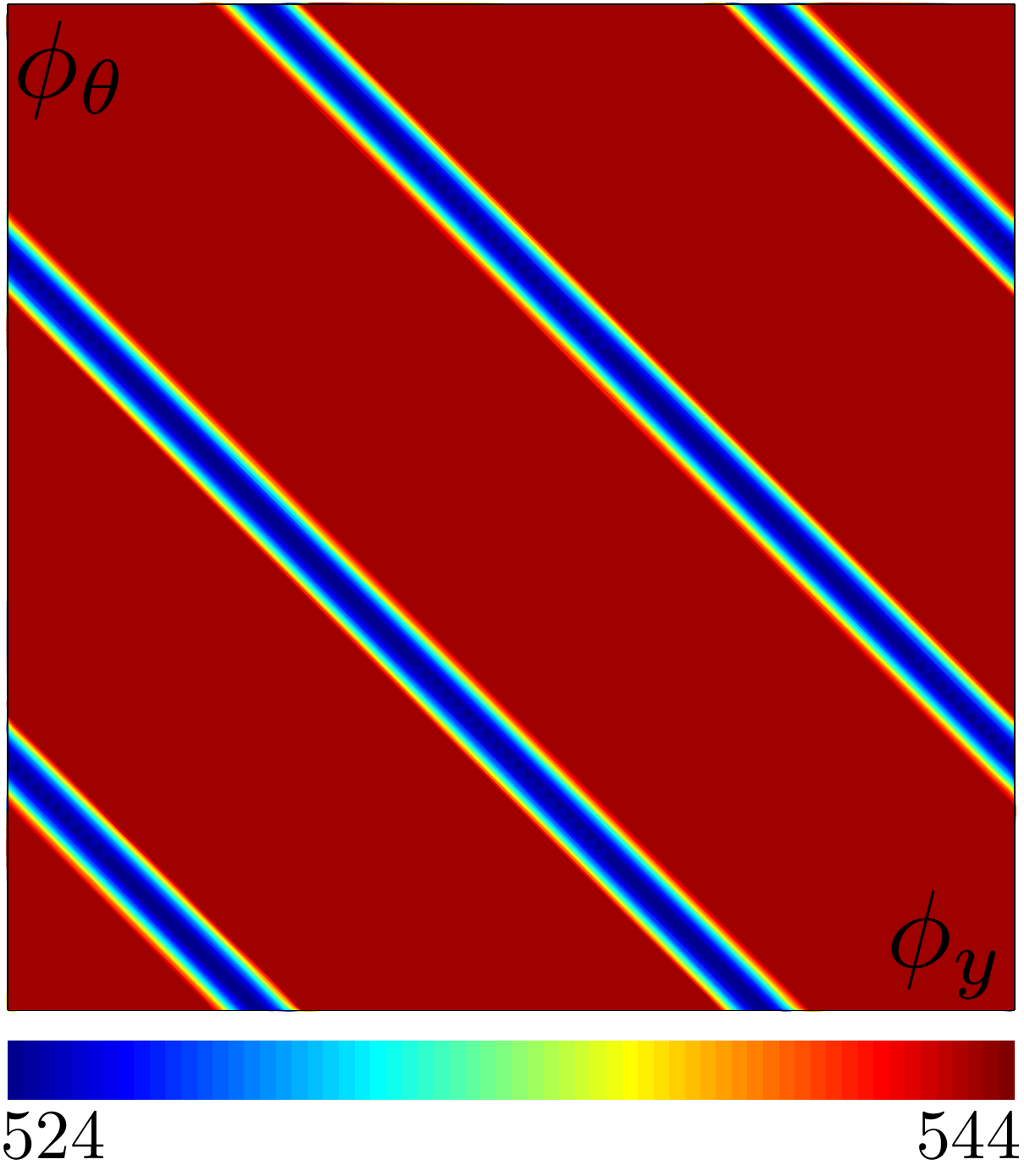}}		
	\end{subfigure}
	\begin{subfigure}[$A_\theta = \pi/8$]{
		\includegraphics[width=0.16\textwidth]{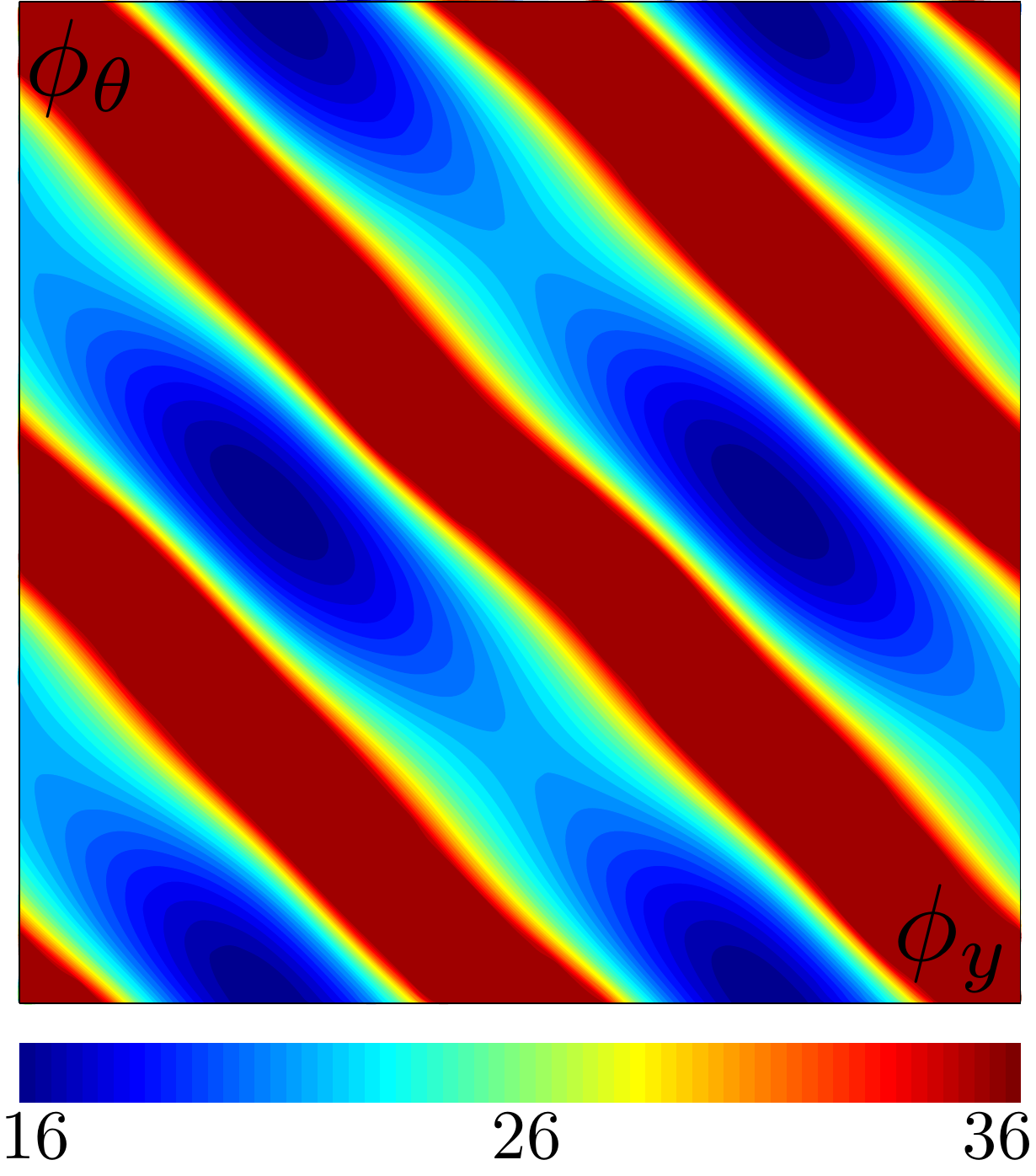}}		
	\end{subfigure}
	\begin{subfigure}[$A_\theta = \pi/4$]{
		\includegraphics[width=0.16\textwidth]{eff_4_1_pi4.pdf}}		
	\end{subfigure}
	\begin{subfigure}[$A_\theta = 3\pi/8$]{
		\includegraphics[width=0.16\textwidth]{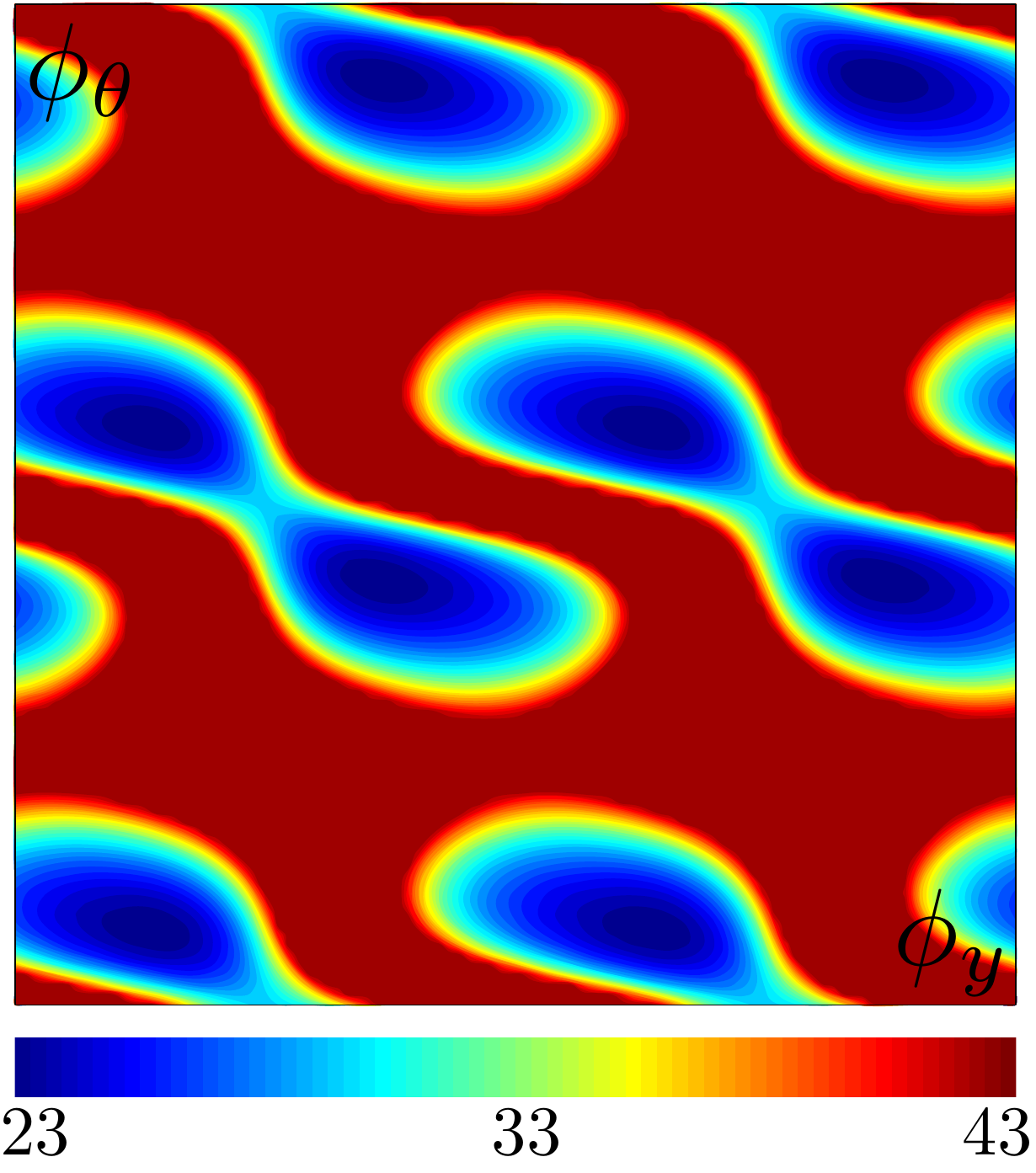}}		
	\end{subfigure}
	\begin{subfigure}[$A_\theta = \pi/2$]{
		\includegraphics[width=0.16\textwidth]{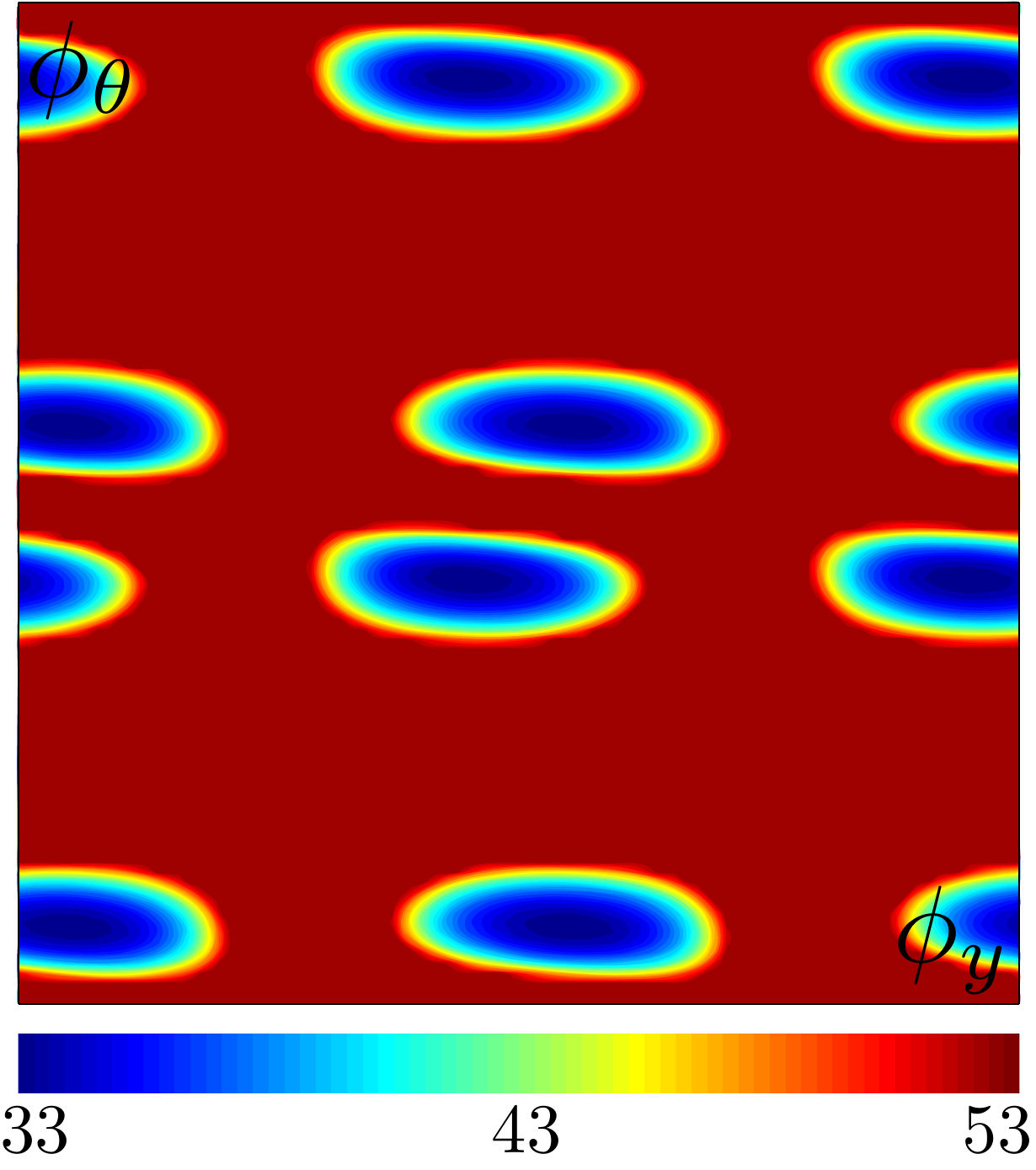}}		
	\end{subfigure}
	\caption{\footnotesize Contour plots of cost of locomotion $e$ for the cases $A_y = 1$, $\gamma = 4$ and various pitching amplitude $A_\theta$. Each plot is evaluated on a $201 \times 201$ mesh in $[-\pi \,,\, \pi] \times [-\pi \,,\, \pi]$ in $(\phi_y , \phi_\theta)$ plane. Lower value of $e$ corresponds to higher efficiency.}
	\label{fig:effvaryAtheta}
\end{figure}

In Figure~\ref{fig:effvarygamma} to \ref{fig:effvaryAtheta}, we examine the dependence of $e$ on $(\phi_y , \phi_\theta)$ by discretizing the domain $[-\pi \,,\, \pi] \times [-\pi \,,\, \pi]$ using a $201 \times 201$ mesh. Contours of $e$ as a function of $(\phi_y , \phi_\theta)$ are depicted for various $\gamma, A_y$ and $A_\theta$. It follows from~(\ref{eq:symmetryx}-\ref{eq:symmetryflap1}) that $e$ has a reflection symmetry about the origin $e(\phi_y, \phi_\theta) = e(-\phi_y, -\phi_\theta)$ and the periodic property $e(\phi_y + \pi, \phi_\theta + \pi) = e(\phi_y, \phi_\theta)$. Therefore, it would have been sufficient to show the dependence of $e$ only on one quarter of the shown domain, say, $[0 \,,\, \pi] \times [0 \,,\, \pi]$.

Note that the parameters that minimize $e$, thus maximize efficiency $\eta$, are approximately $3 \leq \gamma \leq 4, A_y \approx 1.3$, $A_\theta \approx 3\pi/16$ and  $(\phi_y , \phi_\theta) \approx ((m+\frac{1}{2})\pi, n\pi)$, where $m, n = 0, \pm 1, \pm 2, \ldots$ One example of the optimal phases is $\phi_y = \frac{\pi}{2}, \phi_\theta = 0$, with corresponding locomotion shown in Figure~\ref{fig:periodicmotionGamma}. For this optimal motion, the pitching angle is zero when the heaving motion is maximum ($90^\circ$ out of phase), which qualitatively agrees with the results in~\cite{KeKo2006}. The optimal aspect ratio $3\leq \gamma \leq 4$ agrees with the optimal shape aspect ratio obtained in the comprehensive optimization study in~\cite{Eloy2013}, and is representative of the aspect ratio of various Carangiform swimmers such as bass ($\gamma = 3.8$) in~\cite{JaLa1995}, tuna ($\gamma = 3.5$) in~\cite{DoDi2000} and saithe ($\gamma = 4.1$) in~\cite{ViHe1984}. The optimal heave to cord ratio $A_y/a \approx 0.75$ (where $a = \sqrt{\gamma} \approx \sqrt{3}$) and maximum angle $A_\theta \approx 3\pi/16 = 16.875^\circ$ both agree with the optimal motions for the rigid flapping body ($0.75 \leq A_y/a \leq 1, A_\theta \approx 16^\circ$) given in~\cite{TrHoTeYe2005}. This is remarkable given the simplicity of our model in comparison to the models in~\cite{Eloy2013, TrHoTeYe2005}.



\section{Stability of Periodic Locomotion} 
\label{sec:stability_of_periodic_solution}

We study  stability of periodic motion subject to arbitrary perturbations in the surrounding fluid. 
We begin by introducing $\mathbf{q} = [\dot{x} , \dot{y} , \theta , \dot{\theta}]^T$, and rewriting~\eqref{eq:periodiceomX} and~\eqref{eq:periodiceom} as follows
\begin{equation}
	\mathbb{M}(\theta)\dot{\mathbf{q}} = \mathbf{f}(\mathbf{q}) + \mathbf{F}^{\rm flap}, \label{eq:rearrangeeom}
\end{equation}
where detailed expressions for $\mathbb{M}$, $ \mathbf{f}$ and  $\mathbf{F}^{\rm flap}$ are listed in Appendix. In Sections~\ref{sec:problem_setup}--\ref{sec:locomotion}, we prescribed the flapping motion $y(t)$ and $\theta(t)$ according to~\eqref{eq:prescribed} and used~\eqref{eq:xdot} to solve for $x(t)$ and~\eqref{eq:periodiceom} to solve for $F^{\rm flap}$ and $\tau^{\rm flap}$.  The resulting motion $x(t)$, $y(t)$, and $\theta(t)$ as well as the forcing $F^{\rm flap}$ and $\tau^{\rm flap}$ are periodic with period $T$. We let $\mathbf{q}_p$ denote the $\mathbf{q}$ corresponding to such periodic motion. We study the stability of $\mathbf{q}_p$ by introducing a small perturbation $\delta \mathbf{q}$ such that $\mathbf{q} = \mathbf{q}_p + \delta \mathbf{q}$ while keeping $F^{\rm flap}$ and $\tau^{\rm flap}$ the same as that producing the periodic solution. In other words, we account for arbitrary perturbations in the fluid environment while keeping the same flapping forces to check if such perturbations destabilize the periodic trajectory.

\begin{figure}[!tb]
	\centering
	\includegraphics[width=0.85\textwidth]{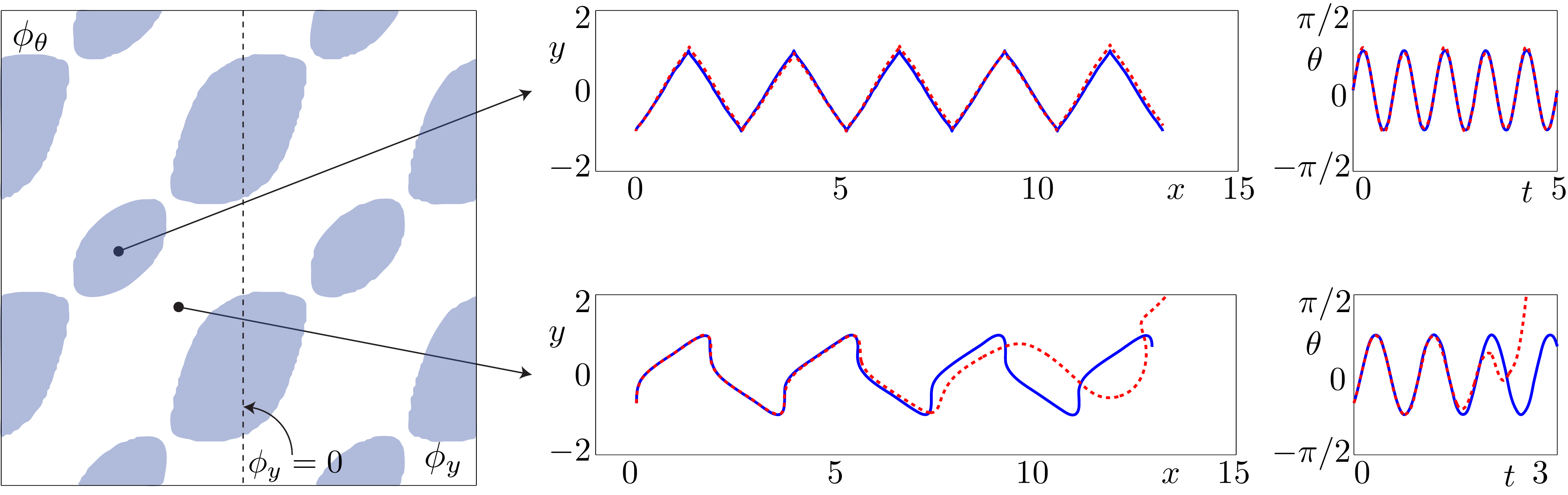}	
	\caption{\footnotesize Left: stability of the case $A_y = 1$, $A_\theta = \pi/4$ and $\gamma = 4$, with phases $(\phi_y , \phi_\theta)$ varied on a $201 \times 201$ mesh in $[-\pi \,,\, \pi] \times [-\pi \,,\, \pi]$. Stable cases correspond to shaded areas, unstable cases are white areas. Middle and right: (top) $(\phi_y , \phi_\theta) = (-\pi/2 , 0)$ is stable. Solid lines correspond to unperturbed periodic solutions, dashed lines correspond to perturbed solutions; (bottom) $(\phi_y , \phi_\theta) = (-\pi/4, -\pi/4)$ is unstable.}
	\label{fig:basecase}
\end{figure}

We linearize equation~\eqref{eq:rearrangeeom} about the periodic trajectory $\mathbf{q}_p(t)$ to get
\begin{equation}
	\delta \dot{\mathbf{q}} = \mathbb{J}(t) \delta \mathbf{q},\label{eq:eomlin}
\end{equation}
where the Jacobian $\mathbb{J}(t)$ is a $4 \times 4$ periodic matrix ($\mathbb{J}(0) = \mathbb{J}(T)$) given by (see Appendix for details),
\begin{equation}
\mathbb{J}(t) = \left.\frac{\partial \mathbf{g}}{\partial \mathbf{q}}\right\vert_{(\mathbf{q}_p, \mathbf{F}^{\rm flap})} \quad \text{where}\quad
\mathbf{g} = \mathbb{M}^{-1}(\mathbf{f} + \mathbf{F}^{\rm flap}).	\label{eq:jacobian}
\end{equation}
Let $\Phi(t)$ denote the {\em fundamental solution matrix} of~\eqref{eq:eomlin}. 
The eigenvalues  $\lambda_i$ of the time-independent matrix,
$B = \Phi(0)^{-1} \Phi(T)$,
are referred to as the {\em characteristic multipliers}. Their locations in the complex plane indicate the stability of the periodic solution $\mathbf{q}_p$:  if at least one characteristic multiplier lies {\em outside} the unit circle, $\mathbf{q}_p$ is unstable; if all $\lambda_i$'s ($i = 1, ..., 4$) are {\em on} the unit circle, then $\mathbf{q}_p$ is regarded to as marginally stable. For a non dissipative system as in our model, these two are the only possible scenarios, namely, unstable or marginally stable (simply referred to as ``stable" hereafter).
One eigenvalue $\lambda_1$ is always 1, reflecting the fact that $\mathbf{q}_p$ is periodic. The remaining eigenvalues may be complex. Complex eigenvalues come in conjugate pairs.

For the case $A_y =1, A_\theta = \frac{\pi}{4}, \gamma = 4$, the stability results are plotted in Figure~\ref{fig:basecase} as a function of the phases $(\phi_y , \phi_\theta)$, again evaluated on a $201 \times 201$ mesh discretizing the domain $[-\pi \,,\, \pi] \times [-\pi \,,\, \pi]$.  The stable regions are shaded areas, and the unstable regions are white areas. Notice the reflection symmetry about $(0,0)$ and periodicity of $\pi$ in both $\phi_y$ and $\phi_\theta$ that we observed in the efficiency analysis is again seen in the stability plot. Two examples with different stability characteristics are shown. The solid lines are unperturbed periodic solutions $\mathbf{q}_p$, and dashed lines are solutions with random initial perturbations with magnitude $|\delta{\mathbf{q}}(t = 0)| \sim O(10^{-3})$. Clearly, the trajectory corresponding to parameters in the stable region remain close to the periodic trajectory for the integration time whereas that corresponding to parameters in the unstable region does not.

\begin{figure}[!tb]
	\centering
		\begin{subfigure}[Real vs. Imaginary parts of $\lambda_i$]{
			\includegraphics[width=0.3\textwidth]{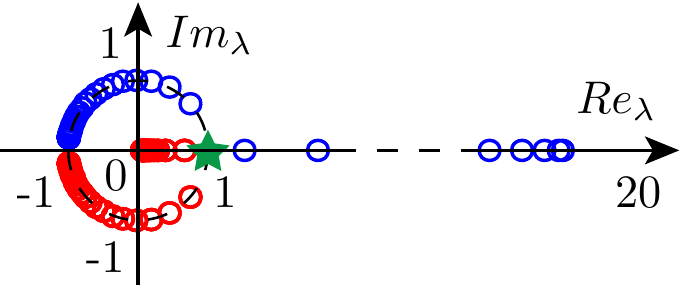}}		
		\end{subfigure}\\
		\begin{subfigure}[Real parts of $\lambda_i$ vs. $\phi_\theta$]{
			\includegraphics[width=0.35\textwidth]{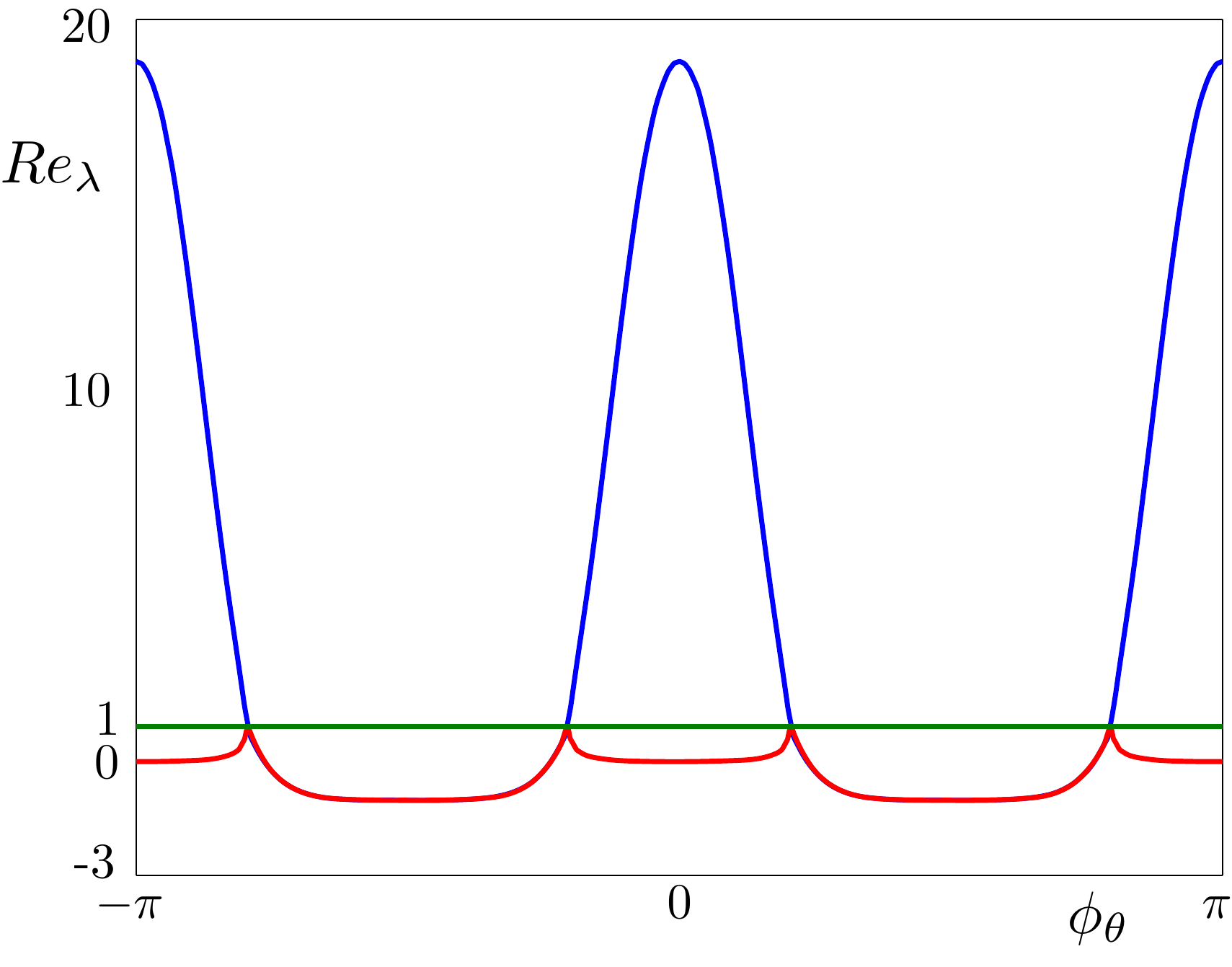}}		
		\end{subfigure}\qquad
		\begin{subfigure}[Imaginary parts of $\lambda_i$ vs. $\phi_\theta$]{
			\includegraphics[width=0.355\textwidth]{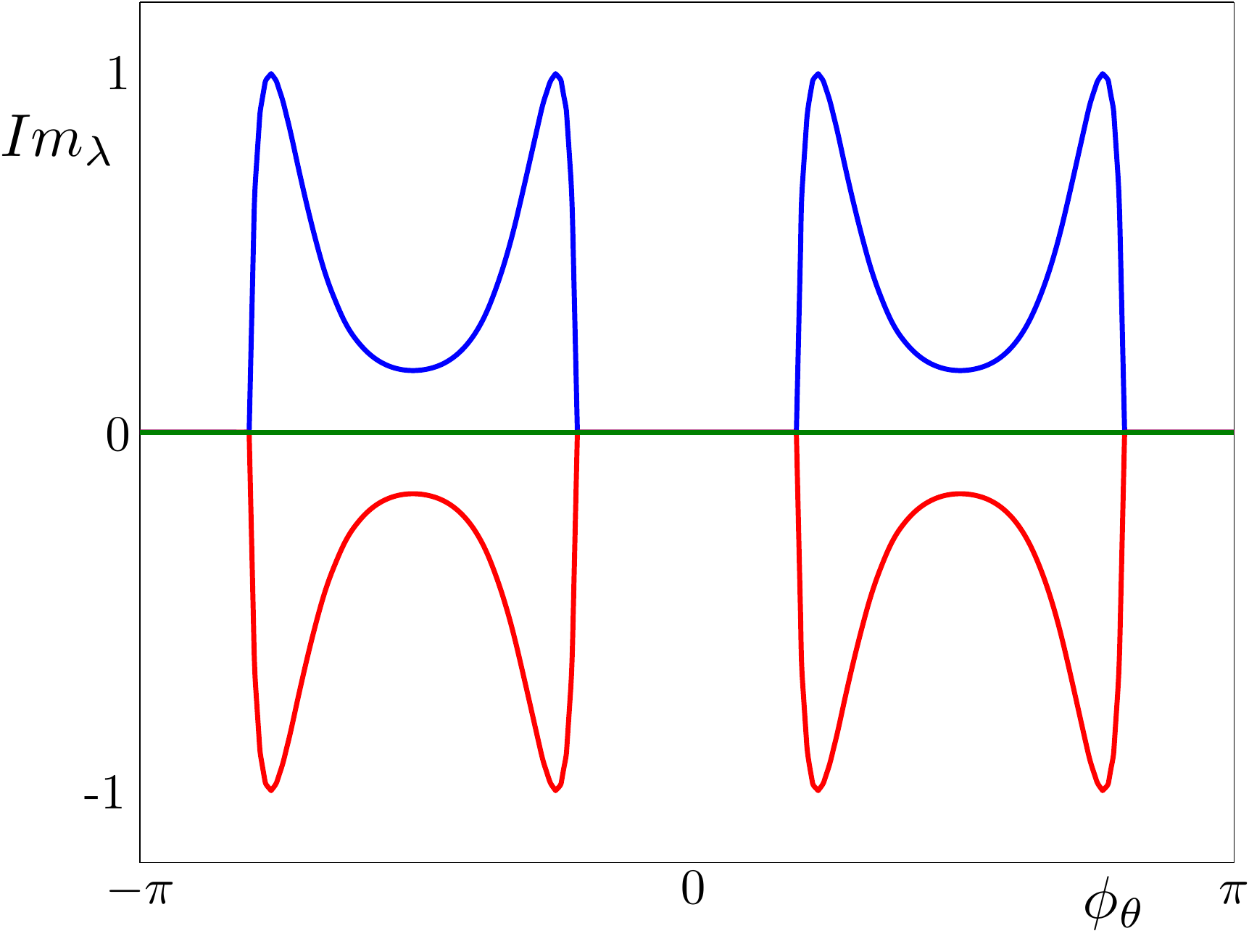}}		
		\end{subfigure}
	\caption{\footnotesize (Color online) Characteristic multipliers of the cases $A_y = 1, A_\theta = \pi/4, \gamma = 4, \phi_y = 0$, and $\phi_\theta$ varied from $-\pi$ to $\pi$: (a) Real and Imaginary parts in complex plane. Two characteristic multipliers always locate at $(1 , 0)$ are represented by $\bigstar$. The two complex conjugates are represented by $\bigcirc$. (b) Real and (c) Imaginary parts of characteristic multipliers. Motion becomes unstable when the two complex conjugates collide at $(1 , 0)$ and split on real axis.}
	\label{fig:eig}
\end{figure}

In Figure~\ref{fig:eig}, we examine the behavior of the real and imaginary parts of the characteristic multipliers as a function of $\phi_\theta$ for $A_y = 1, A_\theta = \pi/4, \gamma = 4$ and $\phi_y = 0$. In other words, we explore the behavior of $\lambda_i$'s for $\phi_y = 0$, along the dashed line in the left plot of Figure~\ref{fig:basecase}. One can see that two characteristic multipliers are always located at $(1 , 0)$ (represented by $\bigstar$). The other two are represented by $\bigcirc$. The dynamics changes from stable to unstable when the two conjugates collide at $(1 , 0)$ and split onto real axis. For the considered parameters, when $\phi_\theta$ varies from $-\pi$ to $\pi$, stability changes from unstable to stable and stable to unstable four times in total.

\begin{figure}[!tb]
	\centering
	\begin{subfigure}[$\gamma = 1.01$]{
		\includegraphics[width=0.17\textwidth]{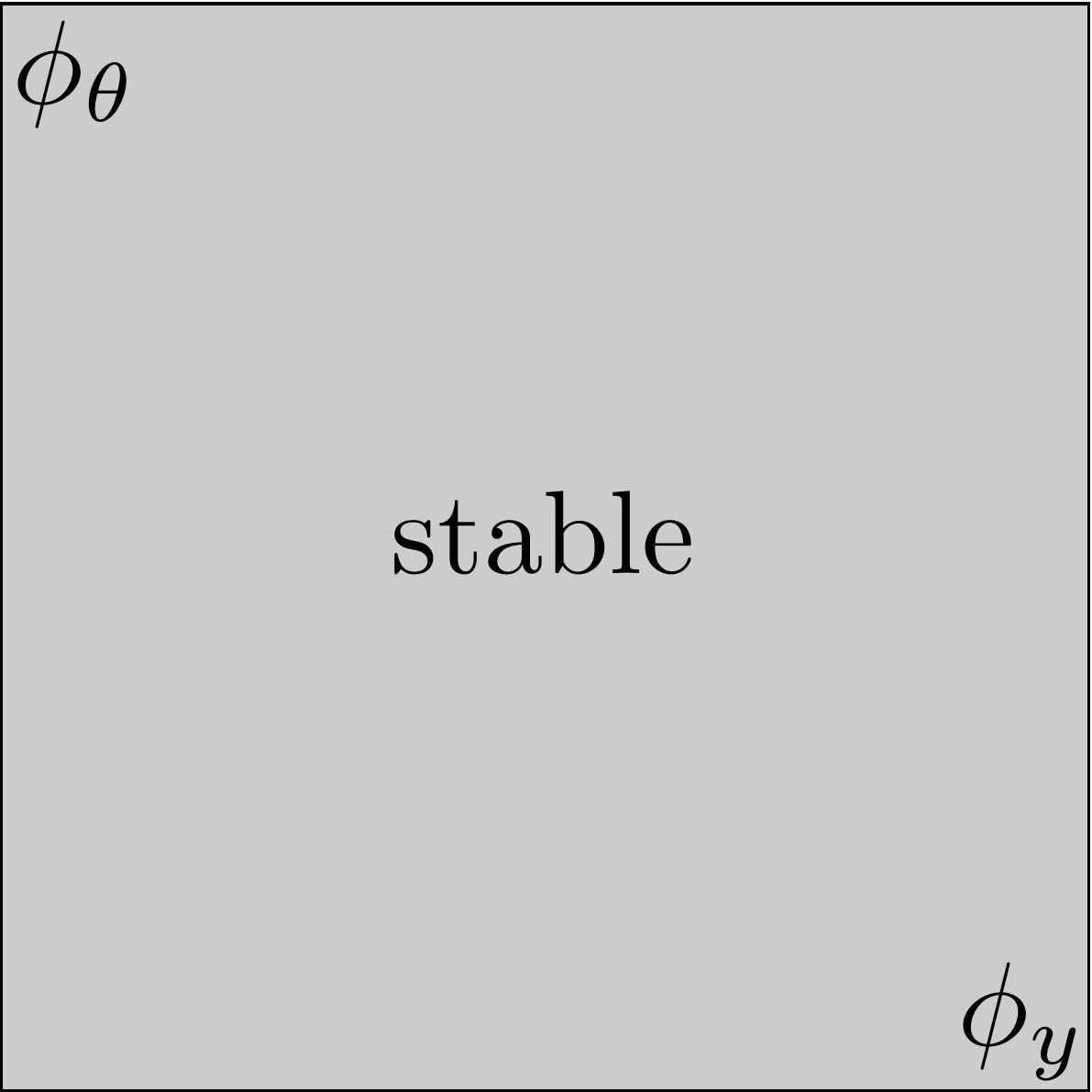}}		
	\end{subfigure}
	\begin{subfigure}[$\gamma = 1.45$]{
		\includegraphics[width=0.17\textwidth]{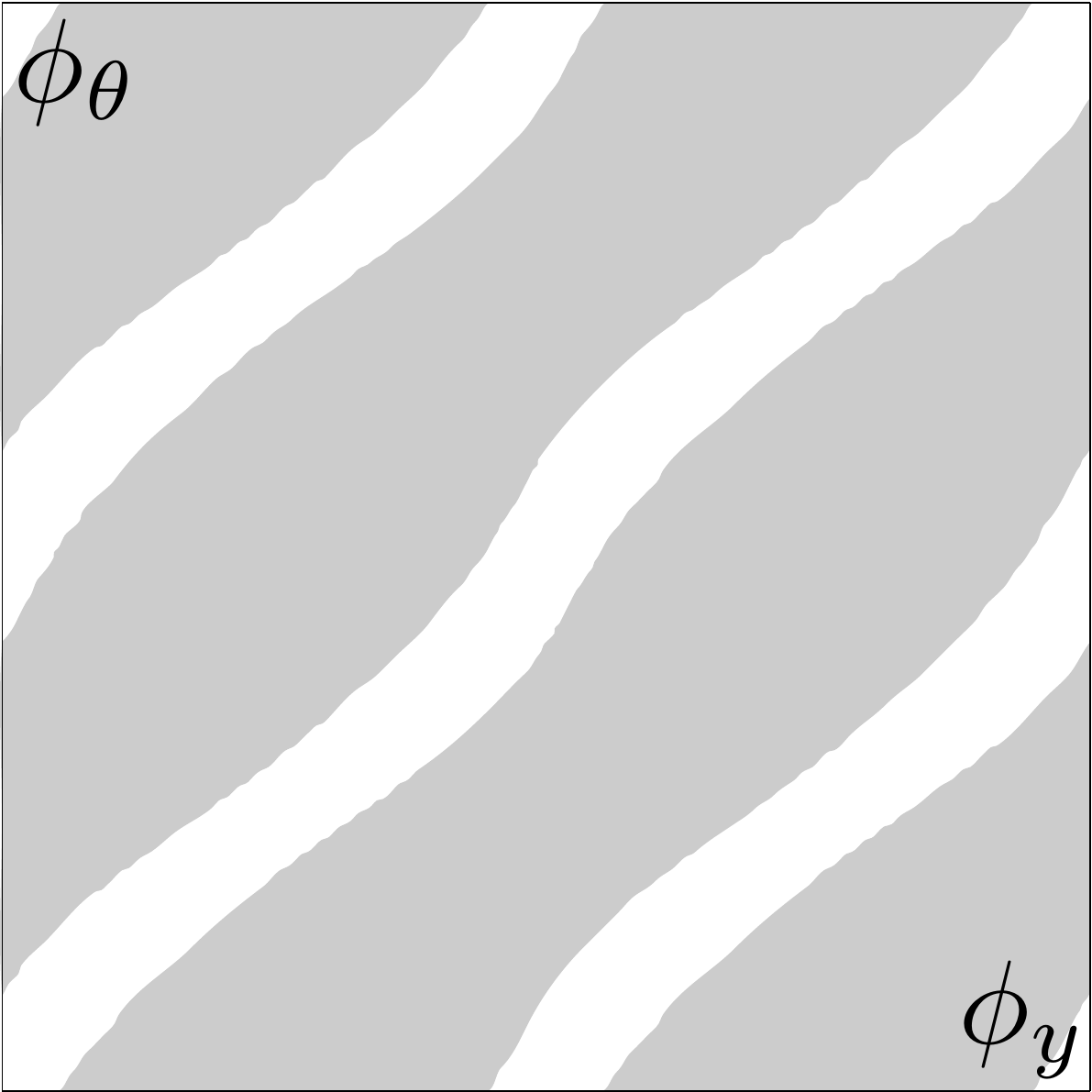}}		
	\end{subfigure}
	\begin{subfigure}[$\gamma = 1.75$]{
		\includegraphics[width=0.17\textwidth]{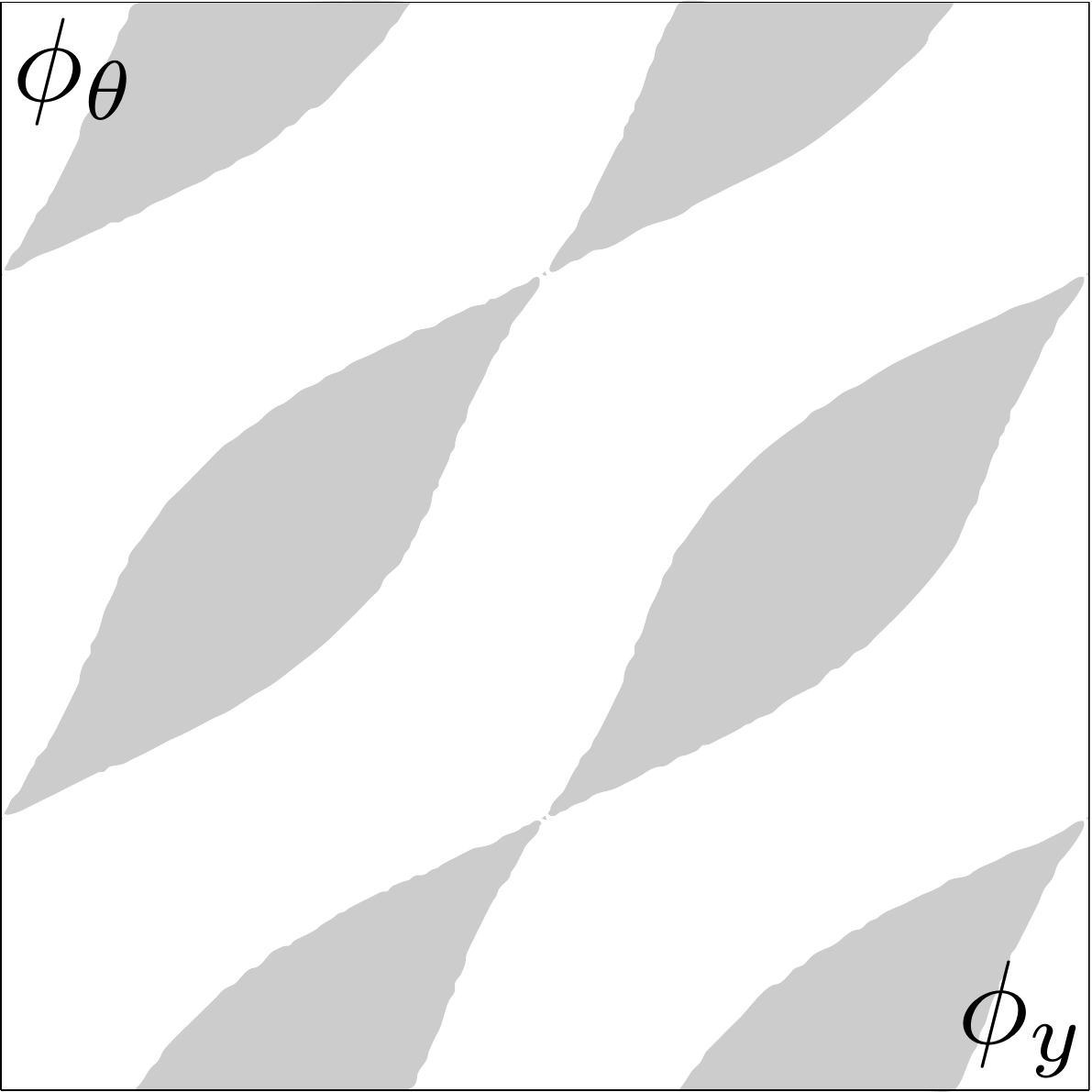}}		
	\end{subfigure}
	\begin{subfigure}[$\gamma = 2.5$]{
		\includegraphics[width=0.17\textwidth]{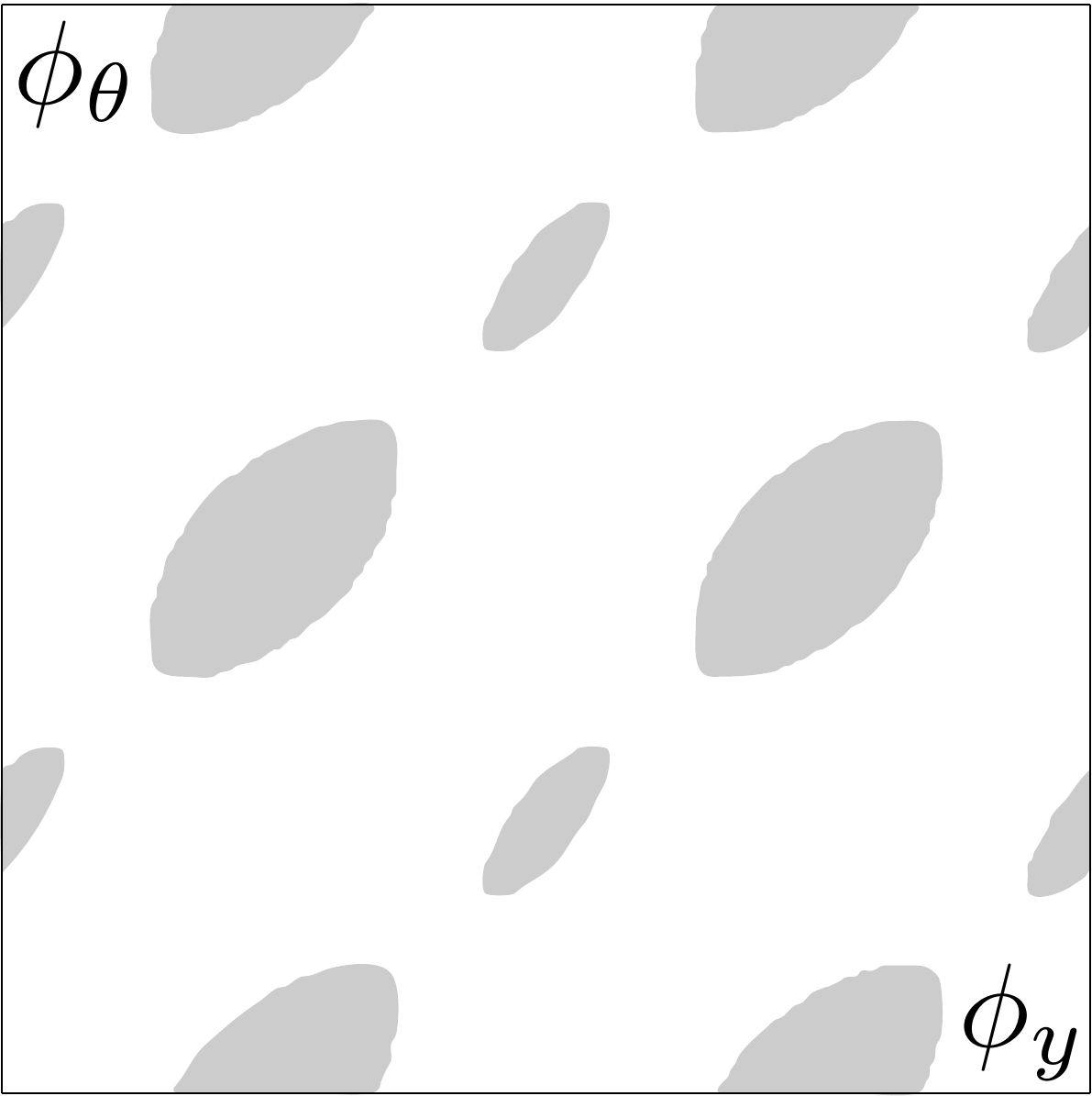}}\\
	\end{subfigure}
	\begin{subfigure}[$\gamma = 4$]{
		\includegraphics[width=0.17\textwidth]{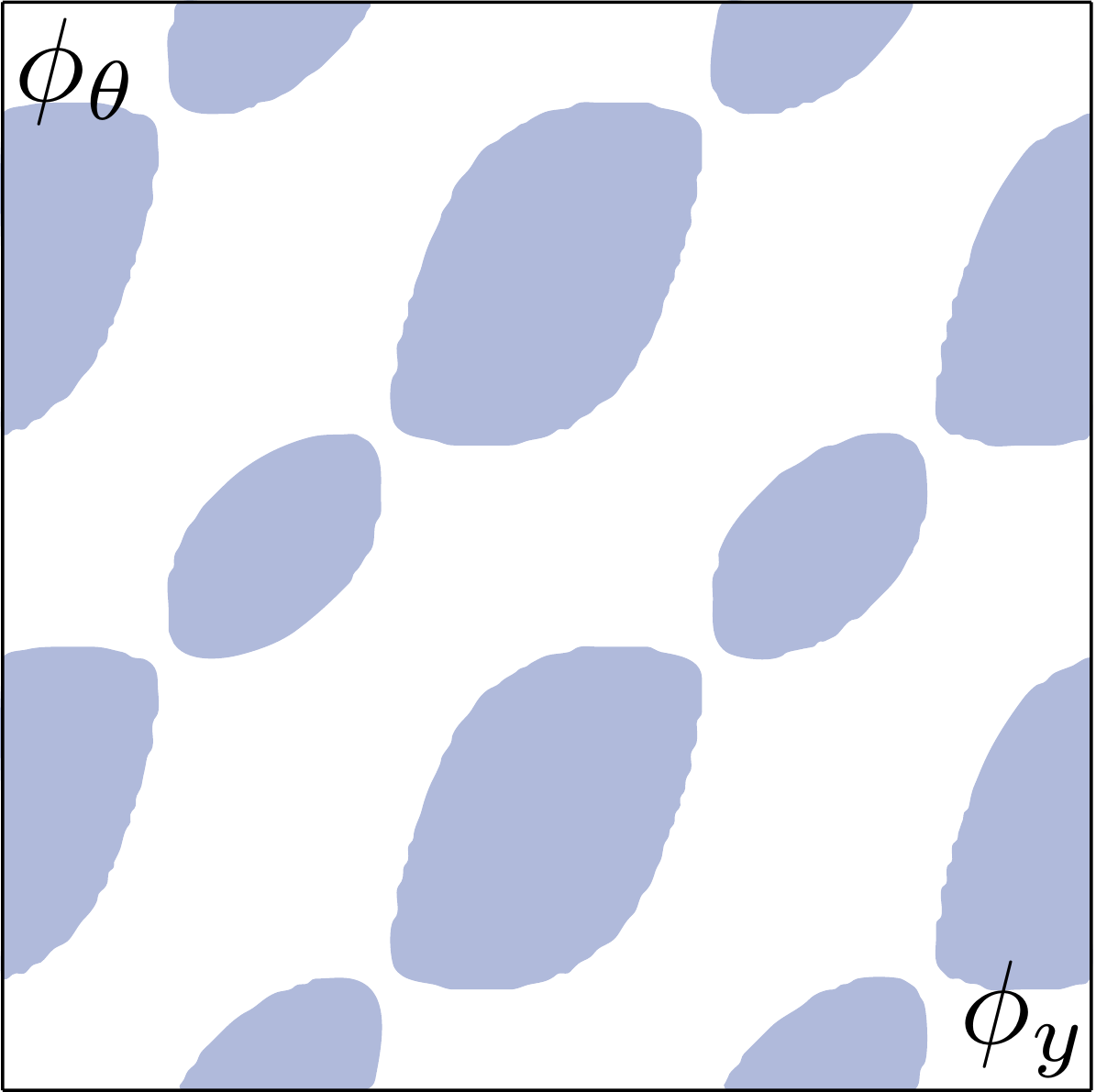}}		
	\end{subfigure}
	\begin{subfigure}[$\gamma = 8$]{
		\includegraphics[width=0.17\textwidth]{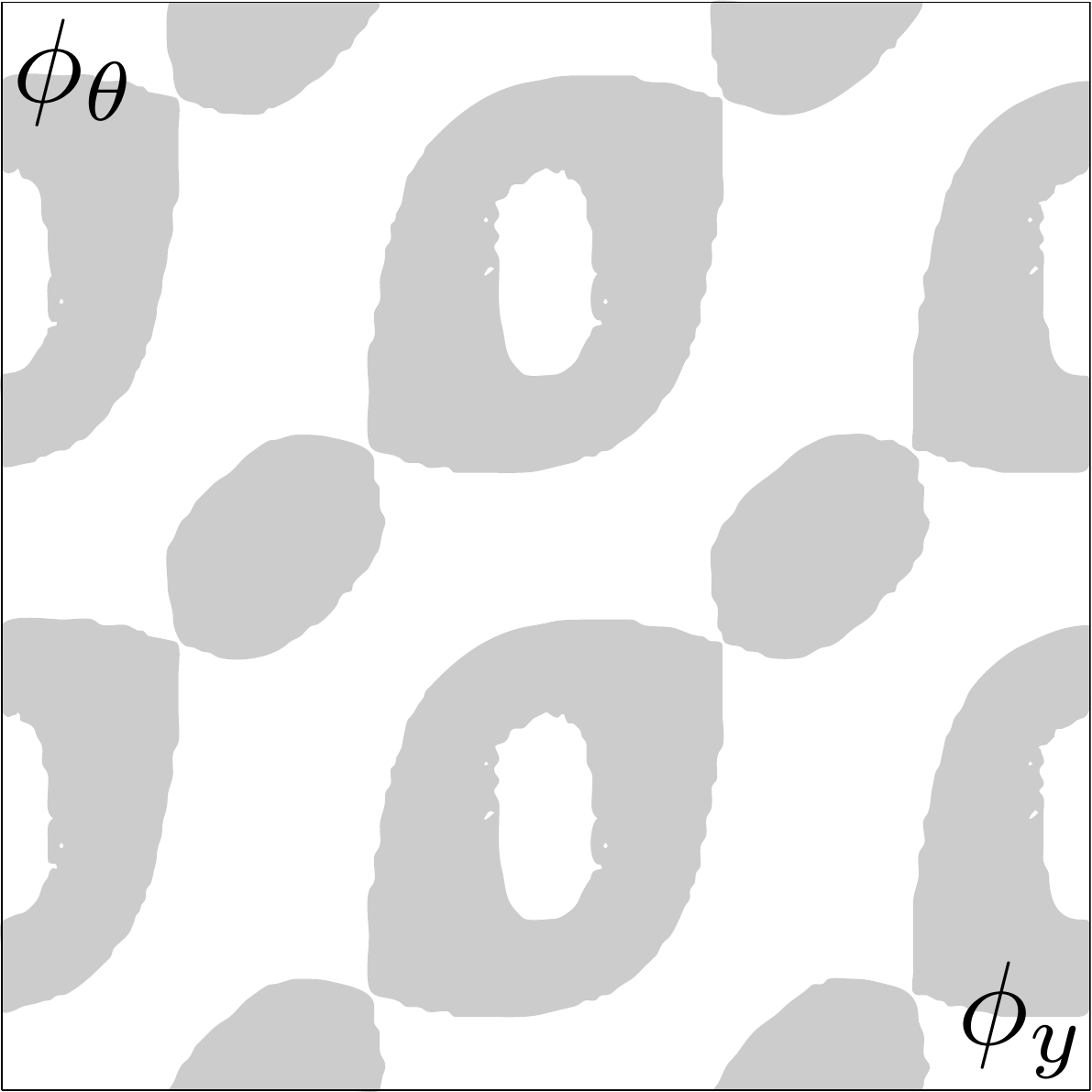}}		
	\end{subfigure}
	\begin{subfigure}[$\gamma = 16$]{
		\includegraphics[width=0.17\textwidth]{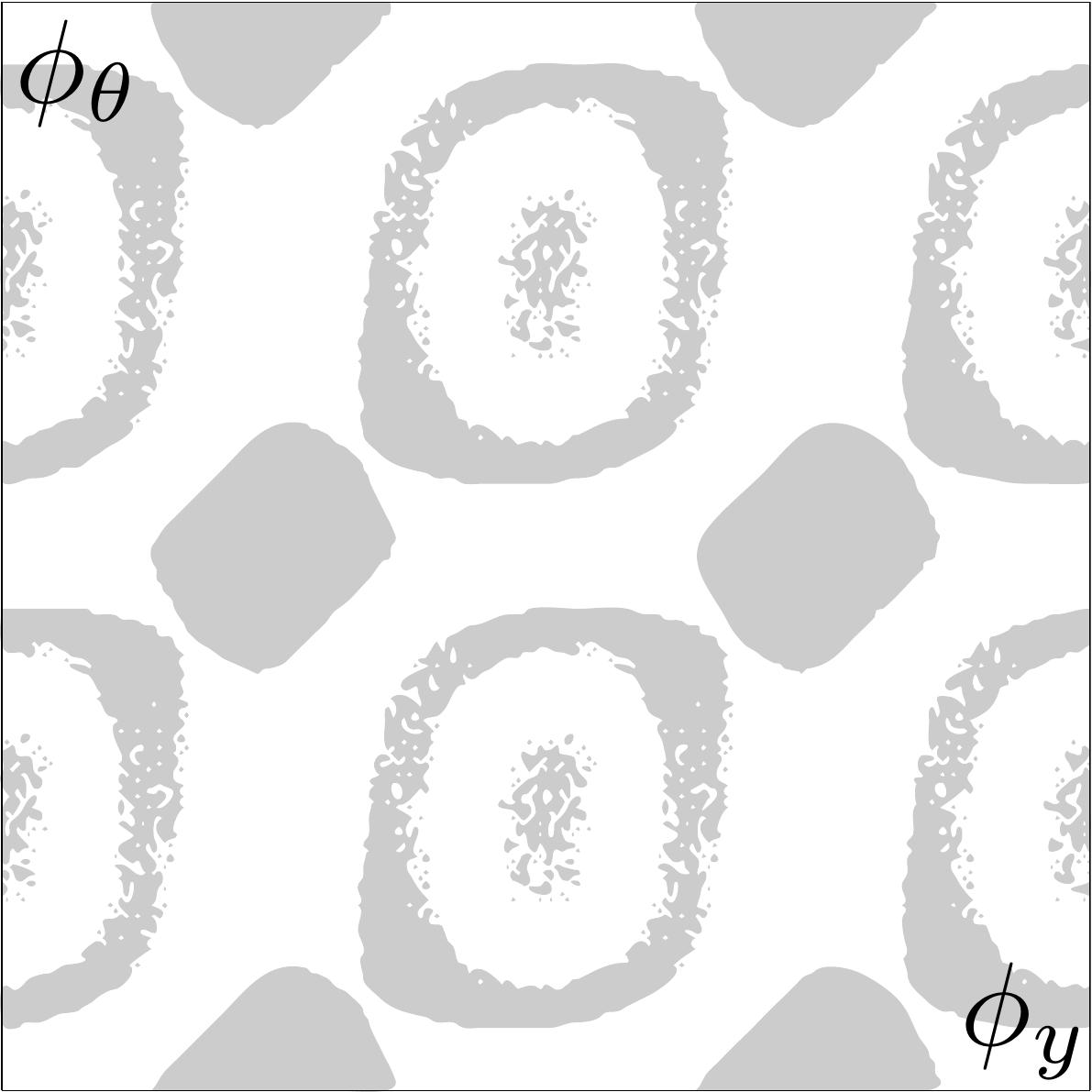}}		
	\end{subfigure}
	\begin{subfigure}[$\gamma = 32$]{
		\includegraphics[width=0.17\textwidth]{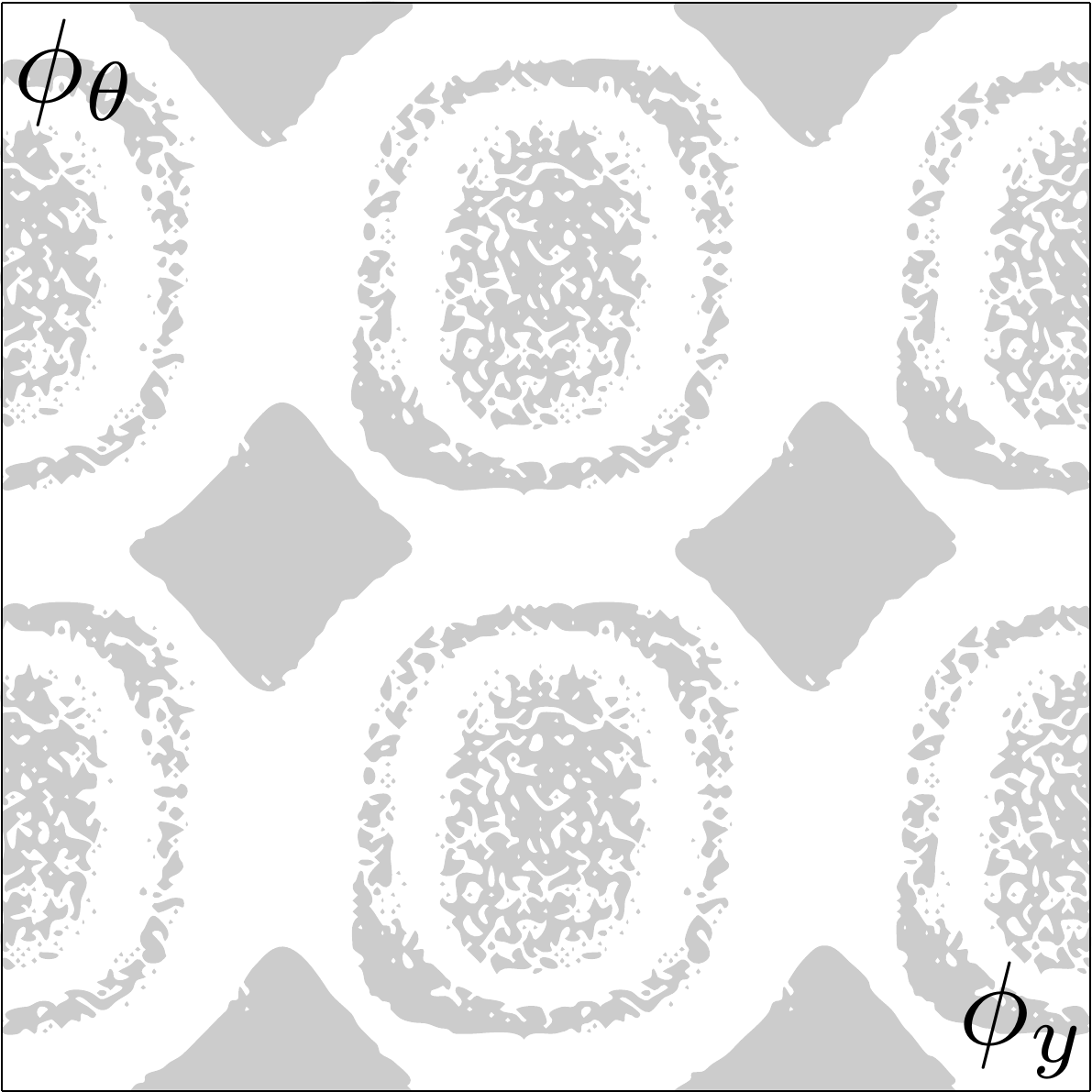}}		
	\end{subfigure}
	\caption{\footnotesize Stability regions for the cases $A_y = 1, A_\theta = \pi/4$ and various aspect ratio $\gamma$. Each plot is evaluated on a $201 \times 201$ mesh in $[-\pi \,,\, \pi] \times [-\pi \,,\, \pi]$ in $(\phi_y , \phi_\theta)$ plane. Shaded areas correspond to stable cases, white areas correspond to unstable cases.}
	\label{fig:varygamma}
\end{figure}

We now examine the stability behavior as we change  $\gamma, A_y$ and $A_\theta$, respectively. Figure~\ref{fig:varygamma} shows stability regions for $A_y = 1, A_\theta = \frac{\pi}{4}$ while varying $\gamma$. For bodies closer to circular shape ($\gamma = 1.01$), the motion is stable for all $(\phi_y , \phi_\theta)$ (but this stability property is not very useful since the net displacement is almost zero). When $\gamma > 1.43$, unstable regions start to appear. As $\gamma$ increases, unstable regions grow while stable regions shrink. The total area of stable regions becomes minimum when $\gamma \approx 2$, and the stable areas around $(\phi_y , \phi_\theta) = ((m+1/2)\pi, n\pi)$ persist. Interestingly, as $\gamma$ continues to increase, new stable regions start to emerge and grow from the previous unstable areas around $(\phi_y , \phi_\theta) = (n\pi , (m+1/2)\pi)$. Then, at these spots, unstable regions emerge and grow from the newly formed stable regions, and so on and so forth. The boundaries between stable and unstable regions around $(n\pi , (m+1/2)\pi)$ become blurry as $\gamma$ becomes larger, and $((m+1/2)\pi, n\pi)$ remain stable. This trend is reminiscent of the phenomenon observed in Spagnolie et al.~\cite{SpMoShZh2010}, in which the authors noticed the motion of an elliptic body subject to prescribed heaving and passive pitching goes through states from ``coherence to incoherence, and back again'' as the aspect ratio changes. Note that the latter studies are in viscous fluid whereas the analysis here is for an inviscid fluid model. Interestingly, this simplified model is able to capture, at least qualitatively, the behavior observed in~\cite{SpMoShZh2010}. 

\begin{figure}[!tb]
	\centering
	\begin{subfigure}[$A_y = 0.01$]{
		\includegraphics[width=0.17\textwidth]{stable.pdf}}		
	\end{subfigure}
	\begin{subfigure}[$A_y = 0.6$]{
		\includegraphics[width=0.17\textwidth]{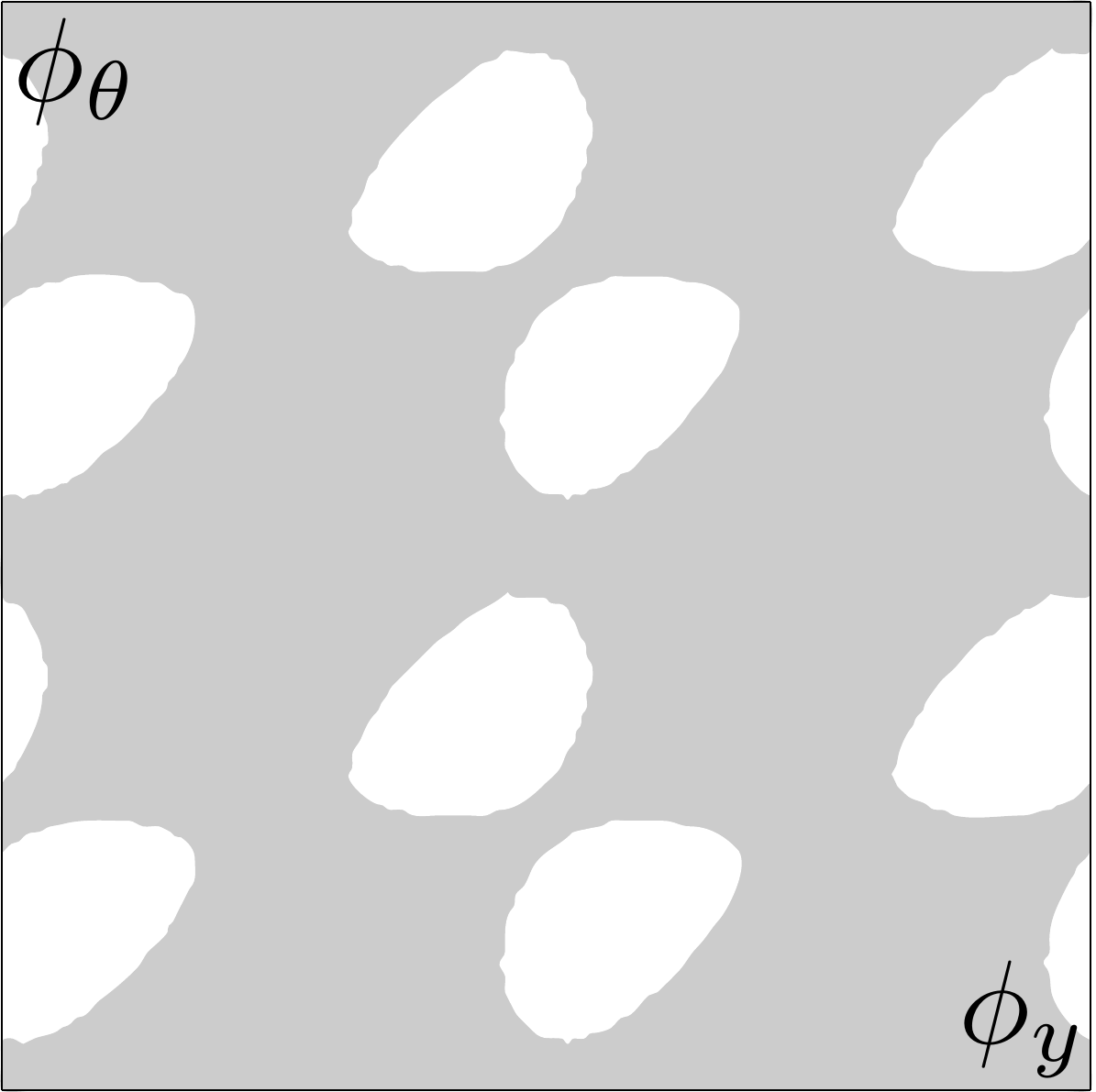}}		
	\end{subfigure}
	\begin{subfigure}[$A_y = 0.625$]{
		\includegraphics[width=0.17\textwidth]{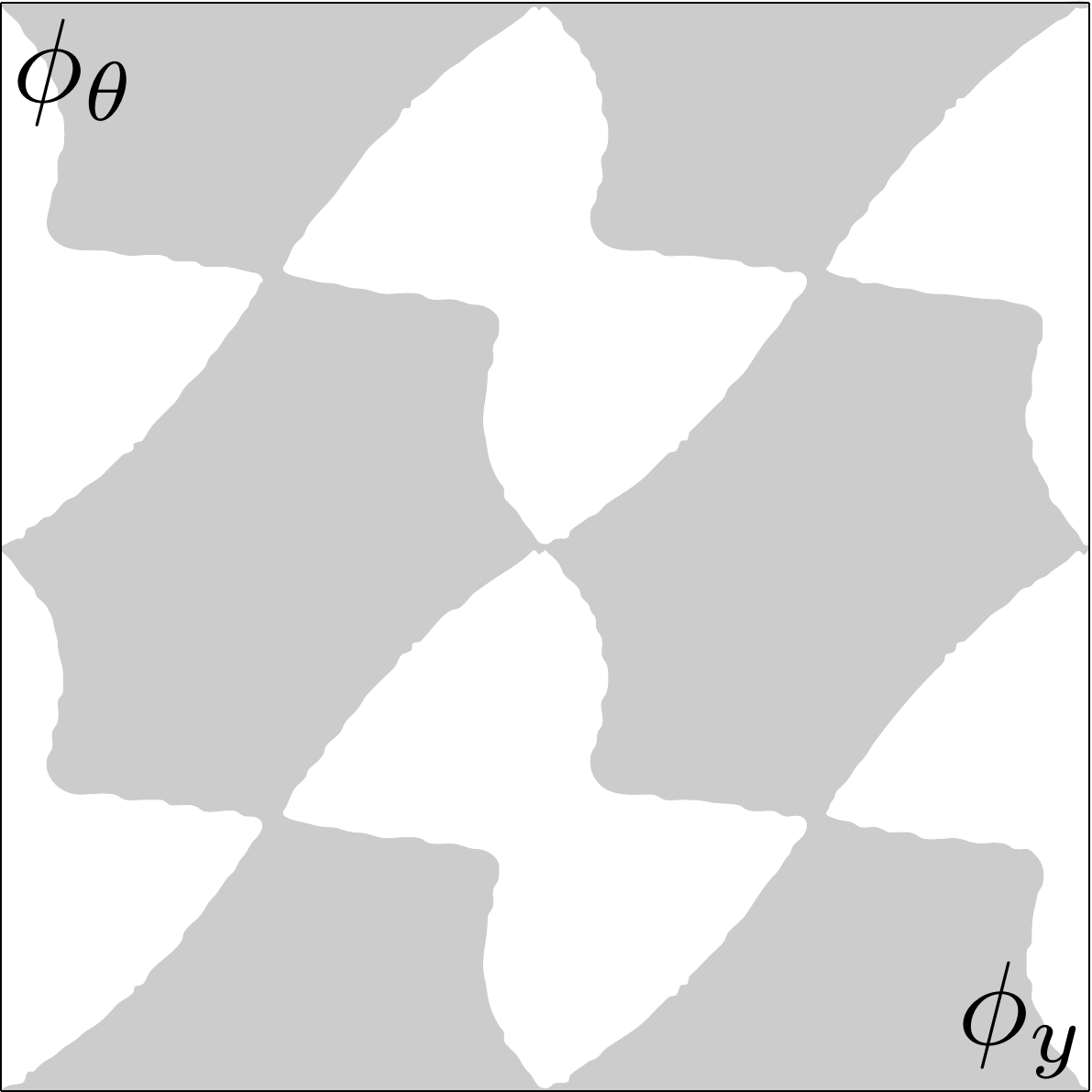}}
	\end{subfigure} 
	\begin{subfigure}[$A_y = 0.75$]{
		\includegraphics[width=0.17\textwidth]{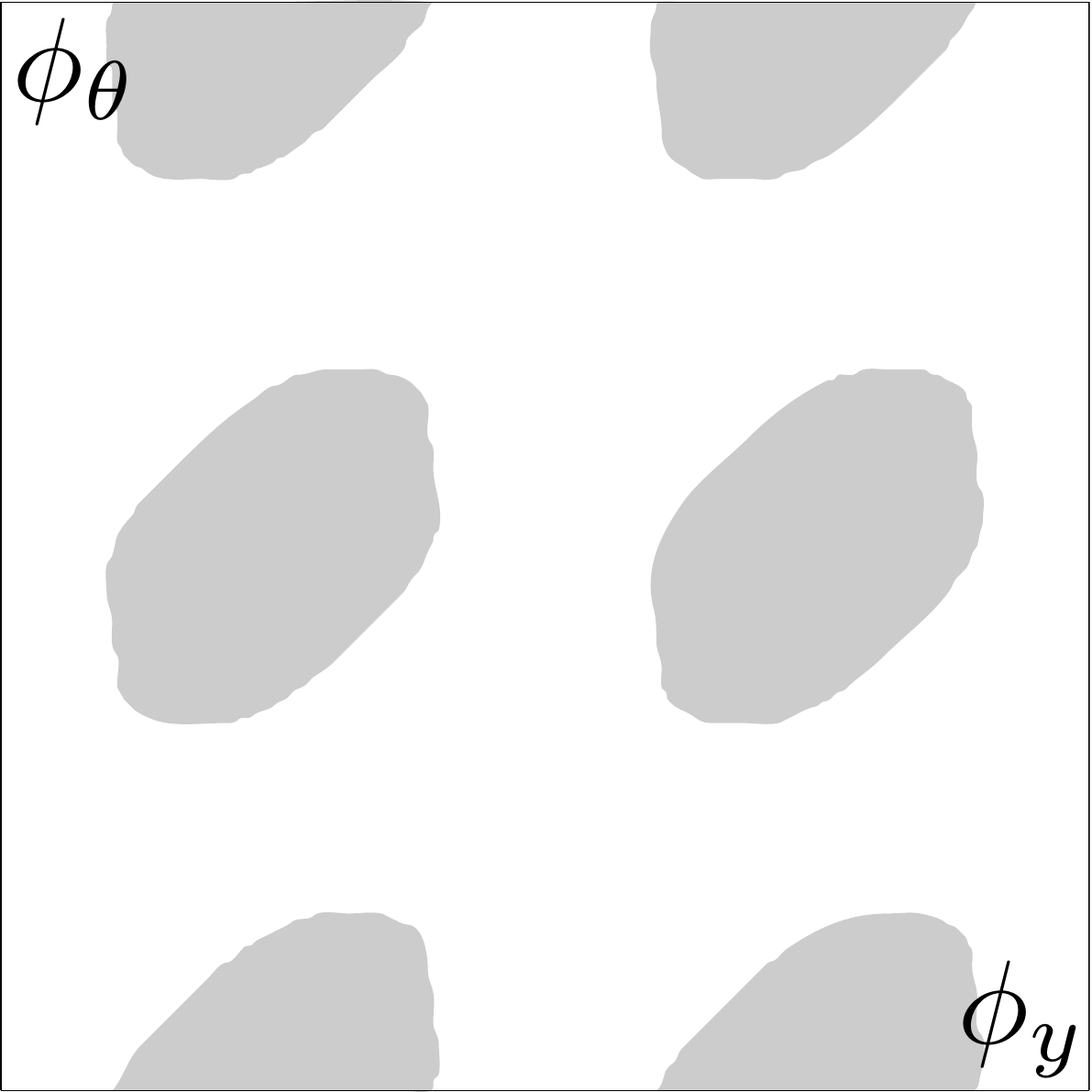}}		
	\end{subfigure}
	\begin{subfigure}[$A_y = 1$]{
		\includegraphics[width=0.17\textwidth]{sr_4_1_pi4.pdf}}
	\end{subfigure}\\
	\begin{subfigure}[$A_y = 1.25$]{
		\includegraphics[width=0.17\textwidth]{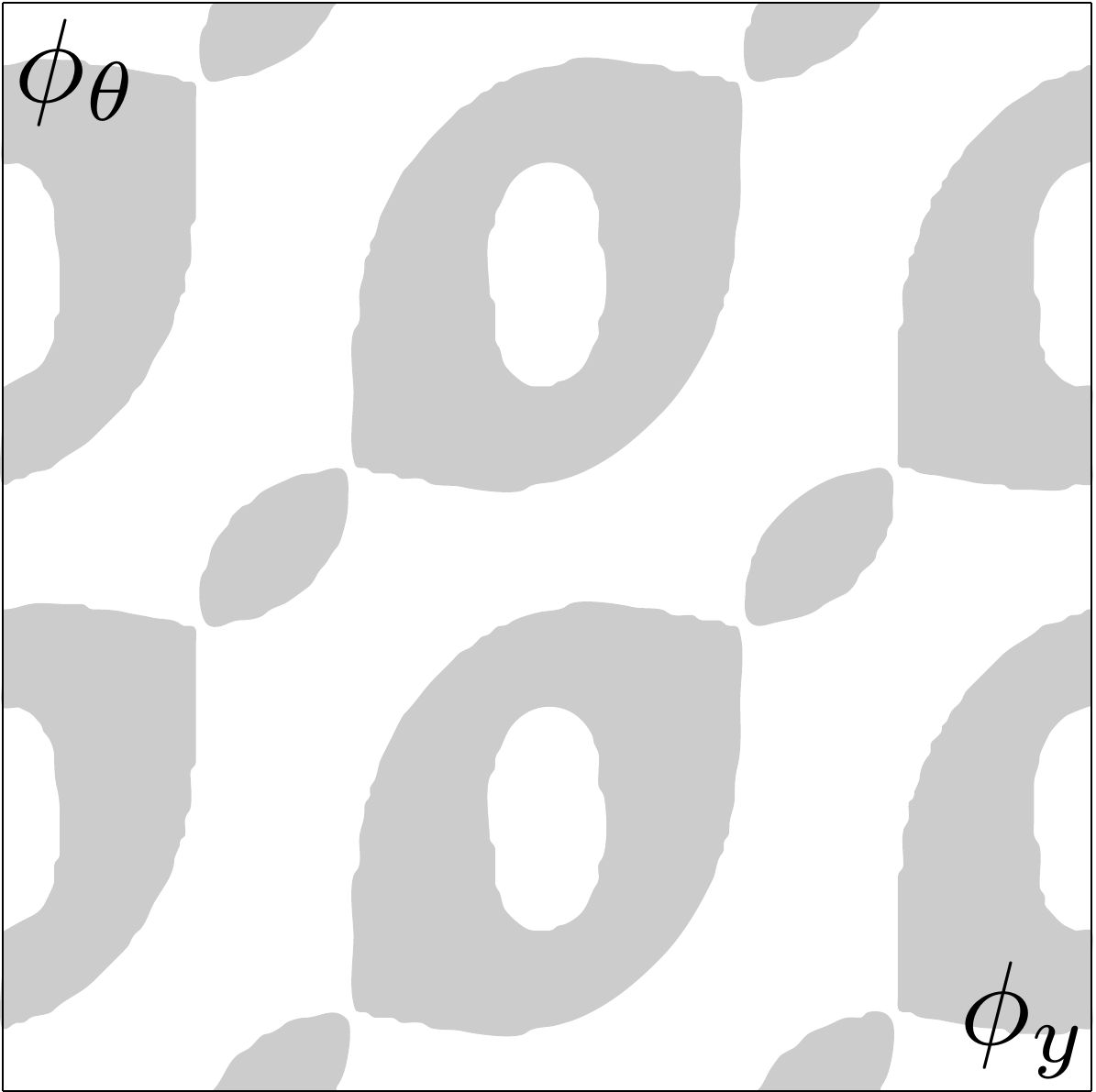}}		
	\end{subfigure}
	\begin{subfigure}[$A_y = 1.5$]{
		\includegraphics[width=0.17\textwidth]{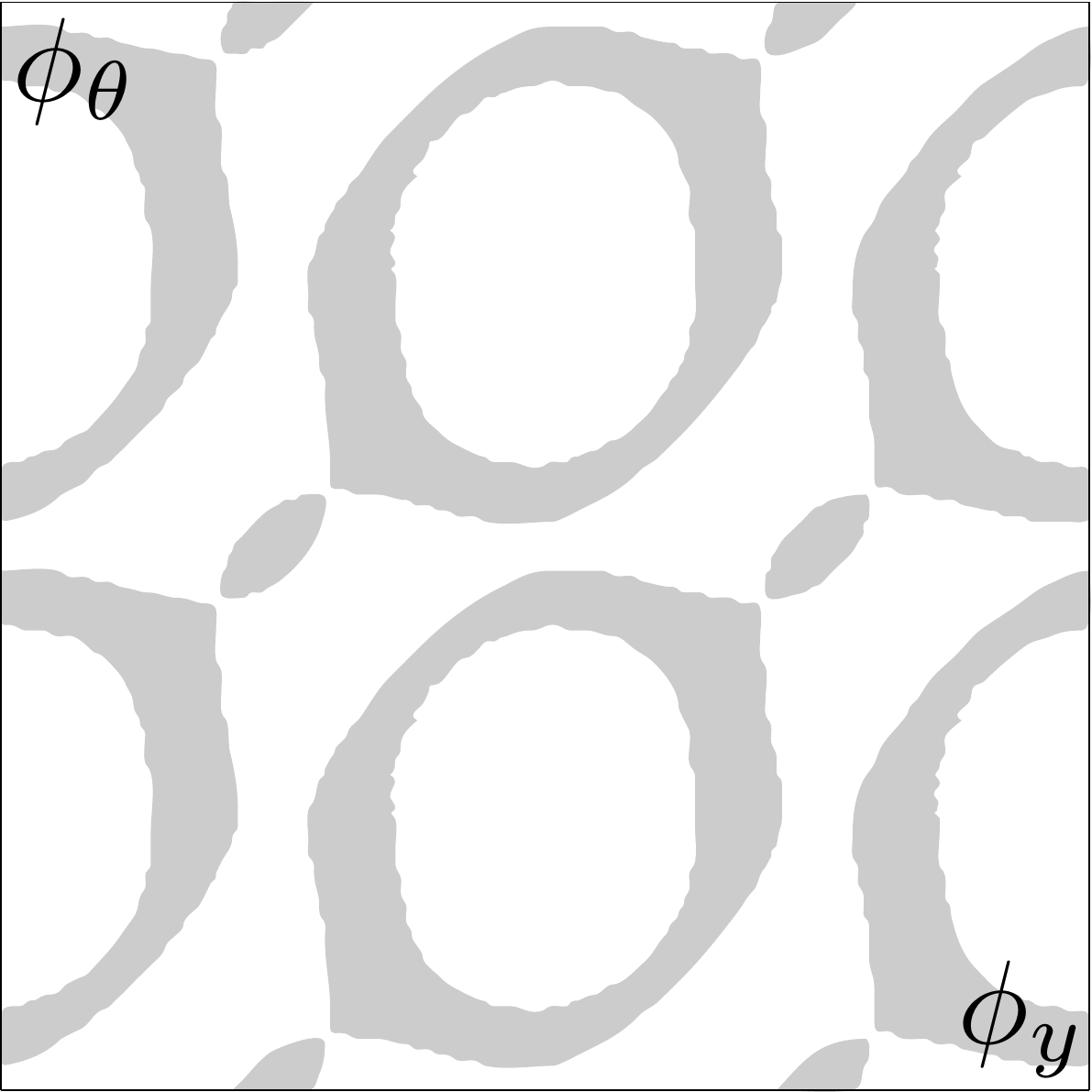}}
	\end{subfigure} 
	\begin{subfigure}[$A_y = 1.75$]{
		\includegraphics[width=0.17\textwidth]{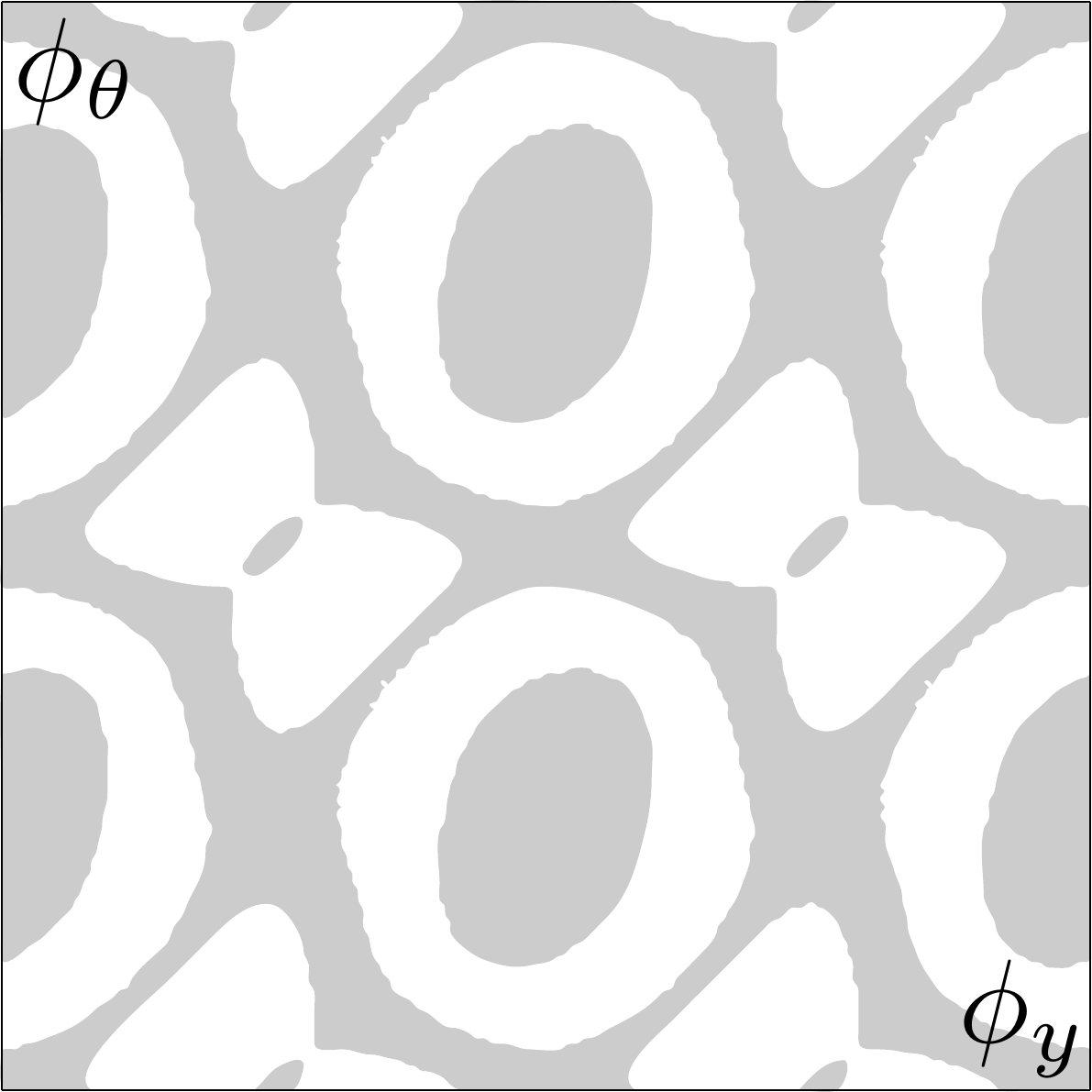}}		
	\end{subfigure}
	\begin{subfigure}[$A_y = 2$]{
		\includegraphics[width=0.17\textwidth]{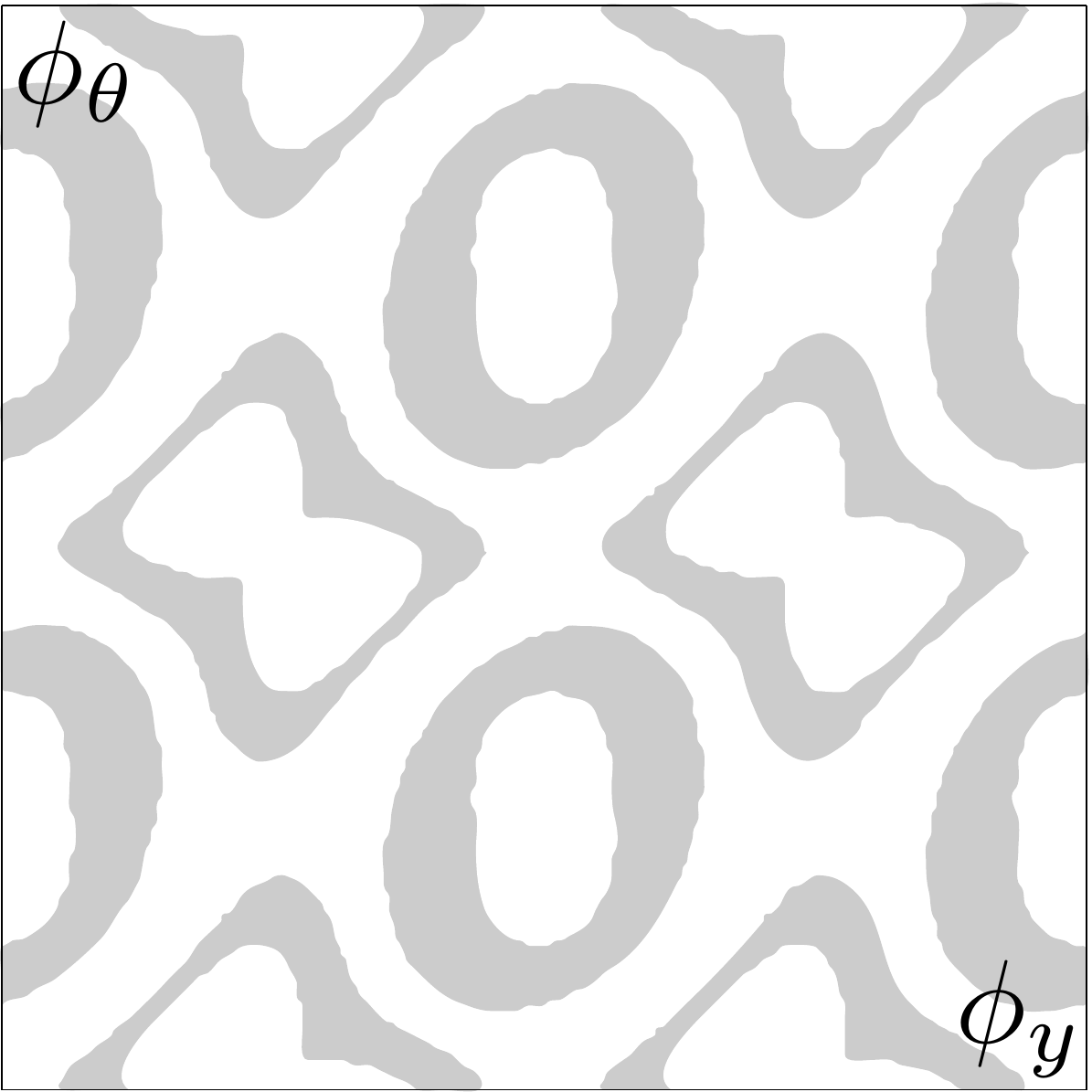}}		
	\end{subfigure}
	\begin{subfigure}[$A_y = 6$]{
		\includegraphics[width=0.17\textwidth]{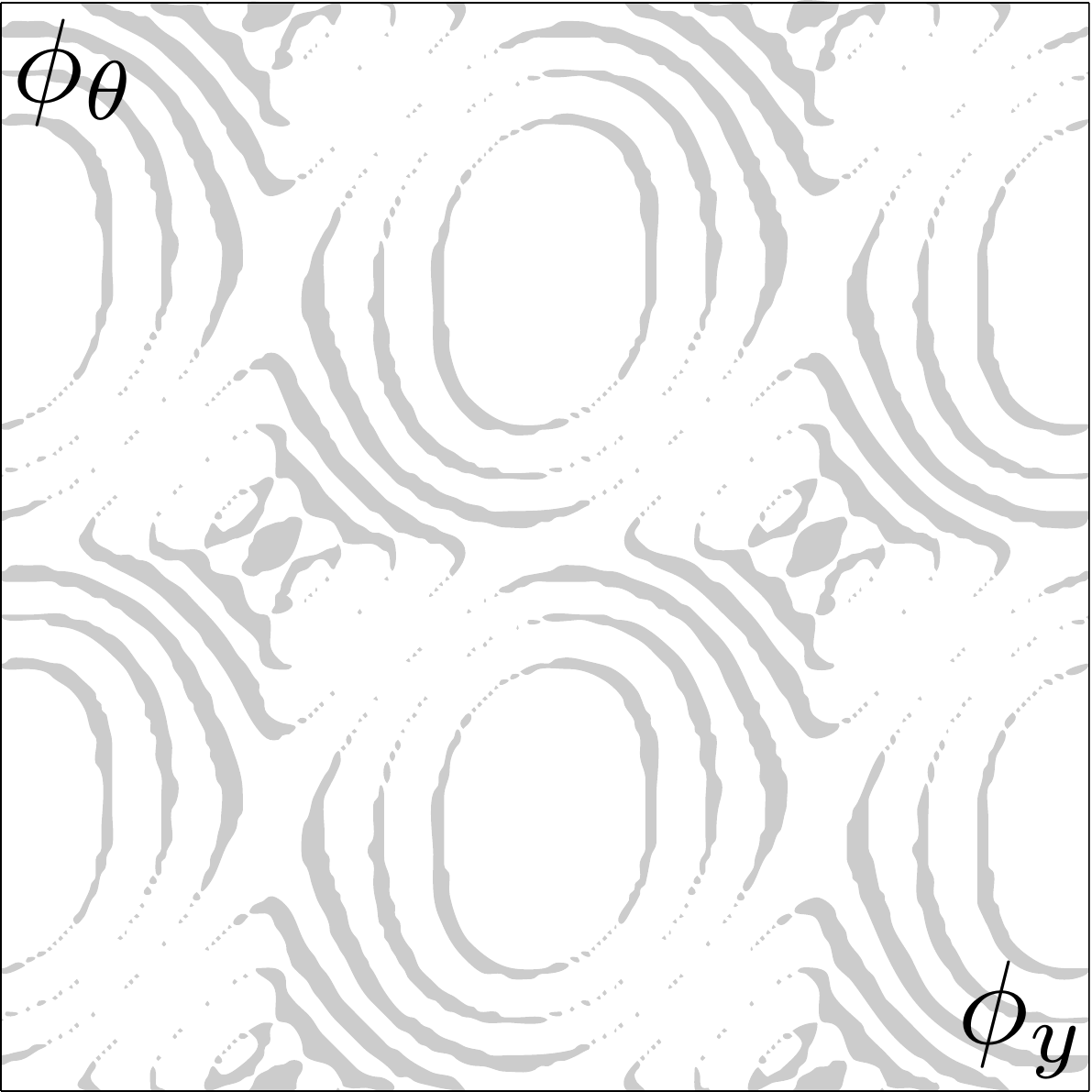}}		
	\end{subfigure}
	\caption{\footnotesize Stability regions for the cases $A_\theta = \pi/4$, $\gamma = 4$ and various heaving amplitude $A_y$. Each plot is evaluated on a $201 \times 201$ mesh in $[-\pi \,,\, \pi] \times [-\pi \,,\, \pi]$ in $(\phi_y , \phi_\theta)$ plane. Shaded areas correspond to stable cases, white areas correspond to unstable cases.}
	\label{fig:varyay}
\end{figure}
\begin{figure}[!tb]
	\centering
	\begin{subfigure}[$A_\theta = 0.01$]{
		\includegraphics[width=0.17\textwidth]{stable.pdf}}
	\end{subfigure}
	\begin{subfigure}[$A_\theta = \pi/8$]{
		\includegraphics[width=0.17\textwidth]{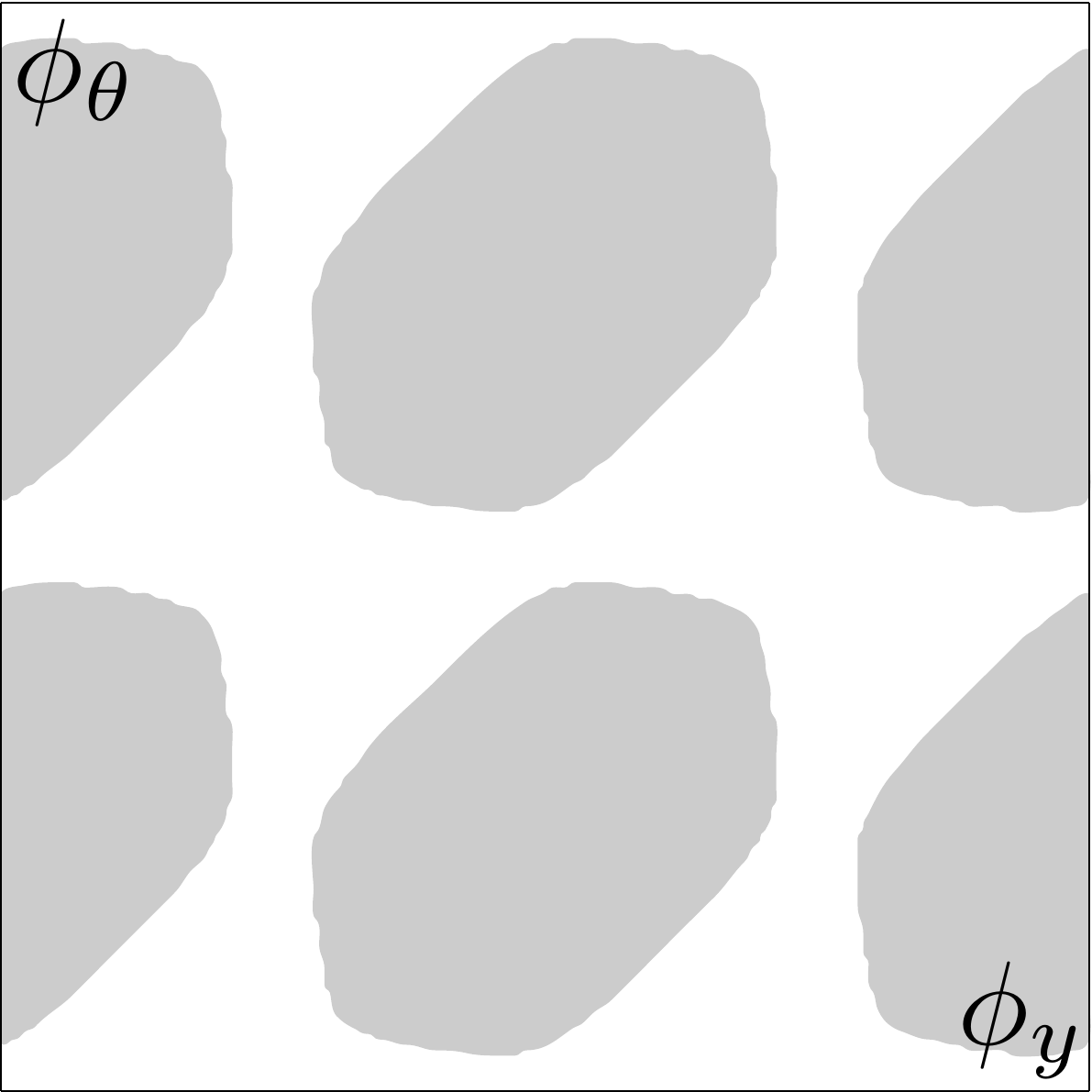}}		
	\end{subfigure}
	\begin{subfigure}[$A_\theta = 3\pi/16$]{
		\includegraphics[width=0.17\textwidth]{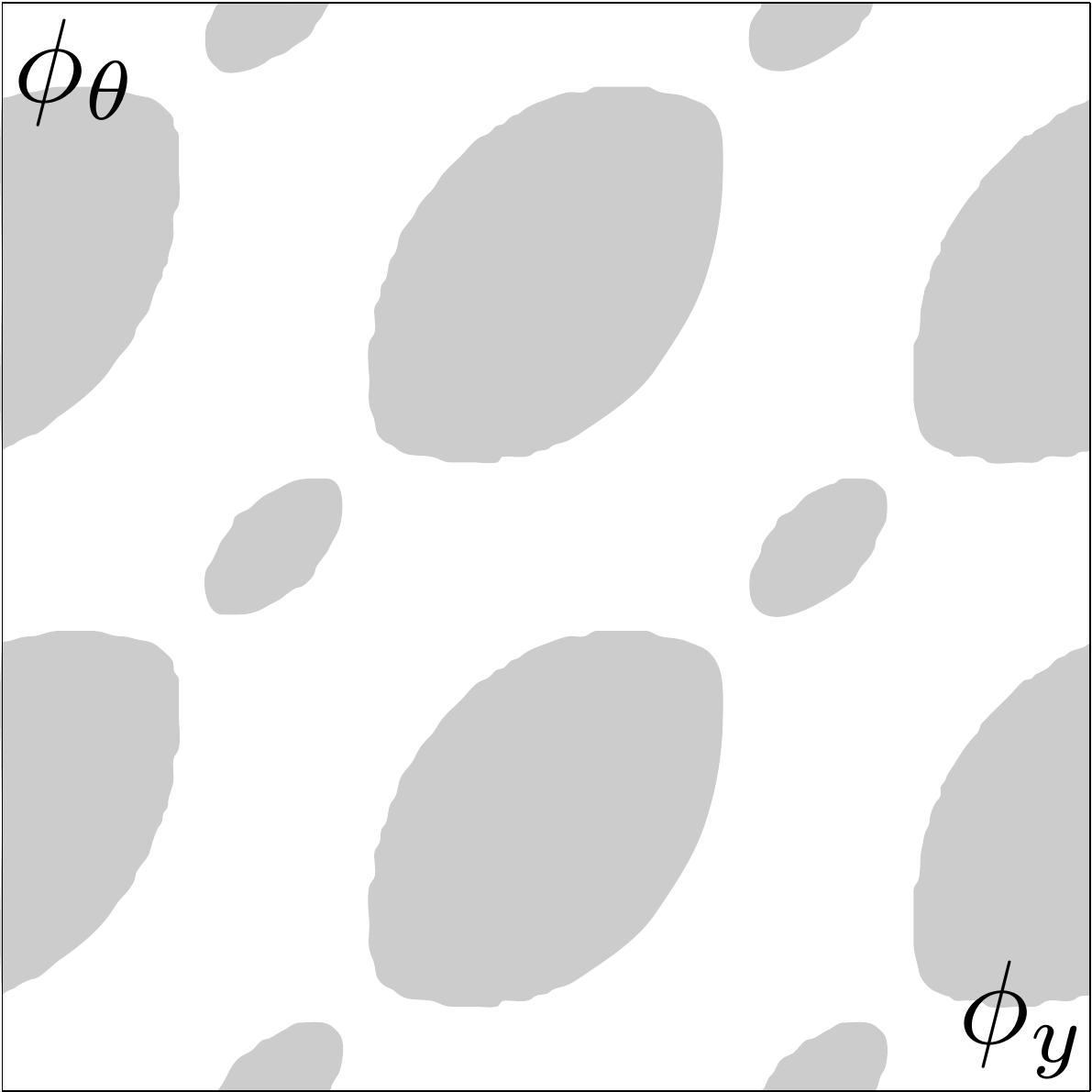}}
	\end{subfigure}
	\begin{subfigure}[$A_\theta = \pi/4$]{
		\includegraphics[width=0.17\textwidth]{sr_4_1_pi4.pdf}}		
	\end{subfigure}\\
	\begin{subfigure}[$A_\theta = 5\pi/16$]{
		\includegraphics[width=0.17\textwidth]{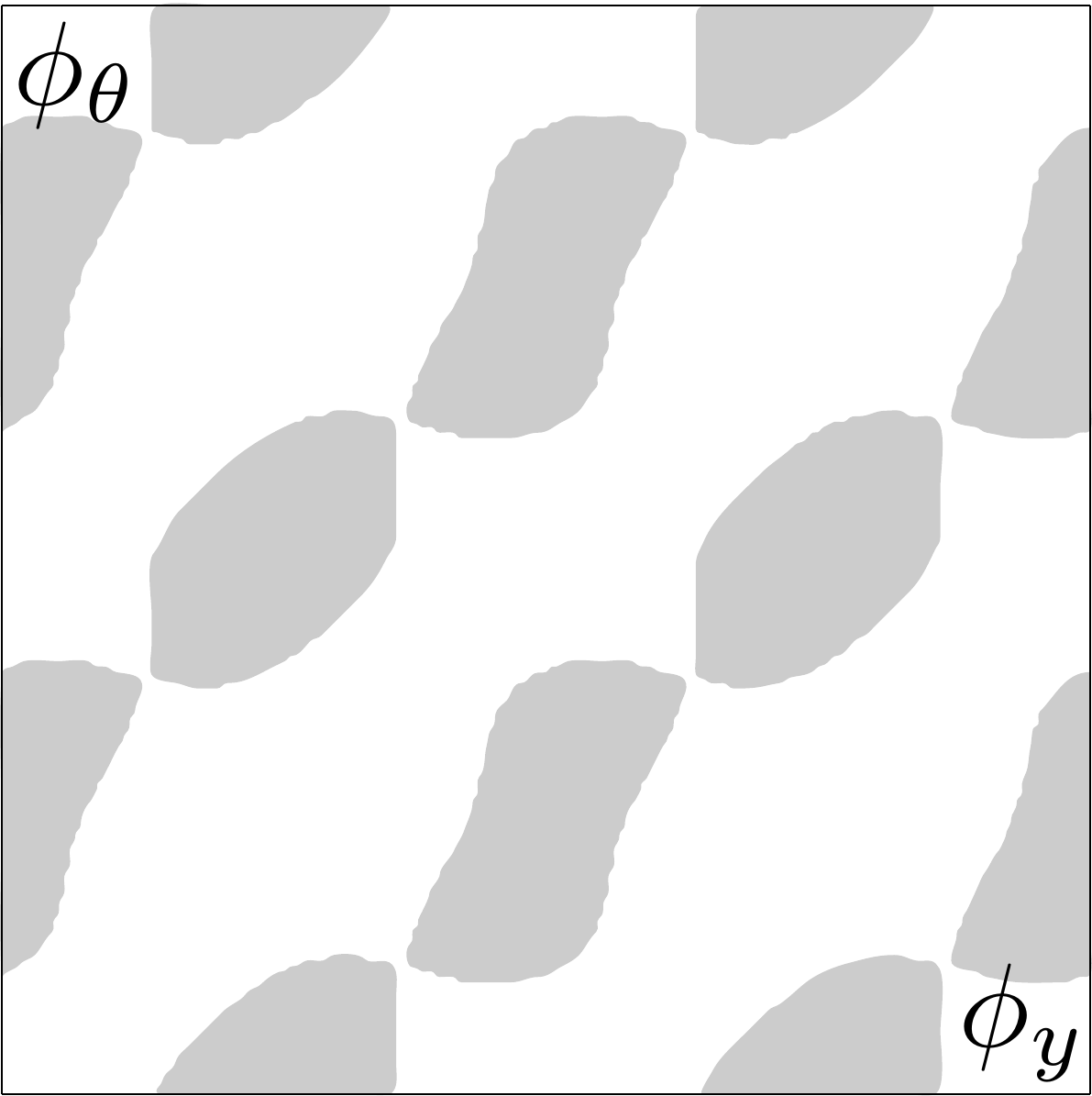}}
	\end{subfigure}
	\begin{subfigure}[$A_\theta = 3\pi/8$]{
		\includegraphics[width=0.17\textwidth]{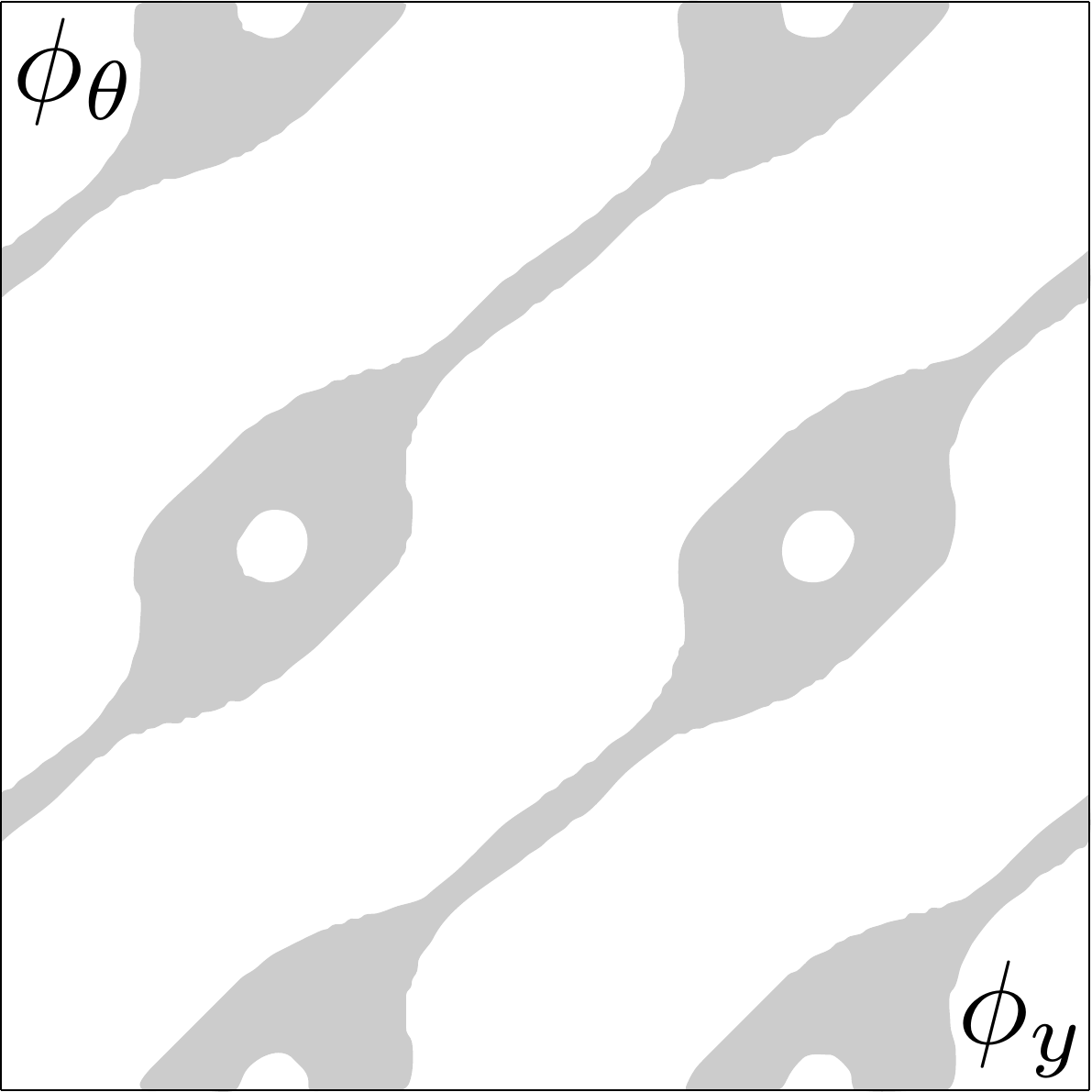}}		
	\end{subfigure}
	\begin{subfigure}[$A_\theta = 7\pi/16$]{
		\includegraphics[width=0.17\textwidth]{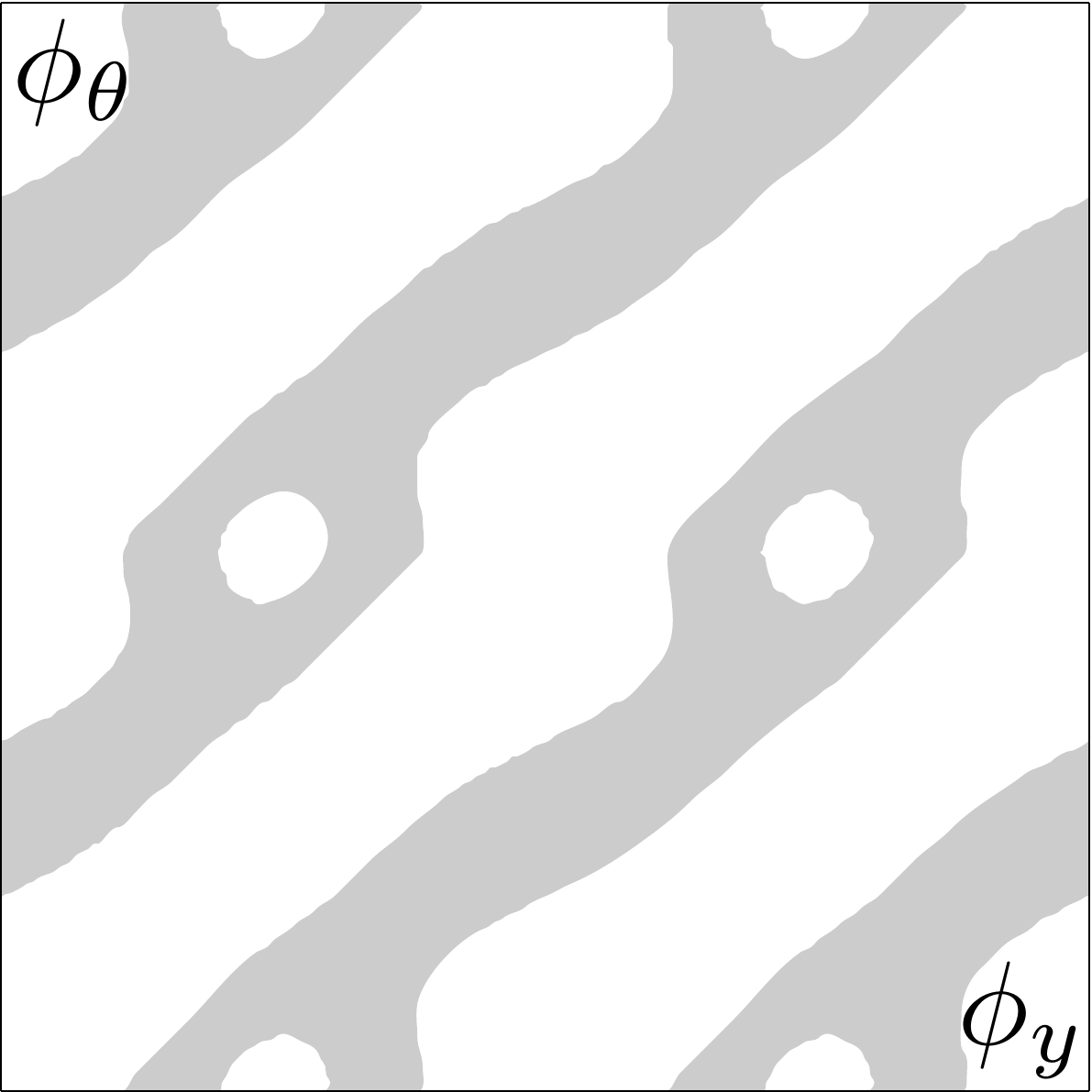}}
	\end{subfigure}
	\begin{subfigure}[$A_\theta = \pi/2$]{
		\includegraphics[width=0.17\textwidth]{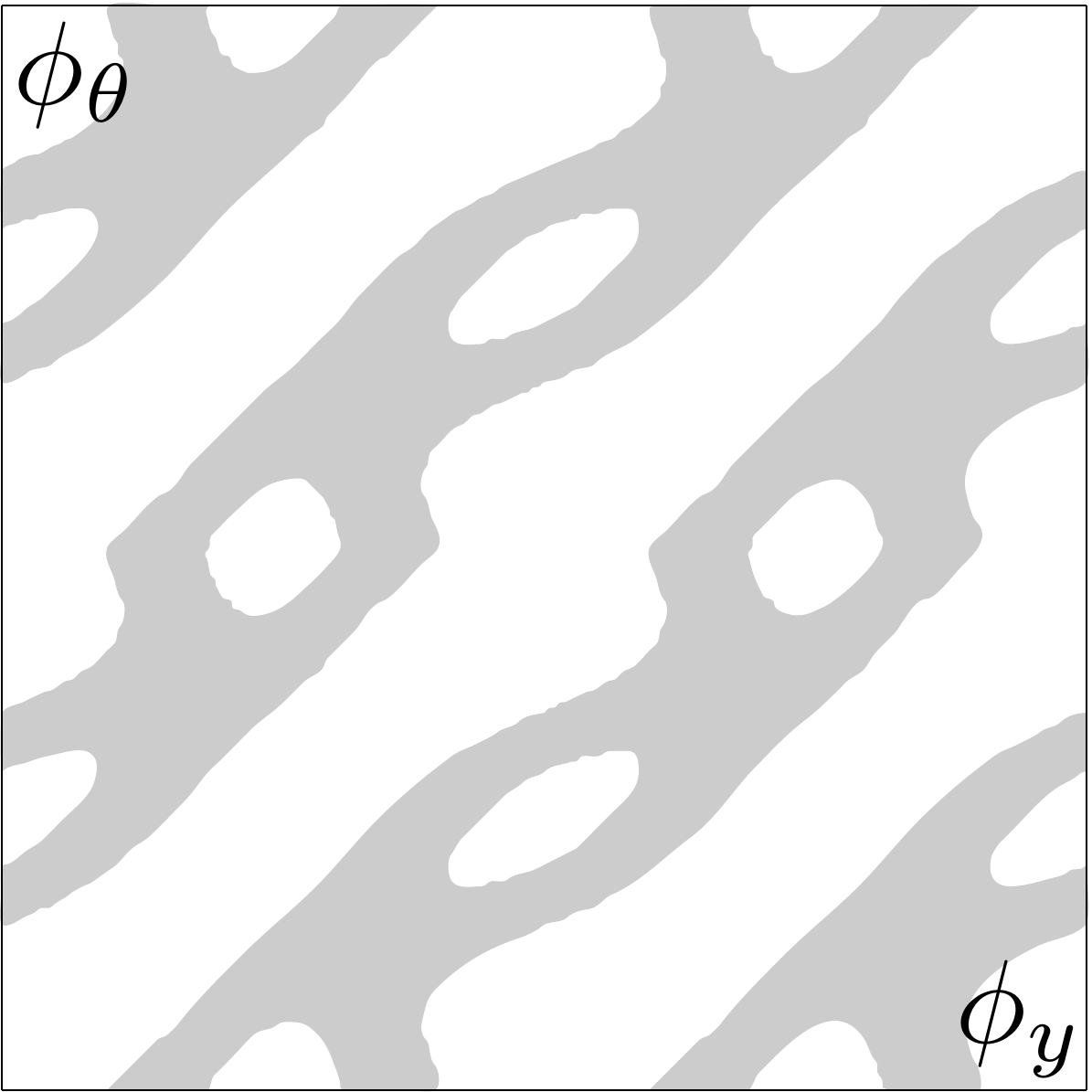}}		
	\end{subfigure}
	\caption{\footnotesize Stability regions for the cases $A_y = 1$, $\gamma = 4$ and various pitching amplitude $A_\theta$. Each plot is evaluated on a $201 \times 201$ mesh in $[-\pi \,,\, \pi] \times [-\pi \,,\, \pi]$ in $(\phi_y , \phi_\theta)$ plane. Shaded areas correspond to stable cases, white areas correspond to unstable cases.}
	\label{fig:varyatheta}
\end{figure}

Figure~\ref{fig:varyay} shows stability regions for $A_\theta = \pi/4, \gamma = 4$ while varying $A_y$. For small $A_y$, the body is mostly rotating, and the motion is stable for all $(\phi_y , \phi_\theta)$ but with no net locomotion. As $A_y$ increases, unstable regions start to form around $(n\pi ,  (m+1/2)\pi)$, as can be seen in Figure~\ref{fig:varyay}(c). As $A_y$ continues to increase, unstable regions grow while stable regions shrink. Then, layers of stable/unstable regions start to form around $(n\pi , (m+1/2)\pi)$. Unlike in Figure~\ref{fig:varygamma}, areas around $((m+1/2) \pi , n\pi)$ do not remain stable. Overall, the total area of stable regions decreases as $A_y$ becomes larger. Interestingly, the area and shape of the stable regions depend nonlinearly on $A_y$ whereas the trajectory of the mass center depends {\em linearly} on $A_y$. 

In Figure~\ref{fig:varyatheta}, we vary $A_\theta$ while keeping $A_y = 1$ and $\gamma = 4$. When $A_\theta$ is small, the whole plane in $(\phi_y , \phi_\theta)$ is stable but again does not result in net locomotion. As $A_\theta$ increases, unstable regions start to emerge and grow, while stable regions shrink but persist around $(n\pi , (m+1/2)\pi)$, and as $A_\theta$ increases further, unstable regions start to emerge within the stable strips.

This trend of switching from stability to instability and back to stability when varying parameters is very interesting. It suggests that such swimmers can change their stability character by changing their flapping motion, and thus can easily switch from stable periodic swimming to an unstable motion (more maneuverable) when they feel the need to, such as when evading a predator. Based on this, one can conjecture that when it comes to live organisms, maneuverability and stability need not be thought of as disjoint properties, rather the organism may manipulate its motion in favor of one or the other depending on the task at hand. Whether live organisms change their stability properties at will is yet to be investigated experimentall. 

\section{Conclusions} 
\label{sec:discussion_and_conclusion}

We studied the locomotion, efficiency and stability of periodic swimming of fish using a simple planar model of an elliptic swimmer undergoing prescribed sinusoidal heaving and pitching in potential flow. We obtained expressions for the locomotion velocity for both small and finite flapping amplitudes, and showed how trajectories depend on key parameters, namely, aspect ratio $\gamma$, amplitudes $A_y$ and $A_\theta$ and phases $\phi_y$ and $\phi_\theta$. Efficiency is defined as the inverse of cost of locomotion $e$. The dependence of $e$ on the parameters were shown for both small and finite amplitude flappings. We observed that the efficiency maximizing parameters are approximately $3 \leq \gamma \leq 4, A_y \approx 1.3, A_\theta \approx \frac{3\pi}{16}, \phi_y \approx (m + \frac{1}{2})\pi$ and $\phi_\theta \approx n\pi$, where $n, m = 0 , \pm1 , \pm2 , \ldots$, whose values are in excellent agreement with results based on experimental and computational motions of flapping fish, see~\cite{Eloy2013,TrHoTeYe2005,KeKo2006} and references therein.

We then studied the stability of periodic locomotion using Floquet theory. To our best knowledge, besides the work of Weihs which uses approximate arguments, this is the first work that rigorously studies the stability of periodic locomotion albeit in a simplified model. We focused on evaluating stability on the whole $(\phi_y , \phi_\theta)$ parameter space, and examined the effect of varying $\gamma, A_y$ and $A_\theta$. We observed that stable and unstable regions in the $(\phi_y , \phi_\theta)$ plane evolved as these parameters change. Particularly noteworthy is the back and forth switching between stability and instability around the spots $((m+\frac{1}{2})\pi , n\pi)$ and $(n\pi , (m+\frac{1}{2})\pi)$.  This switching is reminiscent to the observation in~\cite{SpMoShZh2010} that the motion of a heaving and pitching foil switches from coherence to incoherence and back to coherence when varying the aspect ratio of foil. In our study, we found a similar behavior when varying not only the aspect ratio but also the flapping parameters.
 This indicates that such swimmer can change its stability character by changing its flapping motion, and thus can easily switch from stable periodic swimming to an unstable, yet more maneuverable, state. Based on this, one could conjecture that, when it comes to live organisms, maneuverability and stability are not disjoint properties but may be manipulated depending on the needs of the organism.  Clearly, this statement is speculative until verified by experimental evidence. To date, little is known experimentally on the stability of underwater periodic motions, let alone the stability of biological swimmers. 

Future extensions of this work will include the effects of body deformation and body elasticity, vortex shedding, and frequency of flapping on the observed stability of periodic swimming, as well as on motion efficiency such as in~\cite{JiAl2013}.



\appendix
\section*{Appendix} 
\label{sub:conformal_mapping}

\begin{figure}[!htbp]
	\centering
		\includegraphics[width=0.45\textwidth]{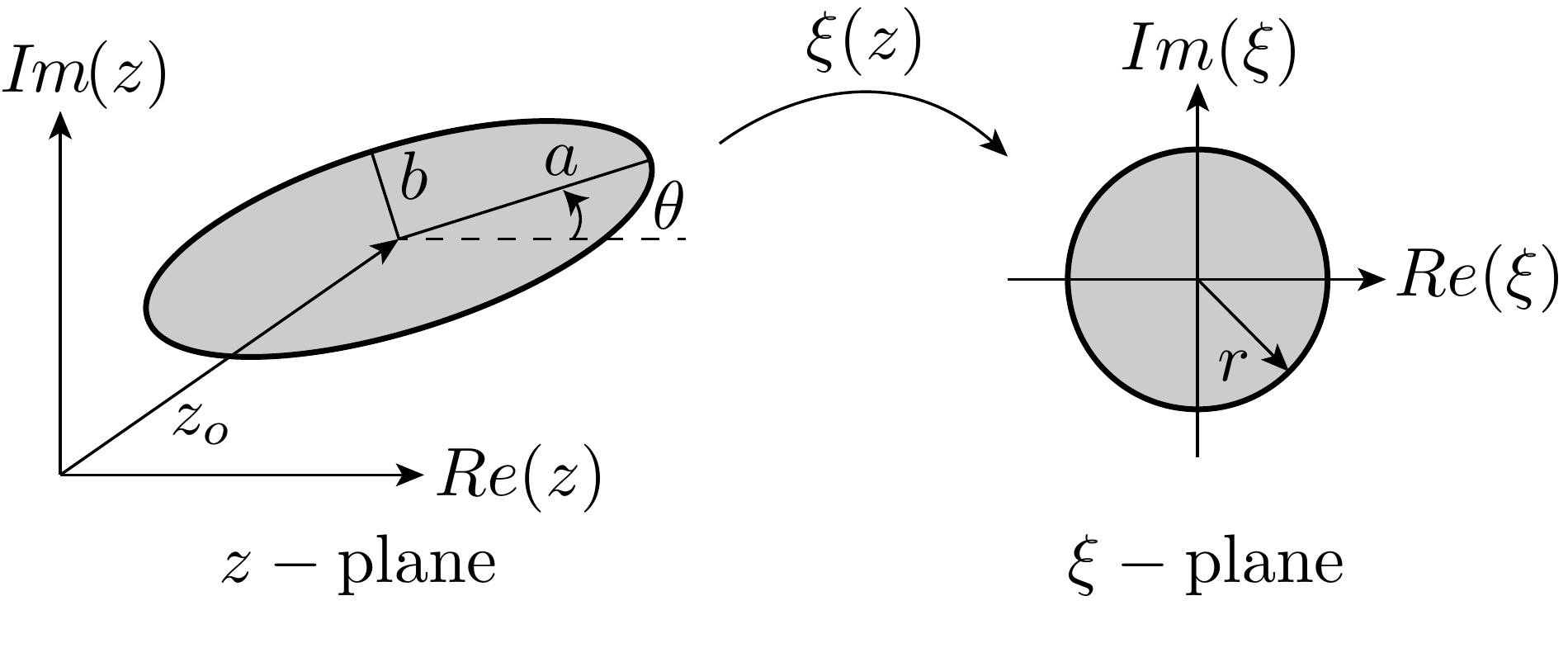}
	\caption{\footnotesize Exterior region of the ellipse of semi-axes $a$ and $b$ in the complex $z$-plane is mapped to the exterior region of a circle with radius $r = (a + b)/2$ in the $\xi$-plane. The mapping is given by~\eqref{eq:mapping}.}
	\label{fig:mapping}
\end{figure}

In potential flow, the fluid forces $F_x, F_y$ and moment $\tau$ can be obtained from the {\em added-mass theory}~\cite{Lamb1932} or from the {\em extended Blasius theorem}~\cite{Lamb1932, Sedov1965, MiTh1968}. In this appendix, we present both derivations and show their equivalence.

The exterior region of the ellipse in the complex $z$-plane ($z = x + i y$) is mapped to the exterior region of a circle with radius $r = (a + b)/2$ in the $\xi$-plane, see Figure~\ref{fig:mapping}. The mapping is given by
\begin{equation}
	z - z_o = \left( \xi + \frac{c^2}{\xi}\right) e^{i\theta}, \quad \text{or equivalently,} \quad \xi = \frac{(z - z_o)e^{-i\theta}}{2} + \frac{\sqrt{(z - z_o)^2 e^{-2i\theta} - 4 c^2}}{2},\label{eq:mapping}
\end{equation}
where $c = \sqrt{a^2 - b^2}/2$. The complex potential of the fluid in $\xi$-plane is given by~\cite{MiTh1968},
\begin{equation}
	W(\xi) = \frac{\overline{U} r^2 - U c^2}{\xi} - \frac{i \dot{\theta}r^2 c^2}{\xi^2},\label{eq:potential}
\end{equation}
where $U = -\dot{\bar{z}}_c e^{i\theta}$ is the velocity of the mass center mapped into the $\xi$-plane. Therefore, the forces and moment exerted by the surrounding fluid on a moving body are given by the {\em extended Blasius theorem}~\cite{Sedov1965}.
In $z$-plane,
\begin{equation}
	\begin{split}
		 F_x + iF_y & = \overline{\frac{i\rho}{2} \oint_{\partial\mathcal{B}} \left(\frac{\text{d}W}{\text{d}z}\right)^2 \text{d}z} + i\rho\frac{\text{d}}{\text{d}t}\left[\oint_{\partial\mathcal{B}} (z - z_o) \frac{\text{d}W}{\text{d}z} \text{d}z\right] + \rho A_{\mathcal{B}} \ddot{z}_o,\\[2ex]
		\tau & = \frac{\rho}{2} \text{Re}\left[ 2 \dot{\bar{z}}_o \oint_{\partial\mathcal{B}} (z - z_o) \frac{\text{d}W}{\text{d}z}\text{d}z - \oint_{\partial\mathcal{B}} (z-z_o)\left(\frac{\text{d}W}{\text{d}z}\right)^2\text{d}z + \frac{\text{d}}{\text{d}t}\left( \oint_{\partial\mathcal{B}} |z - z_o|^2 \frac{\text{d}W}{\text{d}z} \text{d}z\right)\right],
	\end{split}\label{eq:blasius}
\end{equation}
where $A_{\mathcal{B}} = \pi a b$ is the area of the ellipse, $\partial\mathcal{B}$ is the boundary of the body, and the reader is reminded that the densities of the body and fluid are both $\rho$. Notice that the last term in $\tau$ needs to be treated separately. All other integrals are analytic and, using residual theory, can be taken around an infinitely large circle instead of the boundary of the body, which greatly simplifies the calculations. For the last term in moment, since $|z - z_o|^2$ is not analytic, one cannot use this technique. Instead, it needs to be integrated on the boundary. Substituting~\eqref{eq:mapping} and \eqref{eq:potential} into~\eqref{eq:blasius}, one obtains the hydrodynamic forces and moment given by
\begin{equation}\label{eq:hydroforces}
	\begin{split}
		F_x & = \pi\rho \left(- \frac{r^4 + c^4}{r^2} + 2c^2 \cos2\theta\right)\ddot{x} + 2\pi\rho c^2 \ddot{y} \sin2\theta - 4\pi \rho c^2 \dot{\theta} (\dot{x} \sin 2 \theta - \dot{y}\cos 2\theta),\\
		F_y & = \pi\rho \left(- \frac{r^4 + c^4}{r^2} - 2c^2 \cos2\theta\right)\ddot{y} + 2\pi\rho c^2 \ddot{x} \sin2\theta  + 4\pi \rho c^2 \dot{\theta} (\dot{x} \cos 2 \theta + \dot{y} \sin 2\theta),\\
		\tau & = 2\pi \rho c^2 \left(\dot{x}^2\sin 2\theta - \dot{y}^2 \sin 2\theta - 2 \dot{x}\dot{y} \cos 2\theta -  c^2 \ddot{\theta} \right).
	\end{split}
\end{equation}
Since
\[
	2\pi\rho c^2 = \frac{m_2 - m_1}{2}, \quad \pi\rho\frac{r^4 + c^4}{r^2}  = \frac{m_1 + m_2}{2}, \quad 2\pi\rho c^4 = J,
\]
the hydrodynamical forcing terms are equivalent with the expressions given in~\eqref{eq:hydroforcestorque}, which is repeated here for completeness,
\[
	\begin{split}
		F_x & = \frac{1}{2}\left[ - (m_1+m_2) + (m_2-m_1) \cos2\theta\right]\ddot{x} + \frac{1}{2}(m_2-m_1) \ddot{y} \sin2\theta - (m_2-m_1)  (\dot{x} \sin 2 \theta - \dot{y}\cos 2\theta)\dot{\theta},\\[1ex]
		F_y & =   \frac{1}{2}\left[ -(m_1+m_2) - (m_2-m_1)  \cos2\theta\right]\ddot{y}  + \frac{1}{2}(m_2-m_1)  \ddot{x} \sin2\theta + (m_2-m_1)  (\dot{x} \cos 2 \theta + \dot{y} \sin 2\theta)\dot{\theta},\\[1ex]
		\tau & = -J \ddot{\theta}  + \frac{1}{2}(m_2- m_1) \left(\dot{x}^2\sin 2\theta - \dot{y}^2 \sin 2\theta - 2 \dot{x}\dot{y} \cos 2\theta \right).
	\end{split}
\]
And the governing equations are again repeated here
\begin{equation}
	m_b \ddot{x} = F_x, \quad m_b\ddot{y} = F_y + F^{\rm flap}, \quad J_b \ddot{\theta} = \tau + \tau^{\rm flap}.\label{eq:eom}
\end{equation}
When expressed in a body-fixed frame, the hydrodynamic forces and moment take a simpler form in terms of the {\em added mass coefficients} $m_1 , m_2$ and $J$. Roughly speaking, as a body moves through potential flow, the body-fluid system behaves as an augmented body with modified mass and inertia that account for the added mass and  added inertia due to the presence of the fluid. The added mass and inertia depend only on the geometry of the body and direction of motion. 
The Kirchhoff's equations of motion in terms of the body-fixed frame variables are given by
\begin{equation}
	\begin{split}
		(m_b + m_1) \dot{V}_1 & = -(m_b + m_2) V_2 \Omega + F_1,\\
		(m_b + m_2) \dot{V}_2 & = (m_b + m_1) V_1 \Omega + F_2,\\
		(J_b + J) \dot{\Omega} & = (m_1 - m_2) V_1 V_2 + \tau^{\rm flap},
	\end{split}\label{eq:kirchhoff}
\end{equation}
where the body frame velocities and forces are given by
\[
\begin{pmatrix}
	V_1\\
	V_2
\end{pmatrix} = \begin{pmatrix}
	\cos\theta & \sin\theta\\
	-\sin\theta & \cos\theta
\end{pmatrix}
\begin{pmatrix}
	\dot{x}\\
	\dot{y}
\end{pmatrix}, \quad \Omega = \dot{\theta}, \quad \text{and} \quad
\begin{pmatrix}
	F_1\\
	F_2
\end{pmatrix} = \begin{pmatrix}
	\cos\theta & \sin\theta\\
	-\sin\theta & \cos\theta
\end{pmatrix}
\begin{pmatrix}
	F_x\\
	F_y
\end{pmatrix},
\]
and $\tau$ is the same form in inertial frame. Transforming~\eqref{eq:kirchhoff} via a rotation $\theta$ to inertial frame, one obtains the equations given in~\eqref{eq:periodiceomX} and~\eqref{eq:periodiceom}, and the hydrodynamical forcing terms are given by~\eqref{eq:hydroforcestorque}. It is then straightforward to verify that~\eqref{eq:eom} and~\eqref{eq:kirchhoff} are equivalent.


For completeness, we rewrite equation~\eqref{eq:rearrangeeom}
\[
	\mathbb{M}(\theta)\dot{\mathbf{q}} = \mathbf{f}(\mathbf{q}) + \mathbf{F}^{\rm flap},
\]
where
\begin{equation}
	\mathbb{M}(\theta)= \begin{pmatrix}
		\dfrac{r^2}{c^2} - \cos2\theta & -\sin2\theta & 0 & 0\\
		-\sin2\theta & \dfrac{r^2}{c^2} + \cos2\theta & 0 & 0\\
		0 & 0 & 1 & 0\\
		0 & 0 & 0 & \dfrac{r^8 - c^8 + 4r^4c^4}{4 r^4 c^2}
	\end{pmatrix},
\end{equation}
and
\begin{equation}
	\mathbf{f}(\mathbf{q}) = \begin{pmatrix}
			-2\dot{\theta}\dot{x} \sin2\theta + 2 \dot{\theta}\dot{y} \cos2\theta\\[1.2ex]
			2\dot{\theta}\dot{x} \cos2\theta + 2 \dot{\theta}\dot{y} \sin2\theta\\[1.2ex]
			\dot{\theta}\\[1.2ex]
			(\dot{x}^2 - \dot{y}^2) \sin2\theta - 2\dot{x}\dot{y}\cos2\theta
		\end{pmatrix}, \qquad \mathbf{F}^{\rm flap} = \dfrac{1}{2\rho\pi c^2}
		\begin{pmatrix}
			0\\[1.2ex]
			F^{\rm flap}\\[1.2ex]
			0\\[1.2ex]
			\tau^{\rm flap}
		\end{pmatrix}.
\end{equation}
These equations can be rewritten as
\begin{equation}
	\dot{\mathbf{q}} = \mathbf{g} \equiv \mathbb{M}^{-1}(\mathbf{f} + \mathbf{F}^{\rm flap}).
\end{equation}
One can linearize above equation and obtain
\[
	\delta \dot{\mathbf{q}} = \mathbb{J}(t) \delta \mathbf{q},\qquad \text{where} \qquad \mathbb{J}(t)  = \left.\frac{\partial \mathbf{g}}{\partial \mathbf{q}}\right\vert_{(\mathbf{q}_p, \mathbf{F}^{\rm flap})}.
\]
The entries of the Jacobian $\mathbb{J}(t)$ are given by
\begin{equation}
		\mathbb{J}(t)  = \begin{pmatrix}
			\dfrac{2 r^2 c^2 \dot{\theta}\sin2\theta }{-r^4 + c^4} & \dfrac{2 c^2 \dot{\theta}(c^2 + r^2 \cos2\theta) }{r^4 - c^4} & \mathbb{J}_{13} & \mathbb{J}_{14}\\[1.5ex]
			\dfrac{2 c^2 \dot{\theta}(c^2 - r^2 \cos2\theta)}{-r^4 + c^4} & \dfrac{2 r^2 c^2 \dot{\theta}\sin2\theta }{r^4 - c^4} & \mathbb{J}_{23} & \mathbb{J}_{24}\\[1.3ex]
			0 & 0 & 0 & 1\\[1.3ex]
			\mu\alpha & -\mu\beta & \mu\left[(\dot{x}^2 - \dot{y}^2)\cos2\theta + 2\dot{x}\dot{y}\sin2\theta\right] & 0
		\end{pmatrix},
\end{equation}
where
\[
\begin{split}
	\alpha & = \dot{x} \sin2\theta  - \dot{y} \cos2\theta , \quad \beta = \dot{y} \sin2\theta  + \dot{x} \cos2\theta , \quad \mu = 8 r^4 c^2 /(r^8 - c^8 + 4 r^4 c^4),\\
	\mathbb{J}_{13} & = \dfrac{c^2}{\rho\pi (r^4 - c^4)}\left[F^{\rm flap} \cos2\theta - 4 r^2 \rho \pi \beta \dot{\theta}\right], \quad \mathbb{J}_{14} = \frac{2c^4}{r^4 - c^4}\left[-\dot{x} r^2 \sin2\theta + \dot{y} (c^2 + r^2\cos2\theta) \right],\\
	\mathbb{J}_{23} & = \dfrac{c^2}{\rho\pi (r^4 - c^4)}\left[F^{\rm flap} \sin2\theta - 4 r^2 \rho \pi \alpha \dot{\theta}\right], \quad \mathbb{J}_{24} = \frac{2c^4}{r^4 - c^4}\left[\dot{x} (c^2 - r^2\cos2\theta) - \dot{y} r^2 \sin2\theta\right].
\end{split}
\]


\section*{Acknowledgements}
The authors would like to thank Dr. Andrew A. Tchieu and Professor Paul K. Newton for the enlightening discussions. The work of EK is partially supported by the National Science Foundation through the CAREER award CMMI 06-44925 and the grant CCF 08-11480.



\begin{thebibliography}{99}
	
\bibitem{Lighthill1970}
Lighthill MJ (1970) Aquatic animal propulsion of high hydromechanical efficiency. {\em J. Fluid Mech.}, 44({\bf 2}):265--301.
	
\bibitem{Wu2011}
Wu TY (2011) Fish swimming and bird/insect flight. {\em Annu. Rev. Fluid Mech.}, 43({\bf 1}):25–-58.
	
\bibitem{Eloy2013}
Eloy C (2013) On the best design for undulatory swimming. {\em J. Fluid Mech.}, {\bf 717}:48--89.
	
\bibitem{Weihs2002}
Weihs D (2002) Stability versus maneuverability in aquatic locomotion. {\em Integ. and Comp. Biol.},
42({\bf 1}):127--134.

\bibitem{Weihs1993}
Weihs D (1993) Stability of aquatic animal locomotion. {\em Cont. Math.}, {\bf 141}:443--461.

\bibitem{JoSm2007}
Jordan DW, Smith P (2007) {\em Nonlinear ordinary differential equations: an introduction to dynamical systems. (4th ed.)} Oxford Univ. Press, New York.

\bibitem{KaMaRoMe2005}
Kanso E, Marsden JE, Rowley CW, Melli-Huber JB (2005) Locomotion of articulated bodies in a perfect fluid. {\em J. Nonlinear Sci.}, {\bf 15}:255–-289.

\bibitem{Jing2011}
Jing F (2011) {\em Part I-Viscous evolution of point vortex equilibria, Part II-Effects of body elasticity on stability of fish motion.} PhD thesis, University of Southern California, Los Angeles.

\bibitem{JiKa2012}
Jing F, Kanso E (2012) Effects of body elasticity on stability of underwater locomotion. {\em J. Fluid Mech.}, {\bf 690}:461--473.

\bibitem{SpMoShZh2010}
Spagnolie SE, Moret L, Shelley MJ, Zhang J (2010) Surprising behaviors in flapping locomotion with passive pitching. {\em Phys. Fluids}, {\bf 22}:041903.

\bibitem{Newman1977}
Newman JN (1977) {\em Marine hydrodynamics.} The MIT press, Cambridge, MA.

\bibitem{KeKo2006}
Kern S, Koumoutsakos P (2006) Simulations of optimized anguilliform swimming. {\em J Exp. Biol.}, {\bf 209}:4841--4857.

\bibitem{JaLa1995}
Jayne BC, Lauder GV (1995) Red muscle motor patterns during steady swimming in largemouth bass: Effects of speed and correlations with axial kinematics. {\em J. Exp. Biol.}, {\bf 198}:1575--1587.

\bibitem{DoDi2000}
Donley JM, Dickson KA (2000) Swimming kinematics of juvenile kawakawa tuna (Euthynnus affinis) and chub mackerel (Scomber japonicus). {\em J. Exp. Biol.} {\bf 203}:3103--3116.

\bibitem{ViHe1984}
Videler JJ, Hess F (1984) Fast continuous swimming of two pelagic predators, saithe (Pollachius virens) and mackerel (Scomber scombrus): A kinematic analysis. {\em J. Exp. Biol.} {\bf 109}:209--228.

\bibitem{TrHoTeYe2005}
Triantafyllou MS, Hover FS, Techet AH, Yue DKP (2005) Review of hydrodynamic scaling laws in aquatic locomotion and fishlike swimming. {\em Appl. Mech. Rev.} 58({\bf 4}):226--237.

\bibitem{JiAl2013}
Jing F, Alben S (2013) Optimization of two- and three-link snakelike locomotion. {\em Phys. Rev. E} {\bf 87}:022711.

\bibitem{Krasny1986}
Krasny R (1986) A study of singularity formation in a vortex sheet by the point-vortex approximation. {\em J. Fluid Mech.} {\bf 167}:65--93.

\bibitem{BrMi1954}
Brown CE, Michael WH (1954) Effect of leading edge separation on the lift of a delta wing. {\em J. Aero. Sci.} 21({\bf 10}):690--694.

\bibitem{Lamb1932}
Lamb H (1932) {\em Hydrodynamics (6th ed.)} Cambridge Univ. Press, Cambridge.

\bibitem{Sedov1965}
Sedov LI (1965) {\em Two-dimensional problems in hydrodynamics and aerodynamics. (ed. Chu CK, Cohen H, Seckler B, Gillis J)} Interscience Publishers, New York.

\bibitem{MiTh1968}
Milne-Thomson LM (1968) {\em Theoretical hydrodynamics.} Dover Publications, New York.
	
\end{thebibliography}
\end{document}